\documentclass[11pt]{article}

\usepackage{latexsym}
\usepackage{amsmath}
\usepackage{amssymb}
\usepackage{float}
\usepackage{graphicx}

\def\beq{\begin{eqnarray}}
\def\eeq{\end{eqnarray}}

\def\calA{{\cal A}}
\def\calB{{\cal B}}

\def\calX{{\cal X}}

\newfloat{algorithm}{tbph}{log} \floatname{algorithm}{Algorithm}

\begin{document}

\fontsize{11}{14.5pt}\selectfont

\vspace*{12pt}

\begin{center} \Large \bf 
 Modifying Gibbs Sampling to Avoid Self Transitions
\end{center}

\vspace{2pt}

\begin{center} 
  {\Large Radford M. Neal} \\[4pt]
  University of Toronto, Dept.\ of Statistical Sciences \\
  \texttt{https://glizen.com/radfordneal}\\
  \texttt{radford@utstat.utoronto.ca}\\[5pt]
  26 March 2024
\end{center}

\vspace{11pt}

\noindent {\bf Abstract.}  Gibbs sampling is a popular Markov chain
Monte Carlo method that samples from a distribution on $n$ state
variables by repeatedly sampling from the conditional distribution of
one variable, $x_i$, given the other variables, $x_{-i}$, either
choosing $i$ randomly, or updating sequentially using some systematic
or random order for $i$.  When $x_i$ is discrete, a Gibbs sampling
update may choose a new value that is the same as the old value.  A
theorem of Peskun indicates that, when $i$ is chosen randomly, a
reversible method that reduces the probability of such self
transitions, while increasing the probabilities of transitioning to
each of the other values, will decrease the asymptotic variance of
estimates of expectations of functions of the state.  This has
inspired two modified Gibbs sampling methods, originally due to
Frigessi, Hwang, and Younes and to Liu, though these do not always
reduce self transitions to the minimum possible.  Methods that do
reduce the probability of self transitions to the minimum, but do not
satisfy the conditions of Peskun's theorem, have also been devised, by
Suwa and Todo, some of which are reversible and some not.  I review
and relate these past methods, and introduce a broader class of
reversible methods, including that of Frigessi, \textit{et al.}, based
on what I call ``antithetic modification'', which also reduce
asymptotic variance compared to Gibbs sampling, even when not
satisfying the conditions of Peskun's theorem.  A modification of one
method in this class, which I denote as ZDNAM, reduces
self transitions to the minimum possible, while still always reducing
asymptotic variance compared to Gibbs sampling. I introduce another
new class of non-reversible methods based on slice sampling that can
also minimize self transition probabilities.  I provide explicit,
efficient implementations of all these methods, and compare the
performance of Gibbs sampling and these modified Gibbs sampling
methods in simulations of a 2D Potts model, a Bayesian mixture model,
and a belief network with unobserved variables.  The assessments look
at both random selection of $i$, and several sequential update
schemes.  Sequential updates using methods that minimize self
transition probabilities are found to usually be superior, with ZDNAM
often performing best.  There is evidence that the non-reversibility
produced by sequential updating can be beneficial, but no consistent
benefit is seen from the individual updates being done by a
non-reversible method.

\section{\hspace*{-8pt}
  Introduction}\vspace{-11pt}

Gibbs sampling has for some time been widely used to sample from
complex probability distributions in statistics and machine learning
(Geman and Geman 1984; Ackley, Hinton, and Sejnowski 1985; Gelfand and
Smith 1990; Thomas, Spiegelhalter, and Gilks 1992), and has been used
in statistical physics (where it is often called ``Glauber dynamics''
or the ``heatbath'' method) since long before that (see Landau and
Binder (2009) for a review).  Gibbs sampling is easy to implement in
many contexts, and has no adjustable parameters that need tuning.  In
some applications, Gibbs sampling is used alone, but it is also often
combined with other Markov chain Monte Carlo (MCMC) methods, either by
alternating between Gibbs sampling and other updates, or by using
Gibbs sampling for a subset of variables for which it is
well-suited, and other update methods for the remaining variables.
Gibbs sampling or its modifications can also be a component of more
elaborate sampling schemes, such as those aimed at exploitation of
parallel computation (Tjelmeland 2004), or avoidance of backtracking
(Neal 2004).  Improvements to Gibbs sampling that require no additional
computational capabilities are therefore of considerable interest.

I will review several ways of modifying Gibbs sampling to reduce the
probability that a transition leaves the state the same as before, and
introduce two new classes of such methods.  Some of these methods can
be shown to always produce better estimates that Gibbs sampling when
the variable to be updated is chosen randomly, using results discussed
in detail in a companion theoretical paper (Neal and Rosenthal 2023).
I show empirically that these methods can also improve MCMC estimates
in other contexts, such as sequential updating of variables, and that
the methods that reduce self transitions to the minimum possible
generally perform best.

\section{\hspace*{-8pt}
  Review of Gibbs Sampling (GS) and its implementation}\vspace{-11pt}

Gibbs sampling and other Markov chain Monte Carlo methods aim to
sample from some probability distribution, $\pi(x)$, on a state space,
$\calX$, by repeatedly applying updates to the state, each of which
leaves $\pi$ invariant, and eventually converge to $\pi$ regardless of
the initial state.  Gibbs sampling is applicable when the state is
naturally seen as consisting of $n$ variables, with $\calX$ written
as $\calX_1 \times \cdots \times \calX_n$.  A single Gibbs
sampling update of the state $x$ consists of choosing an index $i \in
\{1,\ldots,n\}$ and then replacing $x_i$ with a value drawn from the
conditional distribution for $x_i$ given the remaining variables
\mbox{(denoted as $x_{-i}$).}

I will write $P(x \rightarrow x')$ for the probability that the Markov
chain transitions to state $x'$ when in state $x$.  When only variable
$i$ is updated, I will write $P(x_i \rightarrow x'_i|x_{-i})$ for the
probability of transitioning from the state with $x_i$ to that with
$x'_i$, given that the other variables have values $x_{-i}$ (which
remain unchanged).  For Gibbs sampling, $P(x_i \rightarrow
x'_i|x_{-i})$ is simply $\pi(x'_i|x_{-i})$.  Note that when $\calX$
or $\calX_i$ is finite, it will sometimes be convenient to put the
transition probabilities in a matrix, $P$, with $P_{uv}= P(u
\rightarrow v)$.

A Gibbs sampling update for a single variable, $i$,
is \textit{reversible}, meaning that\vspace{-1pt}
\beq
  \pi(x_i|x_{-i}) P(x_i \rightarrow x'_i|x_{-i}) & = &
    \pi(x'_i|x_{-i}) P(x'_i \rightarrow x_i|x_{-i})\label{eq-rev}
\eeq
as is easily seen by substituting $P(x_i \rightarrow
x'_i|x_{-i})=\pi(x'_i|x_{-i})$.  This reversibility condition is
sufficient (but not necessary) for the update to leave the
conditional distribution \textit{invariant}:
\beq
  \pi(x'_i|x_{-i}) & = & \sum_{x_i\in \calX_i} \pi(x_i|x_{-i})
                             P(x_i \rightarrow x'_i|x_{-i})
\eeq
Since a Gibbs sampling update for variable $i$ does not change $x_{-i}$,
invariance of this conditional distribution
implies invariance of $\pi$ as a whole --- that is,\vspace{-1pt}
\beq
  \pi(x') & = & \sum_{x\in\calX} \pi(x) P(x \rightarrow x')
\eeq

One way to use single-variable Gibbs sampling updates to sample for
the full state is to
randomly select a variable to update each iteration, from some
distribution for $i$ on $\{1,\ldots,n\}$, here assumed uniform.
This produces a chain with the following transition
probabilities:\vspace{-6pt}
\beq
  P(x \rightarrow x') & = & {1 \over n}\, \sum_{i=1}^n\, 
     I(x'_{-i}=x_{-i})\ \pi(x'_i|x_{-i}) \label{eq-rs}
\eeq
(where $I(\cdot)$ is 1 if the enclosed condition is true, and 0 otherwise).
These transitions are easily seen to also be reversible, and hence leave $\pi$
invariant.  However, a disadvantage of this strategy is that some variables
might, by chance, not be updated for a considerable time.

A more common strategy is to perform Gibbs sampling updates for
variables in sequence, with $i$ going from $1$ to $n$.  Since each
such update leaves $\pi$ invariant, the sequence of updates will also
leave $\pi$ invariant, and hence be a suitable Markov chain Monte
Carlo method.  However, such a sequence of updates is not, in general,
reversible.  It can be made reversible by, in each such series of
updates, going through the variables in an order that is randomly
chosen for that series, from a distribution in which any update order
and its reverse are equally likely.  Note, however, that this reversible 
version is not necessarily better than a non-reversible sequence of updates.

For Gibbs sampling to be feasible, it must be possible to sample from
the conditional distribution $\pi(x_i|x_{-i})$ with a reasonable
amount of computation.  This is sometimes possible when $x_i$ is
continuous, or discrete with an infinite number of possible values,
and has a form amenable to sampling.  When $x_i$ takes values from a
finite set, that is not enormous, Gibbs sampling will be feasible as
long as $\pi(x_i|x_{-i})$ can be computed for all $x_i \in \calX_i$,
after which sampling a particular $x_i$ according to these
probabilities can be done in a straightforward way (Devroye 1986, page 85).

In some applications, such as the Potts model, $\pi(x_i|x_{-i})$
depends on $x_{-i}$ only through a function, $e(x_{-i})$, that has
a small number of possible values.  In this case, tables of
conditional probabilities for $x_i$ for all possible values of $e(x_{-i})$
can be pre-computed once, before simulating the Markov chain.  Using the
``alias method'' (Devroye 1986, page 107), tables can then be pre-computed
that allow for sampling from each of these conditional distributions in
time that is independent of the number of possible values for $x_i$.

The probabilities used for Gibbs sampling will usually have a
floating-point representation, and may have been computed with some
round-off error.  I will assume that these probabilities are
guaranteed to be in the interval $[0,1]$, and that their sum is very
close to one, but not necessarily exactly one.  This will be the case
when, as often, these probabilities are first computed in unnormalized
form (guaranteed to be non-negative, and not all zero), and then are
all divided by their (possibly inexact) sum.  I assume the methods for
sampling discussed above can handle probabilities with these
characteristics.  The detailed algorithms for modified Gibbs sampling
that I present here will in turn produce such transition probabilities.

Intuitively, it seems that Gibbs sampling can be inefficient, since it
is quite possible that when updating variable $i$ by sampling from
$\pi(x_i|x_{-i})$, the new value, $x'_i$, will be the same as the
current value, $x_i$.  Since we need the value of the state to move
around in order to explore the distribution, this seems sub-optimal.

Such self transitions are sometimes necessary.  If some value for
$x_i$ has conditional probability greater than $1/2$, this value must
sometimes remain unchanged if its frequency of occurrence
is to match its probability.  In particular, if some value has
probability $p>1/2$, the probability that a transition leaves this
value unchanged must be at least $(2p\!-\!1)\,/\,p > 0$ for
the transitions to leave the distribution invariant.\footnote{\rule{0pt}{10pt}%
Let $u$ be the value with $\pi(u)=p>1/2$.  Then invariance requires
that $p\ = \sum_v \pi(v) P(v\rightarrow u)\ =\ p\,P(u\rightarrow u)
\ +\ \sum_{v\ne u} \pi(v) P(v\rightarrow u)\ \le\ p\,P(u\rightarrow u)\ +\
\sum_{v\ne u} \pi(v)\ =\ p\,P(u\rightarrow u)\ +\ 1-p$, from which it
follows that $P(u \rightarrow u)\ \ge\ (2p\!-\!1)\,/\,p$.}

This paper looks at several methods for modifying Gibbs sampling to
reduce the probability of self transitions.  Some of these methods are
reversible, and never decrease non-self transition probabilities.
These methods can be justified as being superior to Gibbs sampling by
a theorem of Peskun (1973) showing that for such chains the intuition
that self transitions are inefficient is correct.  Some other
reversible methods reduce self transition probabilities and also
reduce some non-self transition probabilities, so Peskun's theorem
does not apply, but they can be justified as improvements to Gibbs
sampling using other theoretical tools (Neal and Rosenthal 2023).
Theoretical analysis of non-reversible methods is more difficult, but
as will be seen empirically, some non-reversible methods often perform
as well or better than reversible methods. However, the reversible
ZDNAM method that I introduce here is a good overall choice, when
employed with a non-reversible sequential updating schedule.

\section{\hspace*{-8pt}
  Asymptotic variance, Peskun-dominance, and efficiency-dominance
  }\label{sec-peskun}\vspace{-11pt}

One fundamental measure of efficiency of an MCMC method designed
to sample from a distribution $\pi$ is the \textit{asymptotic variance} of an 
estimate of the expectation with respect to $\pi$ of some function $f$,
found by averaging over states from the chain:\vspace{-5pt}
\beq
v(f,P) & = & \lim_{K\rightarrow\infty\rule{0pt}{8pt}} \!K\,
\mbox{Var}\Big({1 \over K} \sum_{t=1}^K f(x^{(t)}) \Big) \label{eq-asymvar}
\\[-12pt]\nonumber
\eeq 
Here, $x^{(t)}$ is the state after $t$ transitions of the chain with
transitions $P$, initialized at some state $x^{(0)}$.  When some
large number, $K$, of transitions are simulated, we expect
the variance of the average, $(1/K)\sum f(x^{(t)})$, which is an
estimate for the expectation of $f$, to be approximately $v(f,P)/K$.
(In practice, the early part of a chain is often discarded when estimating 
expectations, but this refinement does not affect the asymptotic variance, 
and will be ignored here.)

Lowering asymptotic variance is an important goal in designing a
Markov chain sampling method, but other criteria such as speed of
convergence to $\pi$ are also important, and can conflict with
minimizing asymptotic variance.\footnote{\rule{0pt}{10pt}%
As an extreme example, it is always easy to define and implement a chain 
that has zero asymptotic variance for estimating the expectation of any
function with respect to the uniform distribution on some easily enumerable 
set, by simply having the chain cycle deterministically through all elements of
this set.  But such a chain is of no practical use, since it gives accurate 
estimates only after having visited every value,
which is impractical for any problem where one would consider using MCMC.
However, examples of this sort are not possible with reversible chains,
which cannot be periodic with period greater than two.
}
One should note in particular that
independent sampling from $\pi$ --- that is, using transition
probabilities $P(x \rightarrow x') = \pi(x')$ --- results in immediate
convergence, but does not minimize asymptotic variance, since lower
variance can be obtained by ``antithetic'' sampling, in which the
transitions induce negative correlations between $f(x^{(t)})$ and
$f(x^{(t+\delta)})$ for some lags $\delta$.

It is not feasible to actually minimize asymptotic variance for the
problems with an enormous state space for which MCMC is used, just as
it is not practical to directly sample from a distribution on such a
state space.  But we can try to improve the asymptotic variance of
less direct methods, such as Gibbs sampling.  For this, we need to
know that an improvement to a component of the method --- such as
sampling a new value for a single variable, with other variables left
unchanged --- will result in an improvement to the overall method.
This is a main topic of a companion paper (Neal and Rosenthal 2023).

Previous work in this area has utilized a theorem of Peskun (1973),
who showed that if two chains with transitions $P$ and $P^{*}$ are
both reversible with respect to $\pi$, and $P^{*}(x \rightarrow x')\
\ge\ P(x \rightarrow x')$ for all $x \ne x'$, then $v(f,P^{*})\ \le\
v(f,P)$, for all functions $f$.

In other words, if a reversible chain is modified to reduce the
probability of some self transitions, and hence necessarily increase
the probability of some non-self transitions, while not reducing the
probability of any other non-self transitions, this will not increase
the asymptotic variance of the estimate for the expectation of any
function.  Typically, such a modification will reduce asymptotic
variance (with some exceptions, such as when the function is constant,
so the variance of the estimate is always zero).

I will say that $P^*$ \textit{Peksun-dominates} $P$ if $P^{*}(x
\rightarrow x')\ \ge\ P(x \rightarrow x')$ for all $x \ne x'$, and
that $P^*$ \textit{efficiency-dominates} $P$ if the asymptotic variance
of the estimate of the expectation of every function is at least as
small when using $P^*$ as when using $P$.  Peksun's theorem then says
that, for reversible chains, Peksun dominance implies efficiency
dominance.

Note that Peskun dominance is only a partial ordering --- it is
possible for two Markov chains to both be reversible with respect to
$\pi$ but for neither to Peskun-dominate the other, since for each
chain there is some non-self transition probability that is larger
than that for the other chain.  Similarly, efficiency-dominance is a
partial order over reversible chains (Neal and Rosenthal 2023, Theorem
10), but not a complete order.  Furthermore, when neither of two
chains Peskun-dominates (or efficiency-dominates) the other, it may be that no
other reversible chain Peskun-dominates (or efficiency-dominates) both of these
chains.\footnote{\rule{0pt}{10pt}%
Consider two chains with
transition probabilities shown below, both reversible with respect to
the uniform distribution on $\{1,2,3,4\}$:\vspace{-11pt}
\beq
\left[\begin{array}{cccc} 
  0 & 0 & { 1 \over 2 } & { 1 \over 2 } \\[3pt]
  0 & 0 & { 1 \over 2 } & { 1 \over 2 } \\[3pt]
{ 1 \over 2 } & { 1 \over 2 } & 0 & 0 \\[3pt]
{ 1 \over 2 } & { 1 \over 2 } & 0 & 0
\end{array}\right]
\ \ \ \ \ \ \ \ \ \
\left[\begin{array}{cccc} 
  0 & { 1 \over 2 } & 0 & { 1 \over 2 } \\[3pt]
{ 1 \over 2 } & 0 & { 1 \over 2 } & 0 \\[3pt]
  0 & { 1 \over 2 } & 0 & { 1 \over 2 } \\[3pt]
{ 1 \over 2 } & 0 & { 1 \over 2 } & 0 
\end{array}\right]\nonumber
\eeq
The first chain has zero asymptotic variance when estimating the expectation
for the function $I(x \in \{1,2\})$, but not for the function 
$I(x \in \{1,3\})$, while the reverse is true for the second chain.
No reversible chain has zero asymptotic variance for both 
functions (a consequence of the fact that periodic reversible chains
must have period two), and hence no reversible chain can 
Peskun-dominate, or efficiency-dominate, both.}

Suppose we modify the Gibbs sampling update for variable $x_i$, for
some particular values of the other variables, say when $x_{-i}=\bar
x_{-i}$, in a way that Peskun-dominates the Gibbs sampling update ---
that is, the probability of changing $x_i$ to any value other than its
current value is greater after the modification than it would be for
Gibbs sampling --- while also being reversible with respect to the
conditional distribution for $x_i$.  It is easy to see that a method
that randomly selects a variable to update, and uses this modified
method when variable $i$ is selected and $x_{-i}=\bar x_{-i}$, will
Peskun dominate Gibbs sampling with random selection of the variable
to update.  The probability of moving from $x$ to $x'$ will be at
least as large when variable $x_i$ is updated and $x_{-i}=\bar
x_{-i}$, and the same otherwise.  Accordingly, by Peskun's theorem,
the overall method using the modified update will efficiency-dominate
Gibbs sampling.  

We can go on to modify the updates for variable $i$ with other values
for $x_{-i}$, and to modify updates for variables other than $i$.  If
each of these local modifications Peskun-dominates Gibbs
sampling, then again the overall method (with random selection of
variable to update) will Peskun-dominate, and hence also
efficiency-dominate, Gibbs sampling.  Peskun dominance thus provides a
way of showing that local efficiency improvements, to updates of
single variables, lead to improvement in the efficiency of the overall
method --- at least, when the variable to be updated is chosen
randomly.

However, the converse of Peskun's theorem is not true.  As I will
discuss later, it is possible, with reversible $P^*$ and $P$, for
$P^*$ to efficiency-dominate $P$ even though some non-self transition
probabilities are greater for $P$ than for $P^*$.  This raises the
question of whether a modification to an update for a single variable
that efficiency-dominates Gibbs sampling, but does not Peskun-dominate
it, will always result in an efficiency improvement when used in an
overall method that randomly selects a variable to update.  A
companion paper (Neal and Rosenthal 2023) shows that this is true,
as will be discussed further later.

Peskun's theorem does not apply if the variables are updated by
a sequential scan for \mbox{$i = 1,\ldots n$}.  One reason is that
this (typically) results in a non-reversible chain (taking a full scan
to be one iteration of the chain).  If the order for updating
variables is randomly selected for each iteration, with any order and
its reverse equally probable, the resulting chain will be
reversible, but Peskun's theorem will still not apply, since it is
possible that a sequence of modified Gibbs sampling updates that
individually Peskun-dominate the corresponding unmodified Gibbs
sampling update will have lower probability of some transition to a
different state, even when this does not happen for any single
modified Gibbs sampling update.\footnote{\rule{0pt}{10pt}%
For example, consider when $\pi$ is uniform over $\calX =
\{0,1\}\times\{0,1\}$. Gibbs sampling updates for both variables give
equal probability to the values $0$ and $1$, and when applied in
either order, the probability of transitioning to any of the four
possible values is $1/4$.  Both Gibbs sampling updates can be modified
so that the value is changed with probability~1, and viewed
individually, these modifications satisfy the condition for Peskun's
theorem.  But when applied sequentially, in any order, even one chosen
at random, these modified updates have probability 0 of moving from
state $(0,0)$ to state $(0,1)$ or to state $(1,0)$, compared to
probability $1/4$ for the unmodified updates.  So Peskun's theorem
does not apply.\vspace*{-8pt}}

It is therefore only when the variable to be updated is selected
randomly that Peskun's theorem provides a guarantee that modifying
Gibbs sampling to increase the probabilities for all non-self
transitions will improve asymptotic variance.  When variables are
updated in sequence, as is the more common practice, Peskun's theorem
provides no guarantee that the modified method will be better.  One
cannot say in general whether updating variables sequentially is
better or worse than updating them at random, as is illustrated by He,
\textit{et al.}  (2016).  However, sequential updating, producing a
non-reversible chain, seems to usually work better in practical
applications, which reduces the practical relevance of Peskun's
theorem.

Nevertheless, Peskun's theorem provides a theoretical motivation for
looking at ways of reducing self transitions in Gibbs sampling.  I
will next describe one method of reducing self transitions for which
Peskun's theorem applies, based on a Metropolis-Hastings modification
of Gibbs sampling.  I will then introduce a more general framework for
improving the efficiency of Gibbs sampling, and discuss several
reversible methods derived in this way.  For some of these, Peskun's
theorem does not apply, but they can still be shown to
efficiency-dominate Gibbs sampling.

\section{\hspace*{-8pt}
  The Metropolis-Hastings Gibbs Sampling (MHGS) 
  method}\label{sec-MHGS}\vspace{-11pt}

Gibbs sampling can be seen as an instance of the Metropolis-Hastings
algorithm (Hastings 1970), in which transition probabilities from a
state $u$ are defined in terms of probabilities, $Q(u \rightarrow v)$, for
proposing to move from $u$ to a state $v$.  After sampling a $v$ according
to these probabilities, $v$ is accepted as the new state with 
probability\vspace{-12pt}
\beq
 \min\!\left(\!1,\ {\pi(v)\,Q(v \rightarrow u) \over \pi(u)\,Q(u \rightarrow v)}
       \right)
\eeq
If $v$ is not accepted, the new state is the same as the old state, $u$.
This transition is reversible with respect to $\pi$, and hence leaves
$\pi$ invariant.

A Gibbs sampling update for component $i$ of state $x$ is obtained by
proposing a new value for the state according to the conditional distribution 
for $x_i$ given the other components of the state.  That is,
\beq
  Q(x \rightarrow x') & = & I(x'_{-i}=x_{-i})\, \pi(x'_i|x_{-i})
\eeq
This proposal is always accepted, since for any proposed $x'$,
\beq
 \!\min\!\left(\!1,\, 
  {\pi(x')\,Q(x'\rightarrow x) \over \pi(x)\,Q(x\rightarrow x')} \right) \, =\,
 \min\!\left(\!1,\, 
  {\pi(x')\,\pi(x_i|x'_{-i}) \over \pi(x)\,\pi(x'_i|x_{-i})} \right) \, =\, 
 \min\!\left(\!1,\, 
  {\pi(x'_{-i})\,\pi(x'_i|x'_{-i})\,\pi(x_i|x'_{-i}) 
    \over \pi_(x_{-i})\,\pi(x_i|x_{-i})\,\pi(x'_i|x_{-i})} \right)
     \, = \, 1\!\!\!
\eeq\vspace{-14pt}

Liu (1996) introduced a method of modifying such a Gibbs sampling update 
by always proposing a value for $x_i$ that is different from the current value, 
with probabilities proportional to the conditional probabilities given
$x_{-i}$.  That is, the proposal probabilities are\vspace{-6pt}
\beq
  Q(x \rightarrow x') & = & I(x'_{-i}=x_{-i})\, I(x'_i \ne x_i)\
     {\pi(x'_i|x_{-i}) \over 1-\pi(x_i|x_{-i})} \label{MHGS-prop}
\eeq
The acceptance probability for such a proposal is\vspace{3pt}
\beq
 \min\!\left(\!1,\ 
  {\pi(x')\,\pi(x_i|x'_{-i})\,/\,(1-\pi(x'_i|x'_{-i}))
    \over \pi(x)\,\pi(x'_i|x_{-i})\,/\,(1-\pi(x_i|x_{-i}))} \right) & \!=\! &
 \min\!\left(\!1,\ 
  {\pi(x'_{-i})\,\pi(x'_i|x'_{-i})\,\pi(x_i|x'_{-i})\,(1-\pi(x_i|x_{-i})
    \over \pi_(x_{-i})\,\pi(x_i|x_{-i})\,\pi(x'_i|x_{-i})\,(1-\pi(x'_i|x'_{-i})}
   \right)\ \ \ \ \\[4pt]
 & \!=\! &
 \min\!\left(\!1,\ 
  {1-\pi(x_i|x_{-i}) \over 1-\pi(x'_i|x_{-i})} \right) \label{MHGS-accept}
  \\[-13pt] \nonumber
\eeq
Note that this is 1 whenever $\pi(x'_i|x_{-i}) \ge \pi(x_i|x_{-i})$.

These proposal and acceptance probabilities give the
following modified non-self transition probabilities:
\beq
  \mbox{when $x' \ne x$},\ \ P^{*}(x \rightarrow x') & = & I(x'_{-i}=x_{-i})\,
   {\pi(x'_i|x_{-i}) \over 1-\pi(x_i|x_{-i})}\,
   \min\!\left(\!1,\ {1-\pi(x_i|x_{-i}) \over 1-\pi(x'_i|x_{-i})} \right)
  \\[5pt]
  & = &
  I(x'_{-i}=x_{-i})\,
    \min\!\left(\!{\pi(x'_i|x_{-i}) \over 1-\pi(x_i|x_{-i})},
                 \ {\pi(x'_i|x_{-i}) \over 1-\pi(x'_i|x_{-i})} \right)
  \label{MHGS-nonself}
\eeq
The modified self transition probability, $P^*(x\rightarrow x)$,
can be found as one minus the sum of non-self transition probabilities 
from $x$.  The self transition probability also equals the
total probability of proposing a value that is not accepted:
\beq
  P^{*}(x \rightarrow x) & = & 
   \sum_{x'_i}\, I\Big(\pi(x'_i|x_{-i})<\pi(x_i|x_{-i}\Big)\
     {\pi(x'_i|x_{-i}) \over 1-\pi(x_i|x_{-i})}\,
     \left(\!1 - {1-\pi(x_i|x_{-i}) \over 1-\pi(x'_i|x_{-i})}\right)
  \label{MHGS-self}
\eeq
It follows that the self transition probability is zero when
$x_i$ has minimal conditional probability, since all proposals from such a 
value are accepted.  So at least one modified self transition
probability is zero.

I will refer to this method as Metropolis-Hastings Gibbs Sampling (MHGS).
Figure~\ref{MHGS-mod} shows an example of how MHGS modifies Gibbs
sampling transition probabilities.

\begin{figure}[t]

{\small
\[
\left[\begin{array}{cccc} 
{ 1 \over 10 } & { 2 \over 10 } & { 3 \over 10 } & { 4 \over 10 } \\[4pt]
{ 1 \over 10 } & { 2 \over 10 } & { 3 \over 10 } & { 4 \over 10 } \\[4pt]
{ 1 \over 10 } & { 2 \over 10 } & { 3 \over 10 } & { 4 \over 10 } \\[4pt]
{ 1 \over 10 } & { 2 \over 10 } & { 3 \over 10 } & { 4 \over 10 }
\end{array}\right]
\ \rightarrow\ 
\left[\begin{array}{cccc} 
{ 0 } & { 2 \over 9 } & { 3 \over 9 } & { 4 \over 9 } \\[4pt]
{ 1 \over 9 } & { 1 \over 72 } & { 3 \over 8 } & { 4 \over 8 } \\[4pt]
{ 1 \over 9 } & { 2 \over 8 } & { 34 \over 504 } & { 4 \over 7 } \\[4pt]
{ 1 \over 9 } & { 2 \over 8 } & { 3 \over 7 } & { 106 \over 504 }
\end{array}\right]
\]
\vspace{-5pt}
}

\caption{An illustration of how MHGS modifies Gibbs sampling 
         transition probabilities. The variable updated in this example
         has four possible values, whose conditional probabilities given
         the current values of other variables are
         1/10, 2/10, 3/10, and 4/10.}
        \label{MHGS-mod}

\end{figure}

As Liu notes, the non-self transition probabilities of
equation~(\ref{MHGS-nonself}) are clearly greater than those for Gibbs
sampling, which are $\pi(x'_i|x_{-i})$, so Peskun's theorem guarantees
that the asymptotic variance of estimates found using MHGS will be
lower than when using GS, when $i$ is selected randomly.

Liu's short paper does not discuss how to implement this method, but
there are two obvious ways.  

First, transitions can be simulated by computing all non-self
transition probabilities from the current value of the state using
equation~(\ref{MHGS-nonself}), then finding the self transition
probability as one minus the sum of these.\footnote{\rule{0pt}{10pt}%
Using equation~(\ref{MHGS-self}) is not recommended, as it may have high
relative error when $1-\pi(x_i|x_{-i})$ is close to zero.}
A new value
can then be sampled according to these probabilities, as discussed
earlier for Gibbs sampling.  This takes expected time asymptotically
proportional to $m$, the number of possible values for~$x_i$, if
probabilities need to be computed for each update, or takes constant time if
probabilities for all possible conditional distributions can be
pre-computed, and the alias method used.  Algorithm~\ref{alg-MHGS}
shows in detail how the needed transitions probabilities can be computed,
including some precautions for avoiding numerical issues.

\begin{algorithm}[t]

\begin{tabbing}

\hspace{1in}\=
\bf Input:\ \ \ \ \= Gibbs sampling probabilities, 
                   $\pi^{\rule{0pt}{1pt}}(i)$, for $i=1,\ldots,m$
\\ \>
                \> The current state value, $k$, in $\{1,\ldots,m\}$
\\[5pt] \>
\bf Output:    \> MHGS transition probabilities, $p(i)$, for $i=1,\ldots,m$
\\
\> \hspace*{20pt} \= \hspace*{20pt} \= \hspace*{20pt} \= \hspace*{20pt} \= 
   \\[-2pt]
\> If $1-\pi(i) \le 0$ for any $i$: \\[3pt]
\>\> \textit{Avoid division by zero by reverting to Gibbs sampling when a} \\
\>\> \textit{probability is 1 (or perhaps very close to 1)}\\[3pt]
\>\> For $i=1,\ldots,m$:\ \ Set $p(i)$ to $\pi(i)$ \\[2pt]
\> Else: \\[2pt]
\>\> \textit{Find non-self transition probabilities, and their sum} \\[3pt]
\>\> Set $s$ to 0 \\[2pt]
\>\> For $i=1,\ldots,m$: \\
\>\>\> If $i \ne k$: \\
\>\>\>\>Set $p(i)$ to 
        $\min\,(1,\ \pi(i)\,/\,(1-\pi(k)),\ \pi(i)\,/\,(1-\pi(i)))$
        \\
\>\>\>\>\ \ \ \ \ \ \ \
        \textit{The min with 1 above guards against error from rounding} \\
\>\>\>\> Add $p(i)$ to $s$ \\[4pt]
\>\> 
\textit{Set the self transition probability, 
        guarding against round-off error producing}
     \\
\>\> \textit{a negative probability (or change $1\!-\!s < 0$ to
             $1\!-\!s < \epsilon$ for some small $\epsilon$ to}
     \\
\>\> \textit{avoid producing tiny probabilities that should be exactly zero)} 
     \\[4pt]
\>\> If $1\!-\!s < 0$: \\
\>\>\> Set $p(k)$ to 0 \\
\>\> Else: \\
\>\>\> Set $p(k)$ to $1\!-\!s$
\end{tabbing}\vspace{6pt}

\caption{Computing MHGS transition probabilities, based on
 equation~(\ref{MHGS-nonself}).  Here, the algorithm is phrased in
 terms of a distribution, $\pi$, for a single variable, but in
 practice, it will be applied to the conditional distribution for one
 variable given values of the others, as would be sampled for
 unmodified Gibbs sampling.}\label{alg-MHGS}

\end{algorithm}

Alternatively, a transition can be simulated by first sampling a
proposal from the distribution defined by equation~(\ref{MHGS-prop}),
and then accepting this proposal with the probability given by
equation~(\ref{MHGS-accept}), or instead rejecting it and retaining
the current value.\footnote{\rule{0pt}{10pt}%
Note that rather than use
equation~(\ref{MHGS-prop}) as written, it may be better to replace the
expression $\pi(x'_i|x_{-i})\ /\ (1-\pi(x_i|x_{-i}))$ with
$\pi(x'_i|x_{-i})\ /\ \sum_{x'_i \ne x_i} \pi(x'_i|x_{-i})$, in order
to mitigate effects of round-off error when $\pi(x_i|x_{-i})$ is close
to 1.}  When Gibbs sampling probabilities have no useful structure,
this procedure also takes time asymptotically proportional to $m$,
since that much time is needed for the computation of $m\!-\!1$
proposal probabilities and their use in sampling of a proposal.

Sometimes, however, the conditional
probabilities given $x_{-i}$ have a form that allows for fast
sampling, which can be modified to sample a proposal according to
equation~(\ref{MHGS-prop}).  One possibility is when sampling can be
done by 
inverting the cumulative distribution function. For example,
suppose the Gibbs sampling probabilities are geometric($\theta$) on
$\{1,\ldots,m\}$, with cumulative distribution function
\beq
  F(a) & = & P(x'_i \le a) \ \ =\ \ { 1 - (1-\theta)^a \over 1 - (1-\theta)^m }
\label{eq-geom-cdf}
\eeq
Inverting the continuous form of this cumulative distribution function,
in which $a$ can be any non-negative real, allows
sampling from this distribution in constant time, independent of $m$
(Devroye 1986, page~87).  For this example,\vspace{-9pt}
\beq
  F^{-1}(u) & = & {\log(1 - u(1-(1-\theta)^m)) \over \log(1-\theta)}
\eeq
and we can generate a value as $\lceil F^{-1}(U) \rceil$, where $U$ is 
drawn from the uniform distribution on $(0,1)$.  

This efficient simulation method can be adapted to MHGS. Given a current 
value of $x_i$, the cumulative 
distribution function of a proposal $x'_i$ from equation~(\ref{MHGS-prop})
will be
\beq
F_{\mbox{\small \it prop}}(a) & = & \left\{\begin{array}{ll} 
     \displaystyle 
     { F(a) \over 1-F(x_i)+F(x_i\!-\!1)\rule{0pt}{10pt}} &
     \mbox{if $a \,<\, x_i\!-\!1$} \\[16pt]
     \displaystyle 
     { F(x_i\!-\!1) \over 1-F(x_i)+F(x_i\!-\!1)\rule{0pt}{10pt}} &
     \mbox{if $x_i\!-\!1 \,\le\, a \,<\, x_i$} \\[16pt]
     \displaystyle 
     { F(a)-F(x_i)+F(x_i\!-\!1) \over 1-F(x_i)+F(x_i\!-\!1)\rule{0pt}{10pt}} &
     \mbox{if $x_i\,\le\,a$}
   \end{array}\right.
\eeq
The corresponding inverse cumulative distribution function is\vspace{-2pt}
\beq 
F^{-1}_{\mbox{\small \it prop}}(u) & = & \left\{\begin{array}{ll}  
   F^{-1}(u\,(1-F(x_i)+F(x_i\!-\!1))
    & \mbox{if this is less than $x_i\!-\!1$}\ \ \ \ \\[8pt]
   F^{-1}(F(x_i)-F(x_i\!-\!1)+u\,(1-F(x_i)+F(x_i\!-\!1)))
    & \mbox{otherwise}
   \end{array}\right.\\[-10pt]\nonumber
\eeq
We can use this to generate a proposal as $\lceil F^{-1}_{\mbox{\small
\it prop}}(U) \rceil$, where $U$ is uniform on $(0,1)$, and then accept
or reject it according to equation~(\ref{MHGS-accept}).

Note that with this technique there is no problem with letting $m$ go
to infinity, to obtain a method that works for a variable with a
distribution on the positive integers.

More generally, if any method for efficient Gibbs sampling is
available (including for a variable with a countably infinite number
of possible values), it can be adapted to sample from the proposal
distribution of equation~(\ref{MHGS-prop}) by sampling repeatedly
until a value for $x'_i$ different from $x_i$ is obtained, which can
then be accepted or rejected according to
equation~(\ref{MHGS-accept}).  However, if the current value, $x_i$,
is such that $\pi(x_i|x_{-i})$ is close to one, a great many
repetitions might be required before a different value is obtained.
This inefficiency can be avoided by reverting to doing a standard
Gibbs sampling update if the maximum value of $\pi(x_i|x_{-i})$ for
all possible $x_i$ is close to one, which preserves reversibility
since this criterion does not depend on the current value. (Of course,
this will slightly increase the probability of a self transition.)

If the value with maximum probability is not easily identifiable, one
can use the following approach: First sample a value as for Gibbs
sampling, then test whether the probability of this value is in
$[\epsilon,\,1-\epsilon]$, for some $\epsilon$ close to zero.  If so,
repeatedly sample (discarding the value just tested) until a value
different from the current value is found, knowing that there is no
problematic value with probability greater than $1-\epsilon$.  If the
probability of the value tested is outside $[\epsilon,\,1-\epsilon]$,
instead revert to Gibbs sampling (again, discarding the value used
for the test).  The probability of reverting to Gibbs sampling if no
value has probability greater than $1-\epsilon$ will be less than
$m\epsilon$.

\section{\hspace*{-8pt}
  Efficiency improvement by Antithetic Modification (AM)
}\label{sec-AM}\vspace{-11pt}

A wide class of methods for modifying Gibbs sampling can be formulated
as applying a sequence of \textit{antithetic modifications} to the
original Gibbs sampling transition matrix.  All these modifications
can be shown to produce a chain that efficiency-dominates Gibbs
sampling, even though many do not Peskun-dominate it.

The concept of an antithetic modification can be applied
to any transition probabilities, $P$, on a finite state space, $\calX$,
that are reversible with respect to some distribution, $\pi$, on
$\calX$.  Two disjoint subsets of $\calX$, $\calA$ and $\calB$, with
$\pi(\calA)$ and $\pi(\calB)$ non-zero, are
specified, along with a value $\delta>0$ that controls the magnitude
of the modification, which must satisfy $P(a \rightarrow a') \ge
\delta \pi(a')\pi(\calB)/\pi(\calA)$ for all $a,a' \in \calA$, and
$P(b \rightarrow b') \ge \delta \pi(b')\pi(\calA)/\pi(\calB)$ for all
$b,b' \in \calB$.

The modification will alter only transitions to and from
values that are both in $\calA \cup \calB$,
with modified transition probabilities, $P^*$, as follows:
\beq\begin{array}{lclcll}
  P^*(a \rightarrow a')\! & = & P(a \rightarrow a')\!\!\! & - & 
                              \delta \pi(a') \pi(\calB) / \pi(\calA),\ \ \
          & \mbox{if $a \in \calA$ and $a' \in \calA$} \\[4pt]
  P^*(a \rightarrow b')\! & = & P(a \rightarrow b')\!\!\!   & + & 
                              \delta \pi(b'),
          & \mbox{if $a \in \calA$ and $b' \in \calB$} \\[4pt]
  P^*(b \rightarrow b')\! & = & P(b \rightarrow b')\!\!\! & - &
                              \delta \pi(b') \pi(\calA)/\pi(\calB),
          & \mbox{if $b \in \calB$ and $b' \in \calB$} \\[4pt]
  P^*(b \rightarrow a')\! & = & P(b \rightarrow a')\!\!\!   & + &
                              \delta \pi(a'),
          & \mbox{if $b \in \calB$ and $a' \in \calA$} \\[4pt]
  P^*(u \rightarrow v')\! & = & P(u \rightarrow v'),\!\!\!
       &&& \mbox{if $u \notin \calA \cup \calB$ or $v' \notin \calA \cup \calB$}
\end{array}\label{AM-def}\eeq
One can easily verify that the modified transitions probabilities from
each value are non-negative and sum to one, and that these transition
probabilities are reversible with respect to $\pi$.

If $P^*$ can be derived from $P$ by applying a sequence of zero or
more antithetic modifications, then I will say that $P^*$ \textit{is
antithetically derivable from} $P$.  It is easy to see that this is a
partial order, since the only non-trivial condition for this is
antisymmetry, which holds because an antithetic modification
always reduces some self transition probabilities, and never
increases any self transition probabilities, so it is not possible
for $P$ and $Q$ to be antithetically derivable from each other
unless they are equal.

Figure~\ref{am-mods} shows an example in which three
antithetic modifications are applied starting from an initial
transition probability matrix with all rows equal to $\pi$.

For any $P^*$ and $P$ that are reversible with respect to $\pi$, if $P^*$
Peskun-dominates $P$, then $P^*$ must also be antithetically derivable from
$P$ by a sequence of modifications in which $\calA$ and $\calB$ are
singleton sets.  If \mbox{$P^*(a \rightarrow b) > P(a \rightarrow b)$,} an
antithetic modification with $\calA=\{a\}$, $\calB=\{b\}$, and \mbox{$\delta
\ =\ (P^*(a \rightarrow b) - P(a \rightarrow b))\,/\, \pi(b)$} will change
the transition probabilities between $a$ and $b$ from those of $P$ to
those of $P^*$, without altering transition probabilities involving
values other than $a$ and $b$.  By a sequence of such modifications,
$P^*$ is antithetically derivable from~$P$.  

\begin{figure}[t]
\beq
  \!P\, = \left[ \begin{array}{cccc}
    {1\over2} & {1\over4} & {1\over6} & {1\over12} \\[4pt]
    {1\over2} & {1\over4} & {1\over6} & {1\over12} \\[4pt]
    {1\over2} & {1\over4} & {1\over6} & {1\over12} \\[4pt]
    {1\over2} & {1\over4} & {1\over6} & {1\over12}
  \end{array}\!\right] \rightarrow 
  \left[ \begin{array}{cccc}
    \mathbf{7\over24} & \mathbf{3\over8} & \mathbf{3\over12} & {1\over12}\\[4pt]
    \mathbf{3\over4} & \mathbf{1\over10} & \mathbf{1\over15} & {1\over12}\\[4pt]
    \mathbf{3\over4} & \mathbf{1\over10} & \mathbf{1\over15} & {1\over12}\\[4pt]
    {1\over2} & {1\over4} & {1\over6} & {1\over12}
  \end{array}\!\right] \rightarrow
  \left[ \begin{array}{cccc}
    {7\over24} & {3\over8} & {3\over12} & {1\over12}\\[4pt]
    {3\over4} & {1\over10} & {1\over15} & {1\over12}\\[4pt]
    {3\over4} & {1\over10} & \mathbf{1\over30} & \mathbf{7\over60}\\[4pt]
    {1\over2} & {1\over4} & \mathbf{7\over30} & \mathbf{1\over60}
  \end{array}\!\right] \rightarrow
  \left[ \begin{array}{cccc}
    \mathbf{283\over1176} & \mathbf{23\over56} 
                          & \mathbf{23\over84} & \mathbf{11\over147}
\\[4pt]
    \mathbf{23\over28}    & \mathbf{1\over20} 
                          & \mathbf{1\over30} & \mathbf{2\over21}
\\[4pt]
    \mathbf{23\over28}    & \mathbf{1\over20} 
                          & \mathbf{0} & \mathbf{27\over210}
\\[4pt]
    \mathbf{22\over49}    & \mathbf{2\over7} 
                          & \mathbf{27\over105} & \mathbf{6\over735}
  \end{array}\!\right] =\, P^*\ \ \
\nonumber\eeq
\hspace*{1.05in}$\calA=\{1\},\ \calB=\{2,3\}$ 
\hspace*{0.23in}$\calA=\{3\},\ \calB=\{4\}$
\hspace*{0.21in}$\calA=\{1,4\},\ \calB=\{2,3\}$
\\[3pt]
\hspace*{1.45in}$\delta={1\over2}$
\hspace*{1.21in}$\delta={2\over5}$
\hspace*{1.19in}$\delta={1\over7}$

\caption{Changes to a transition probability matrix through three successive
antithetic modifications.  On the left, $P$ has all rows equal to $\pi$.
Three antithetic modifications are then applied, with $\calA$, $\calB$, and
$\delta$ as shown, to obtain $P^*$.  For each modification, probabilities that
change are shown in bold.}\label{am-mods}

\end{figure}

However, an antithetic modification in which $\calA$ and/or $\calB$ have more
than one element can change $P$ to a $P^*$ that efficiency-dominates
$P$, but does not Peskun-dominate it.  The first modification in 
Figure~\ref{am-mods} provides an example:\ \ There, $P(2\rightarrow3)$ decreases
from $1/6$ to $1/15$ after the first modification with $\calA=\{1\}$
and $\calB=\{2,3\}$, while $P(1\rightarrow2)$ increases from $1/4$ to
$3/8$, so neither the original nor the modified transition matrix 
Peskun-dominates the other.

However, whenever $P^*$ is antithetically derivable from $P$ it \textit{does}
efficiency-dominate $P$.  This follows from Theorem~9 of the companion
paper (Neal and Rosenthal 2023), which states that if $P$ and $Q$ are
reversible irreducible Markov chains on a finite state space, then $P$
efficiency-dominates $Q$ if and only if the matrix $Q-P$ has only
non-negative eigenvalues. (See also (Mira and Geyer 1999)).

For a general antithetic modification, as defined by~(\ref{AM-def}),
the difference matrix, $P-P^*$, will be zero except for the submatrix
corresponding to states in $\calA\cup\calB$.  If we order states in
$\calA=\{a_1,a_2,\ldots\}$ before those in $\calB=\{b_1,b_2,\ldots\}$,
with any other states following, $P-P^*$ will look like this:\vspace{4pt}
\beq P-P^* & = &
  \left[\begin{array}{cccccccc}
    \delta\pi(a_1){\pi(\calB)\over\pi(\calA)} & 
    \delta\pi(a_2){\pi(\calB)\over\pi(\calA)} & \cdots &
    -\delta\pi(b_1) & -\delta\pi(b_2) & \cdots & 0 & \cdots\ \\[10pt]
    \delta\pi(a_1){\pi(\calB)\over\pi(\calA)} & 
    \delta\pi(a_2){\pi(\calB)\over\pi(\calA)} & \cdots &
    -\delta\pi(b_1) & -\delta\pi(b_2) & \cdots & 0 & \cdots\ \\[6pt]
    \vdots & \vdots & & \vdots & \vdots & & \vdots \\[6pt]
    -\delta\pi(a_1) & -\delta\pi(a_2) & \cdots &
    \delta\pi(b_1){\pi(\calA)\over\pi(\calB)} & 
    \delta\pi(b_2){\pi(\calA)\over\pi(\calB)} & \cdots & 0 & \cdots\ \\[10pt]
    -\delta\pi(a_1) & -\delta\pi(a_2) & \cdots &
    \delta\pi(b_1){\pi(\calA)\over\pi(\calB)} & 
    \delta\pi(b_2){\pi(\calA)\over\pi(\calB)} & \cdots & 0 & \cdots\ \\[6pt]
    \vdots & \vdots & & \vdots & \vdots & & \vdots \\[6pt]
    0 & 0 & \cdots & 0 & 0 & \cdots & 0 & \cdots\ \\[6pt]
    \vdots & \vdots & & \vdots & \vdots & & \vdots
  \end{array}\right]
\eeq
This is a rank-one matrix, since all rows are equal to the first row, or
equal the first row times $-\pi(\calA)/\pi(\calB)$, or are zero.  
If we let $D$ be the diagonal matrix with entries
$\pi(a_1), \pi(a_2), \ldots, \pi(b_1), \pi(b_2), \ldots, 0, \ldots$ on 
its diagonal, and let\vspace{-10pt}
\beq
   v & = & \Bigg[ \
    \sqrt{\pi(\calB)\over\pi(\calA)}\ \ \
    \sqrt{\pi(\calB)\over\pi(\calA)}\ \
    \cdots\ \
    -\sqrt{\pi(\calA)\over\pi(\calB)}\ \ \ 
    -\sqrt{\pi(\calA)\over\pi(\calB)}\ \
    \cdots\ \ \ 0\ \ \cdots\
   \Bigg]^T 
\eeq
then we can write\vspace{-11pt}
\beq 
   P-P^* & = & \delta\, v\, v^T D
\eeq
Since $v^TDv\ =\ 
  \sum_{a\in\calA} \pi(a) \pi(\calB)/\pi(\calA)\ +\ 
  \sum_{b\in\calB} \pi(b) \pi(\calA)/\pi(\calB)\ =\ \pi(\calA)+\pi(\calB)$, 
we see that
$(P-P^*)\,v \ =\ \delta\,(\pi(\calA)+\pi(\calB))\,v$, so 
$\delta\,(\pi(\calA)+\pi(\calB))$ is an eigenvalue of $P-P^*$, with
all other eigenvalues being zero (since $P-P^*$ is of rank one).
Since this eigenvalue is positive, Theorem 9 from (Neal and Rosenthal 2023)
shows that $P^*$ efficiency-dominates $P$.

Since efficiency-dominance is transitive, it follows that the result
of any sequence of antithetic modifications will efficiency-dominate the
original transition matrix.

A wide variety of improved methods can be derived using antithetic
modifications.  In this paper, I will focus on generic methods, in
which nothing is known that distinguishes one state from another,
except for their probabilities under $\pi$.  However, antithetic
modifications can also be designed in a way that exploits some known
structure of the state space as a guide to how to choose the subsets
$\calA$ and $\calB$.

For example, suppose it is beneficial for the value chosen from
$\calX=\{ 1,\ldots,m \}$ to be far from the current value.  If
$m=2^j$, we can try to flip from the current value to one in the other
half of $\calX$, which will on average be more distant than a value
chosen from all of $\calX$.  Failing that, we could try to flip from
the current value to one in the other quarter of the same half, and so
forth.  To do this, we can modify the probabilities for
independent sampling (all rows equal to $\pi$) by applying an antithetic
modification with
$\calA=\{1,\ldots,2^{j-1}\}$ and $\calB=\{2^{j-1}\!+1,\ldots,2^j\}$.
If $\pi(\calA)\ge\pi(\calB)$, we use $\delta=\pi(\calB)/\pi(\calA)$,
which results in all transition probabilities amongst values in
$\calB$ being zero.  Otherwise, we use $\delta=\pi(\calA)/\pi(\calB)$,
and all transition probabilities amongst values in $\calA$ will be
zero.  We then apply another antithetic modification, in the
first case using $\calA=\{1,\ldots,2^{j-2}\}$ and
$\calB=\{2^{j-2}\!+1,\ldots,2^{j-1}\}$, which partitions the previous
$\calA$, and in the second case using
$\calA=\{2^{j-1}\!+1,\ldots,2^{j-1}\!+2^{j-2}\}$ and
$\calB=\{2^{j-1}\!+2^{j-2}+1,2^j\}$, partitioning the previous
$\calB$, in both cases with $\delta$ chosen to make transition
probabilities within either $\calA$ or $\calB$ zero.  This continues
until $\calA$ and $\calB$ are singleton sets.

Here is an example with $m=4$:
\beq
  \left[ \begin{array}{cccc}
    {1\over4} & {3\over10} & {1\over5} & {1\over4} \\[4pt]
    {1\over4} & {3\over10} & {1\over5} & {1\over4} \\[4pt]
    {1\over4} & {3\over10} & {1\over5} & {1\over4} \\[4pt]
    {1\over4} & {3\over10} & {1\over5} & {1\over4}
  \end{array}\!\right] \rightarrow 
  \left[ \begin{array}{cccc}
    {10\over121} & {12\over121} & {4\over11} & {5\over11} \\[4pt]
    {10\over121} & {12\over121} & {4\over11} & {5\over11} \\[4pt]
    {5\over11} & {6\over11} & {0} & {0} \\[4pt]
    {5\over11} & {6\over11} & {0} & {0}
  \end{array}\!\right] \rightarrow
  \left[ \begin{array}{cccc}
    {0} & {2\over11} & {4\over11} & {5\over11} \\[4pt]
    {5\over33} & {1\over33} & {4\over11} & {5\over11} \\[4pt]
    {5\over11} & {6\over11} & {0} & {0} \\[4pt]
    {5\over11} & {6\over11} & {0} & {0}
  \end{array}\!\right]
\eeq
Note that only the row of this matrix for transition probabilities from 
the current value need be computed.

This procedure is equivalent to one used for the No-U-Turn Sampler by
Hoffman and Gelman (2014, Section 3.1.2) to select from amongst states
found by simulating a trajectory using Hamiltonian dynamics.

When the goal is to improve Gibbs sampling, antithetic modifications
can be applied to the Gibbs sampling transition matrix for updating a
particular variable, when other variables have particular values.
When the variable to be updated is selected randomly, each such
modification will improve the efficiency of the overall chain, and
hence so will a set of antithetic modifications for updates to every
variable, for every combination of values for other variables.

To see this is detail, suppose that there are two state variables,
so $\calX=\calX_1\times\calX_2$, with $X_1=\{1,2\}$ and $X_2=\{1,2,3\}$,
and that $\pi((1,1))=1/8$, $\pi((1,2))=1/4$, $\pi((1,3))=1/8$, $\pi((2,1))=1/4$,
$\pi((2,2))=1/8$, and $\pi((2,3))=1/8$.  With states ordered 
lexicographically (i.e., with $\calX_1$ changing more slowly), 
the transition matrices for Gibbs sampling updates of 
the first and second variables will be
\beq
  P_1\ =\ \left[\begin{array}{cccccc}
  1/3 & 0 & 0 & 2/3 & 0 & 0 \\
  0 & 2/3 & 0 & 0 & 1/3 & 0 \\
  0 & 0 & 1/2 & 0 & 0 & 1/2 \\
  1/3 & 0 & 0 & 2/3 & 0 & 0 \\
  0 & 2/3 & 0 & 0 & 1/3 & 0 \\
  0 & 0 & 1/2 & 0 & 0 & 1/2
  \end{array}\right],\ \ \ \ \
  P_2\ =\ \left[\begin{array}{cccccc}
  1/4 & 1/2 & 1/4 & 0 & 0 & 0 \\
  1/4 & 1/2 & 1/4 & 0 & 0 & 0 \\
  1/4 & 1/2 & 1/4 & 0 & 0 & 0 \\
  0 & 0 & 0 & 1/2 & 1/4 & 1/4 \\
  0 & 0 & 0 & 1/2 & 1/4 & 1/4 \\
  0 & 0 & 0 & 1/2 & 1/4 & 1/4
  \end{array}\right]\\[6pt]
\mbox{\small\ \ \ \ State order:\hspace{10pt}
(1,1)\hspace{2.8pt}
(1,2)\hspace{2.8pt}
(1,3)\hspace{2.8pt}
(2,1)\hspace{2.8pt}
(2,2)\hspace{2.8pt}
(2,3)\hspace{2.8pt}
\hspace{70pt}
(1,1)\hspace{2.8pt}
(1,2)\hspace{2.8pt}
(1,3)\hspace{2.8pt}
(2,1)\hspace{2.8pt}
(2,2)\hspace{2.8pt}
(2,3)\hspace{2.8pt}\ \ 
}\nonumber
\eeq
If the order of states were changed so that $\calX_2$ changed more
slowly, $P_1$ would change to
\beq
  \widetilde P_1 \ =\ \left[\begin{array}{cccccc}
  1/3 & 2/3 & 0 & 0 & 0 & 0 \\
  1/3 & 2/3 & 0 & 0 & 0 & 0 \\
  0 & 0 & 2/3 & 1/3 & 0 & 0 \\
  0 & 0 & 2/3 & 1/3 & 0 & 0 \\
  0 & 0 & 0 & 0 & 1/2 & 1/2 \\
  0 & 0 & 0 & 0 & 1/2 & 1/2
  \end{array}\right] \\[6pt]
\mbox{\small\ \ \ \ State order:\hspace{12pt}
(1,1)\hspace{2.8pt}
(2,1)\hspace{2.8pt}
(1,2)\hspace{2.8pt}
(2,2)\hspace{2.8pt}
(1,3)\hspace{2.8pt}
(2,3)\hspace{2.8pt}\ \
}\nonumber
\eeq
So, with a suitable order, both Gibbs sampling updates have block-diagonal
transition matrices, with each block being the transition matrix 
for an update of that one variable, which is reversible
with respect to the conditional distribution for that variable given
the current value of the other variable (of all other variables, when
there are more than two variables).  If the variable to update is selected
uniformly at random, the transition probability matrix for the entire
chain is $P\ =\ (1/2)(P_1+P_2)$. More generally, when there are $n$
variables, $P\ =\ (1/n)(P_1+\cdots+P_n)$.

We can apply an antithetic modification that affects only a single block,
in the update for one variable.  For example, applying equations~(\ref{AM-def}),
the block for updating the second variable in the above example when the
first variable has the value 2 can be antithetically modified with
$\calA=\{(2,1)\}$, $\calB=\{(2,2),(2,3)\}$, and $\delta=2$, giving the
following modified version of $P_2$:
\beq
  P^*_2\ =\ \left[\begin{array}{cccccc}
  1/4 & 1/2 & 1/4 & 0 & 0 & 0 \\
  1/4 & 1/2 & 1/4 & 0 & 0 & 0 \\
  1/4 & 1/2 & 1/4 & 0 & 0 & 0 \\
  0 & 0 & 0 & 0 & 1/2 & 1/2 \\
  0 & 0 & 0 & \ 1\  & 0 & 0 \\
  0 & 0 & 0 & \ 1\  & 0 & 0
  \end{array}\right]\\[6pt]
\mbox{\small State order:\hspace{12pt}
(1,1)\hspace{2.7pt}
(1,2)\hspace{2.7pt}
(1,3)\hspace{2.2pt}
(2,1)\hspace{2.2pt}
(2,2)\hspace{2.7pt}
(2,3)\hspace{2.7pt}\ \ 
}\nonumber
\eeq
Note that this can also be viewed as an antithetic modification to just the 
lower-right block, regarding it as a transition matrix that is
reversible with respect to a conditional distribution for that variable:
\beq
\left[\begin{array}{ccc}
  1/2 & 1/4 & 1/4 \\
  1/2 & 1/4 & 1/4 \\
  1/2 & 1/4 & 1/4 
\end{array}\right]\ \rightarrow\ 
\left[\begin{array}{ccc}
  0 & 1/2 & 1/2 \\
  1 & 0 & 0 \\
  1 & 0 & 0 
\end{array}\right]
\eeq
For this block modification, $\pi$ is the conditional distribution, 
with probabilities $1/2$, $1/4$, $1/4$, and \mbox{$\delta=1$.}

As discussed above, the eigenvalues of the difference between the
original transition matrix $P_2$ and the antithetically-modified
matrix $P^*_2$ will all be non-negative.  But we cannot conclude that
$P^*_2$ efficiency-dominates $P_2$, because neither of these are
irreducible --- on their own, they cannot move over the full state
space.  We \textit{can} conclude that the modified overall transition
matrix with random selection of variable to update,
$(1/2)(P_1+P^*_2)$, efficiency-dominates the original overall
transition matrix, $(1/2)(P_1+P_2)$, provided these are both
irreducible, by using Theorem~12 of (Neal and Rosenthal 2023).  More
generally, $P^*=(1/n)(P^*_1+\cdots+P^*_n)$ efficiency-dominates
$P=(1/n)(P_1+\cdots+P_n)$ if each of the differences
$P_k-P^*_k$ has only non-negative eigenvalues, provided $P$ and $P^*$
are irreducible and the $P_k$ and $P^*_k$ are reversible.

Accordingly, when antithetic modification is used to improve the
efficiency of individual Gibbs sampling updates, this improvement
extends to an overall method that randomly selects a variable to
update.  Note, however, that this guarantee does not apply when
variables are updated in some systematic order, even though, as will
be seen later in the empirical evaluations, this is often better than
random updates.

An antithetic modification may produce transition probabilities that
do not converge to $\pi$, but instead are periodic, flipping between
different distributions at even and odd iterations.  Seen in
isolation, averages from such an update will nevertheless be correct
estimates of expectations.  When such transitions are used to update
single variables in a Gibbs sampling framework, with the variable to
update chosen randomly, such exact periodicity is 
possible,\footnote{\rule{0pt}{10pt}%
Let $\pi$ be uniform over
$\calX=\{0,1\}\times\{0,1\}\times\{0,1\}$. There is an antithetic
modification of Gibbs sampling for each variable that flips the value
with probability one. With random selection of the variable to update,
the number of 1s will alternate in periodic fashion between an even
number and an odd number when this modified method is
used.\label{footp}} though rare, but averages will still be correct
even with periodicity.  When variables are updated in some systematic
order, rather than randomly, it is possible for periodicity of
individual updates to produce incorrect estimates.\footnote{\rule{0pt}{10pt}%
With the same example as in footnote~\ref{footp}, flipping values in 
a systematic
scan starting at state $(0,0,0)$ will produce the cycle $(0,0,0),\,(1,0,0),\,
(1,1,0),\,(1,1,1),\,(0,1,1),\,(0,0,1),\,(0,0,0),\,\ldots$, in which the
values $(0,1,0)$ and $(1,0,1)$ never appear.} Though
this problem is rare in practice, if necessary it can be avoided by
occasionally doing an unmodified Gibbs sampling update.

\section{\hspace*{-8pt}
  Nested Antithetic Modification (NAM) methods}\label{sec-NAM}\vspace{-11pt}

I will now look at methods in which a sequence of $m-1$ antithetic
modifications are applied to a transition matrix in which all rows are
the same, as for a Gibbs sampling update.  These antithetic
modifications will use subsets of states, $\calA_i$ and $\calB_i$, for
$i=1,\ldots,m\!-\!1$, in which each $\calA_i=\{a_i\}$ is a singleton
set contained in $\calB_{i-1}$ and $\calB_i =
\calB_{i-1}\!\setminus\calA_i$, with $\calB_0=\calX$.  I call these
\textit{nested antithetic modification (NAM)} methods, since
$\calX=\calB_0 \supset \calB_1 \supset \calB_2 \supset \cdots \supset
\calB_{m-1}$ are nested sets of states.  Different NAM methods result
from different ways of choosing which element of $\calB_{i-1}$ is
chosen as $a_i$ --- what I will call the \textit{focal value} for that
stage in the sequence.  For all these antithetic modifications, $\delta$
will be chosen to be as large as possible.

When focal values $a_1,a_2,\ldots,a_{m-1}$ are chosen to have non-decreasing
probability under $\pi$, the resulting NAM method turns out to be
equivalent to a method described by Frigessi, Hwang, and Younes
(1992), and later independently by Tjelmeland (2004).  In this case,
the modified transitions Peskun-dominate Gibbs sampling.  This is not
true for all NAM methods, but, as discussed in the previous section,
they, like all AM methods, do efficiency-dominate Gibbs sampling.

In this section, I look at NAM methods in general, for any selection
of $a_1,a_2,\ldots,a_{m-1}$, and present efficient implementations of
these methods.  For notational simplicity, I will describe how these
methods would be applied to modify transitions for the entire state,
but in practice they will modify Gibbs sampling probabilities for a
single state variable, with $\pi$ being the conditional distribution
for that variable given the current values of other variables.

I will present NAM methods assuming that the state space is $\calX =
\{1,\ldots,m\}$.  The choice of focal values can then be represented using a
permutation, $\sigma$, on $1,\ldots,m$, with $a_i = \sigma(i)$.  
The antithetic modifications will produce successive transition probability
matrices $P_0, P_1, \ldots, P_{m-1}$, where $P_0$ has all rows equal
to $\pi$ (i.e., the Gibbs sampling probabilities), and $P_i$ is the
result of applying an antithetic modification to $P_{i-1}$ with
$\calA_i=\{\sigma(i)\}$ and $\calB_i=\{ \sigma(j)\ :\ j=i\!+\!1,\ldots,m\}$.
In some cases, all modified transition probabilities, $P_i(b \rightarrow b')$,
for $b,b'\in\calB_i$ will be zero at stage $i$, in which case the
procedure is terminated at that point, with $P_i$ being the final
modified transition matrix.

The submatrix of $P_i$ with rows and columns in $\calB_i$ will always
have all rows the same, with row elements proportional to $\pi(b')$
for $b'\in\calB_i$.  This is obvious for $P_0$, and can be seen below
to carry over from $P_{i-1}$ to~$P_i$, since the changes from
$P_{i-1}(b \rightarrow b')$ to $P_i(b \rightarrow b')$ for
$b,b'\in\calB_i$ do not depend on $b$ and are proportional to
$\pi(b')$.  Define the ratio of the sum of a row of this submatrix to
the sum of probabilities for its values as follows:\vspace{-9pt}
\beq
   r_i & = & { 1  \over \pi(\calB_i) }\,\sum_{b'\in\calB_i} P_i(b\rightarrow b')
\eeq
This will be the same for any $b\in \calB_i$.  Since $\calB_0=\calX$, we will
have $r_0=1$.  We can express the transition probabilities in this sub-matrix
as\vspace{-9pt}
\beq
   P_i(b\rightarrow b') & = & r_i\,\pi(b') \label{eq-rprop}
\eeq
for any $b$ and $b'$ in $\calB_i$.

Stage $i$ of the NAM procedure will operate differently depending on whether
or not $\pi(\calA_i)<\pi(\calB_i)$. If $\pi(\calA_i)=\pi(a_i)=\pi(\sigma(i))
< \pi(\calB_i)$, we set 
$\delta_i\,=\,r_{i-1}\,\pi(a_i)\,/\,\pi(\calB_i)$. 
Using~(\ref{AM-def}), this gives $P_i$ as follows:\vspace{-3pt}
\beq\begin{array}{lclcll}
  P_i(a_i \rightarrow a_i) & = & 0 \\[2pt]
  P_i(a_i \rightarrow b')\! & = & 
   P_{i-1}(a_i \rightarrow b')\!\!\!   & + & \delta_i \pi(b'),
          & \mbox{if $b' \in \calB_i$} \\[2pt]
  P_i(b \rightarrow b')\! & = & 
   P_{i-1}(b \rightarrow b')\!\!\! & - &
                              \delta_i \pi(b') \pi(a_i)/\pi(\calB_i),\ \
          & \mbox{if $b \in \calB_i$ and $b' \in \calB_i$} \\[2pt]
  P_i(b \rightarrow a_i)\! & = & P_{i-1}(b \rightarrow a_i)\!\!\!   & + &
                              \delta_i \pi(a_i),
          & \mbox{if $b \in \calB_i$} \\[2pt]
  P_i(u \rightarrow v')\! & = & P_{i-1}(u \rightarrow v'),\!\!\!
       &&& \mbox{if $u \notin \calA_i \cup \calB_i$ 
                 or $v' \notin \calA_i \cup \calB_i$}
\end{array}\label{eq-NAM-a}\eeq
We can see that the value of $P_i(b\rightarrow b')$ above will be positive
using~(\ref{eq-rprop}) twice, along with $\pi(a_i)<\pi(\calB_i)$:
\beq
  \delta_i\pi(b')\pi(a_i)/\pi(\calB_i)\ <\ \delta_i\pi(b') 
  \ =\ r_{i-1}\pi(b')\pi(a_i)/\pi(\calB_i)
  \ =\ P_{i-1}(b \rightarrow b')\pi(a_i)/\pi(\calB_i)
  \ <\ P_{i-1}(b \rightarrow b')\ \
\eeq
so $P_i(b \rightarrow b')\,=\,P_{i-1}(b \rightarrow b')\,-\,
\delta_i\pi(b')\pi(a_i)/\pi(\calB_i)$ is positive.


When instead $\pi(\calA_i)=\pi(a_i)\ge\pi(\calB_i)$, we make use 
of~(\ref{AM-def})
with $\delta_i\,=\,r_{i-1}\,\pi(\calB_i)\,/\,\pi(a_i)$ and obtain
\beq\begin{array}{lclcll}
  P_i(a_i \rightarrow a_i) & = & P_{i-1}(a_i \rightarrow a_i)\!
                                 & - & \delta_i \pi(\calB_i) \\[2pt]
  P_i(a_i \rightarrow b')\! & = & 
   P_{i-1}(a_i \rightarrow b')\!\!\!   & + & \delta_i \pi(b'),\ \ 
          & \mbox{if $b' \in \calB_i$} \\[2pt]
  P_i(b \rightarrow b')\! & = & 0,
          & & & \mbox{if $b \in \calB_i$ and $b' \in \calB_i$} \\[2pt]
  P_i(b \rightarrow a_i)\! & = & P_{i-1}(b \rightarrow a_i)\!\!\!   & + &
                              \delta_i \pi(a_i),
          & \mbox{if $b \in \calB_i$} \\[2pt]
  P_i(u \rightarrow v')\! & = & P_{i-1}(u \rightarrow v'),\!\!\!
       &&& \mbox{if $u \notin \calA_i \cup \calB_i$ 
                 or $v' \notin \calA_i \cup \calB_i$}
\end{array}\label{eq-NAM-b}\eeq
$P_i(b \rightarrow b')=0$ because applying equation~(\ref{eq-rprop})
to its expression in~(\ref{AM-def}), and noting that $\calA_i=\{a_i\}$, gives%
\beq
 P_{i-1}(b \rightarrow b') \, -\, \delta_i\pi(b')\pi(\calA_i)/\pi(\calB_i)
 \, = \, P_{i-1}(b \rightarrow b') \, -\, r_{i-1}\pi(b')
 \, = \, P_{i-1}(b \rightarrow b') \, -\, P_{i-1}(b \rightarrow b')
 \, =\, 0\
\eeq
$P_i(a_i \rightarrow a_i)$ is guaranteed to be non-negative because, 
using $\pi(a_i)\ge\pi(\calB_i)$ and equation~(\ref{eq-rprop}),
\beq
  \delta_i\pi(\calB_i) 
  \ \le \ \delta_i\pi(a_i)\ =\ r_{i-1} \pi(\calB_i)
  \ \le \ r_{i-1}\pi(a_i)
  \ =\ P_{i-1}(a_i \rightarrow a_i)
\eeq
so $P_i(a_i \rightarrow a_i)
\ =\ P_{i-1}(a_i \rightarrow a_i)\ -\ \delta_i \pi(\calB_i)$
is non-negative.
Since a modification in which $\pi(a_i)\ge\pi(\calB_i)$ results in
$P_i(b\rightarrow b')$ being zero for all $b,b'\in\calB_i$, 
the NAM procedure is terminated at this point, with $P_i$ being the
final result.

Figures~\ref{NAM-mod1} and~\ref{NAM-mod2} shows two examples of Nested
Antithetic Modification, with different orderings, $\sigma$, of focal
values.  The example in Figure~\ref{NAM-mod2} ends after the second
stage, when the probability of the focal value ($a_2=4$) is as large as the
probability of the remaining values (in $\calB_2=\{1,2\}$).  Subsequent
stages would operate on an all-zero sub-matrix, and hence do nothing.

\begin{figure}[t]

\hspace*{0.7in}$r_0=1$
\hspace*{1.27in}$r_1={80\over81}$
\hspace*{1.2in}$r_2={400\over441}$
\hspace*{1.25in}$r_3={25\over63}$

\vspace{10pt}

\hspace*{0.2in}$
\left[\begin{array}{cccc} 
{ 1 \over 10 } & { 2 \over 10 } & { 3 \over 10 } & { 4 \over 10 } \\[4pt]
{ 1 \over 10 } & { 2 \over 10 } & { 3 \over 10 } & { 4 \over 10 } \\[4pt]
{ 1 \over 10 } & { 2 \over 10 } & { 3 \over 10 } & { 4 \over 10 } \\[4pt]
{ 1 \over 10 } & { 2 \over 10 } & { 3 \over 10 } & { 4 \over 10 }
\end{array}\right]
\ \ \rightarrow\ \
\left[\begin{array}{cccc} 
{ 0 } & { 2 \over 9 } & { 3 \over 9 } & { 4 \over 9 } \\[4pt]
{ 1 \over 9 } & { 16 \over 81 } & { 24 \over 81 } & { 32 \over 81 } \\[4pt]
{ 1 \over 9 } & { 16 \over 81 } & { 24 \over 81 } & { 32 \over 81 } \\[4pt]
{ 1 \over 9 } & { 16 \over 81 } & { 24 \over 81 } & { 32 \over 81 }
\end{array}\right]
\ \ \rightarrow\ \
\left[\begin{array}{cccc} 
{ 0 } & { 2 \over 9 } & { 3 \over 9 } & { 4 \over 9 } \\[4pt]
{1 \over 9} & { 0 } & { 24 \over 63 } & { 32 \over 63 } \\[4pt]
{1 \over 9} & { 16 \over 63 } & { 120 \over 441 } & { 160 \over 441 } \\[4pt]
{1 \over 9} & { 16 \over 63 } & { 120 \over 441 } & { 160 \over 441 }
\end{array}\right]
\ \ \rightarrow\ \
\left[\begin{array}{cccc} 
{ 0 } & { 2 \over 9 } & { 3 \over 9 } & { 4 \over 9 } \\[4pt]
{ 1 \over 9 } & { 0 } & { 24 \over 63 } & { 32 \over 63 } \\[4pt]
{ 1 \over 9 } & { 16 \over 63 } & { 0 } & { 40 \over 63 } \\[4pt]
{ 1 \over 9 } & { 16 \over 63 } & { 30 \over 63 } & { 10 \over 63 }
\end{array}\right]
$

\vspace{11pt}

\hspace*{1.15in}$\calA_1=\{1\},\ \calB_1=\{2,3,4\}$ 
\hspace*{0.09in}$\calA_2=\{2\},\ \calB_2=\{3,4\}$
\hspace*{0.36in}$\calA_3=\{3\},\ \calB_3=\{4\}$
\\[3pt]
\hspace*{1.55in}$\delta_1={1\over9}$
\hspace*{1.21in}$\delta_2={160\over567}$
\hspace*{1.22in}$\delta_3={300\over441}$

\vspace{7pt}

\caption{An example of the Nested Antithetic Modification (NAM)
         method, with $m=4$ values having probabilities 
         $1/10,\,2/10,\,3/10,\,4/10$, ordered by $\sigma(i)=i$.
         The arrows show transition probabilities being modified
         at each stage, using~(\ref{eq-NAM-a}), since 
         here $\pi(a_i)<\pi(\calB_i)$ at every stage.
         Compare with the MHGS modification in
         Figure~\ref{MHGS-mod}. Note that the final result has
         all off-diagonal transition probabilities smaller
         than in the original, and hence Peskun-dominates it.}\label{NAM-mod1}

\end{figure}

\begin{figure}[t]

\vspace*{0.35in}

\hspace*{1.7in}$r_0=1$
\hspace*{1.3in}$r_1={40\over49}$

\vspace{10pt}

\hspace*{1.2in}$
\left[\begin{array}{cccc} 
{ 1 \over 10 } & { 2 \over 10 } & { 3 \over 10 } & { 4 \over 10 } \\[4pt]
{ 1 \over 10 } & { 2 \over 10 } & { 3 \over 10 } & { 4 \over 10 } \\[4pt]
{ 1 \over 10 } & { 2 \over 10 } & { 3 \over 10 } & { 4 \over 10 } \\[4pt]
{ 1 \over 10 } & { 2 \over 10 } & { 3 \over 10 } & { 4 \over 10 }
\end{array}\right]
\ \ \rightarrow\ \
\left[\begin{array}{cccc} 
{ 4 \over 49 } & { 8 \over 49 } & { 3 \over 7 } & { 16 \over 49 } \\[4pt]
{ 4 \over 49 } & { 8 \over 49 } & { 3 \over 7 } & { 16 \over 49 } \\[4pt]
{ 1 \over 7 }  & { 2 \over 7 }  & { 0 }         & { 4 \over 7 } \\[4pt]
{ 4 \over 49 } & { 8 \over 49 } & { 3 \over 7 } & { 16 \over 49 }
\end{array}\right]
\ \ \rightarrow\ \
\left[\begin{array}{cccc} 
{ 0 }          & { 0 }          & { 3 \over 7 } & { 28 \over 49 } \\[4pt]
{ 0 }          & { 0 }          & { 3 \over 7 } & { 28 \over 49 } \\[4pt]
{ 1 \over 7 }  & { 2 \over 7 }  & { 0 }         & { 4 \over 7 } \\[4pt]
{ 7 \over 49 } & { 14 \over 49 } & { 3 \over 7 } & { 7 \over 49 }
\end{array}\right]
$

\vspace{11pt}

\hspace*{2.15in}$\calA_1=\{3\},\ \calB_1=\{1,2,4\}$ 
\hspace*{0.10in}$\calA_2=\{4\},\ \calB_2=\{1,2\}$
\\[3pt]
\hspace*{2.55in}$\delta_1={3\over7}$
\hspace*{1.21in}$\delta_2={30\over49}$

\vspace{7pt}

\caption{The same example as in Figure~\ref{NAM-mod1}, except with
         $\sigma(1)=3$ and $\sigma(2)=4$, so for the second stage, 
         $\pi(a_i)>\pi(\calB_i)$,
         and hence the modification is done using~(\ref{eq-NAM-b}).
         Since this sets the $\{1,2\}$ sub-matrix to all zeros, the
         procedure ends after this stage.  Note that the final result
         does not Peskun-dominate the original, since $P(1\rightarrow2)$
         and $P(2\rightarrow1)$ decrease to zero, but the new matrix
         does efficiency-dominate the original, as discussed in
         Section~\ref{sec-AM}.
         }\label{NAM-mod2}

\end{figure}

When simulating a Markov chain, we need only the row of the transition
matrix giving the transition probabilities from the current state
value, $k$.  Algorithm~\ref{alg-NAM} computes just these
probabilities, given a particular order, $\sigma$, of focal values,
taking time proportional to the number of possible values, $m$.

The algorithm considers successive focal values, $a_i=\sigma(i)$ for
$i=1,2,\ldots$, but rather than compute the whole transition matrix,
for each focal value it computes only the single transition
probability from the current value, $k$, to that focal value, until $k$
itself is the focal value.  Once $k$ is the focal value, the
transition probabilities from $k$ to all remaining values are computed.
Note that once transition probabilities to and from a focal value are
computed at some stage, they are not modified by later stages, so
there is no need to consider further focal values past $k$.

\begin{algorithm}[p]

\begin{tabbing}

\hspace{0.6in}\=
\bf Input:\ \ \ \ \= Gibbs sampling probabilities, 
                   $\pi^{\rule{0pt}{1pt}}(i)$, for $i=1,\ldots,m$
\\ \>
                \> A permutation, $\sigma$, on $\{1,\ldots,m\}$ giving the
                   order of focal values
\\ \>
                \> The current state value, $k$, in $\{1,\ldots,m\}$
\\[5pt] \>
\bf Output:    \> 
     NAM transition probabilities from $k$, as $p(i)$ for $i=1,\ldots,m$
\\
\> \hspace*{20pt} \= \hspace*{20pt} \= \hspace*{20pt} \= \hspace*{20pt} \= 
   \hspace*{120pt} \= \\[-3pt]
\> Set $s$ to 1
   \>\>\>\textit{The sum of probabilities for values that have not yet
                 been focal} \\[4pt]
\> Set $f$ to 1 
   \>\>\>\textit{The sum of transition probabilities from k to values that 
                 have not yet been focal} \\[7pt]
\> \textit{Find modified transition probabilities from the current value to
           successive focal values,} \\
\> \textit{until the focal value is the current value}\\[4pt]
\> Set $i$ to 1 \\[4pt]
\> While $\sigma(i) \ne k$: \\[4pt]
\>\> \textit{After seeing a focal value with probability at least as large as 
             remaining values, just store} \\
\>\> \textit{zeros (can change $f \le 0$ to $f \le \epsilon$
             for small $\epsilon$ to avoid tiny
             probabilities from rounding)}
     \\[4pt]
\>\> If $f \le 0$: \\[4pt]
\>\>\> Set $p({\sigma(i)})$ to 0 \\
\>\> Else: \\[2pt]
\>\>\> \textit{Let $q$ be the probability of the focal value; 
             update $s$ to be the sum}\\
\>\>\> \textit{of probabilities for remaining non-focal values} \\[4pt]
\>\>\> Set $q$ to $\pi({\sigma(i)})$ \\
\>\>\> Subtract $q$ from $s$ \>\>\> \textit{Sets variable s to $s_i$} \\[4pt]
\>\>\> \textit{Compute the transition probability from the current value, k, to 
               the focal value,}\\
\>\>\>\textit{and find the new total probability for transitions to remaining 
              values}
       \\[4pt]
\>\>\> If $q \ge s$: \\
\>\>\>\> Set $p({\sigma(i)})$ to $f$ \\
\>\>\>\> Set $f$ to 0 \\
\>\>\> Else: \\
\>\>\>\> Set $p({\sigma(i)})$ to $(q / s) f$ 
   \>\> \textit{Guarantees $p(\sigma(i))\le f\le 1$, even with rounding} \\
\>\>\>\> Subtract $p({\sigma(i)})$ from $f$ 
        \>\> \textit{Sets variable f to $f_i$, was previously $f_{i-1}$} \\[4pt]
\>\> Add 1 to $i$ \\[4pt]
\>\textit{Compute modified transition probabilities from the current value, k,
          which is now focal, to} \\
\>\textit{values that have not previously been focal, as well as the 
          self transition probability for $k$} \\[4pt]
\> If $f \le 0$: \\[4pt]
\>\> Set $p(k)$ to 0 \\
\>\> For $j=i\!+\!1,\ldots,m$:\ \ Set $p({\sigma(j)})$ to 0 \\[4pt]
\> Else: \\[4pt]
\>\> Set $q$ to $\pi(k)$ \\
\>\> Subtract $q$ from $s$ \\[4pt]
\>\> If $q > s$: \\
\>\>\> Set $p(k)$ to $((q\!-\!s)/q) f$ 
   \>\>\> \textit{Guarantees $p(k)\le f\le 1$, even with rounding} \\
\>\>\> For $j=i\!+\!1,\ldots,m$:\ \ Set $p({\sigma(j)})$ 
       to $\min\,(f,\,(\pi({\sigma(j)})/q)f)$\ \ \ \
       \textit{Min in case of rounding} \\
\>\> Else: \\
\>\>\> Set $p(k)$ to 0 \\
\>\>\> For $j=i\!+\!1,\ldots,m$:\ \ Set $p({\sigma(j)})$ 
       to $\min\,(f,\,(\pi({\sigma(j)})/s) f)$\ \ \ \
       \textit{Min in case of rounding} \\[-20pt]
\end{tabbing}\vspace{6pt}

\caption{Computation of modified transition probabilities from the current
         value by the NAM method.\vspace*{-10pt}}\label{alg-NAM}

\end{algorithm}

Algorithm~\ref{alg-NAM} incrementally maintains two sums:
\beq
 s_i & = & \!\! \sum\limits_{j=i+1}^m\!\! \pi(\sigma(j))\ \ = \ \ \pi(\calB_i)
 \\[4pt]
 f_i & = & \!\! \sum\limits_{j=i+1}^m\!\! P_i(k \rightarrow \sigma(j))
\eeq
Starting from $s_0=1$ and $f_0=1$, these are updated by\vspace{-4pt}
\beq
  s_i & = & s_{i-1}\ -\ \pi(\sigma(i)) \label{eq-s-update} \\[4pt]
  f_i & = & f_{i-1}\ -\ P_i(k \rightarrow \sigma(i)) \label{eq-f-update}
\eeq
for $j=1,2,\ldots$ until $\sigma(j)=k$. Note that $r_i = f_i/s_i$, and
that the values of $f_i$, $s_i$, and $r_i$ do not actually depend on the value
of $k$.

The stage $i$ computation for the transition
probability from the current value, $k$, to a focal value, $a_i=\sigma(i)$,
when $\pi(a_i)<\pi(\calB_i)=s_i$, 
can be re-written from its form in~(\ref{eq-NAM-a}) as follows, 
using equation~(\ref{eq-rprop}):
\beq
P_i(k \rightarrow a_i) 
 & = & P_{i-1}(k \rightarrow a_i)\ +\ \delta_i \pi(a_i) \\[4pt]
 & = & r_{i-1}\pi(a_i)\ +\ r_{i-1}\,(\pi(a_i)/\pi(\calB_i))\,\pi(a_i) \\[4pt]
 & = & {f_{i-1}\over s_{i-1}}\pi(a_i)\,\left[1+\pi(a_i)/\pi(\calB_i)\right]
       \\[4pt]
 & = & {f_{i-1}\over s_{i-1}}\pi(a_i)\,{\pi(a_i)+\pi(\calB_i)\over\pi(\calB_i)}
       \\[4pt]
 & = & {f_{i-1}\over s_{i-1}}\pi(a_i)\,{s_{i-1}\over s_i}\ \ =\ \ 
       {f_{i-1}\over s_i}\pi(a_i) \label{eq-fstrans}
\eeq
This is the method that Algorithm~\ref{alg-NAM} uses to compute $P_i(k
\rightarrow a_i)$, which is $p(\sigma(i))$ in the program, after which
it updates $f_{i-1}$ to $f_i$ by subtracting the result, as in
equation~(\ref{eq-f-update}).

If $\pi(a_i)\ge\pi(\calB_i)=s_i$ at some stage before $k$ becomes the
focal value,~(\ref{eq-NAM-b}) gives
\beq
P_i(k \rightarrow a_i) 
 & = & P_{i-1}(k \rightarrow a_i)\ +\ \delta_i \pi(a_i) \\[4pt]
 & = & r_{i-1}\pi(a_i)\ +\ r_{i-1}(\pi(\calB_i)/\pi(a_i))\pi(a_i) \\[4pt]
 & = & r_{i-1}\,[\pi(a_i)\ +\ \pi(\calB_i)\ \\[4pt]
 & = & {f_{i-1}\over s_{i-1}}\,s_{i-1}\ \ =\ \ f_{i-1}
\eeq
and transition probabilities from $k$ to all remaining values are zero.

When instead $\pi(a_i)=\pi(\sigma(i))$ is less than $\pi(\calB_i)$ for all
stages prior to when $k$ becomes the focal value, the transition
probabilities from $k$ to the remaining values are found once
$k$ is the focal value using 
either~(\ref{eq-NAM-a}) or~(\ref{eq-NAM-b}).  When $k$ becomes the 
focal value at stage $i$,
so $k=a_i=\sigma(i)$, then if $\pi(k)=\pi(a_i)<\pi(\calB_i)=s_i$, 
using~(\ref{eq-NAM-a}) gives $P_i(k \rightarrow k) = 0$, and for
$b'\in\calB_i$,
\beq
  P_i(k \rightarrow b') 
    & = & P_{i-1}(k \rightarrow b')\ +\ \delta_i \pi(b') \\[4pt]
    & = & r_{i-1} \pi(b')\ +\ r_{i-1}\,(\pi(k)/\pi(\calB_i))\,\pi(b') \\[4pt]
    & = & {f_{i-1}\over s_{i-1}} \pi(b')\,
          \left[1 + \pi(k)/\pi(\calB_i))\right] \\[4pt]
    & = & {f_{i-1}\over s_{i-1}} \pi(b')\,
          {\pi(k)+\pi(\calB_i) \over \pi(\calB_i)} \\[4pt]
    & = & {f_{i-1}\over s_{i-1}} \pi(b')\,{s_{i-1} \over s_i}
          \ \ =\ \ {f_{i-1}\over s_i} \pi(b')
\eeq
Whereas, if $\pi(k)=\pi(a_i)\ge\pi(\calB_i)=s_i$, using~(\ref{eq-NAM-b}) gives
\beq
  P_i(k \rightarrow k) & = & P_{i-1}(k \rightarrow k)
                             \ -\ \delta_i \pi(\calB_i) \\[4pt]
  & = & r_{i-1}\pi(k)\ -\ r_{i-1} (\pi(\calB_i)/\pi(k)) \pi(\calB_i) \\[4pt]
  & = & {f_{i-1} \over s_{i-1}}\, [\pi(k) - \pi(\calB_i)^2/\pi(k)] \\[4pt]
  & = & {f_{i-1} \over \pi(k)+s_i}\,{\pi(k)^2 -s_i^2\over\pi(k)}
   \ \ =\ \ f_{i-1}\, {\pi(k)-s_i \over \pi(k)}
\eeq
and for $b' \in \calB_i$,\vspace{-8pt}
\beq
  P_i(k \rightarrow b') 
  & = & P_{i-1}(k \rightarrow b')\ +\ \delta_i \pi(b') \\[4pt]
  & = & r_{i-1} \pi(b')\ +\ r_{i-1} (\pi(\calB_i)/\pi(k)) \pi(b') \\[4pt]
  & = & {f_{i-1} \over s_{i-1}} \pi(b')\,[1 + \pi(\calB_i)/\pi(k)] \\[4pt]
  & = & {f_{i-1} \over s_{i-1}} \pi(b')\,{\pi(k) + \pi(\calB_i) \over \pi(k)}
        \\[4pt]
  & = & {f_{i-1} \over s_{i-1}} \pi(b')\,{s_{i-1} \over \pi(k)}
  \ \ =\ \ {f_{i-1} \over \pi(k)} \pi(b')
\eeq
These formulas are used for the computations at the end of
Algorithm~\ref{alg-NAM}.

As presented, Algorithm~\ref{alg-NAM} computes all transition
probabilities from the current value of the state, which will
subsequently be used to sample the value for the next state. This is
inefficient when many of these transition probabilities are zero, as
occurs when at some point $\pi(a_i) \ge s_i$.  The algorithm
could be modified to return only the non-zero transition
probabilities, which also saves time when sampling.  Note also that
when $\pi(\sigma(1))$ is $1/2$ or more, any value other than
$\sigma(1)$ has probability one of transitioning to $\sigma(1)$, so in
this case there is no need to generate a random variate except when
the current value is $\sigma(1)$.

It would also be possible to modify Algorithm~\ref{alg-NAM} so that,
rather than returning transition probabilities from state $k$, it
instead returns a value randomly sampled according to these
probabilities.  Since Algorithm~\ref{alg-NAM} computes transition
probabilities in the order $\sigma$, and does not change them once
they are first computed, this could be done by sampling a number, $U$,
uniformly distributed on $[0,1]$, maintaining the cumulative sum of
transition probabilities computed so far, and returning the value just
considered once this cumulative sum exceeds $U$.  This would save some
computation time, though the savings would not be dramatic in the
typical case where computing the normalized probabilities, $\pi$,
requires looking at all $m$ values in any case.

When $\pi(\sigma(i))<s_i$ at every step, we can visualize the full 
matrix of transition probabilities computed by this algorithm (a row
at a time) as illustrated below, for $m=5$ and $\sigma(i)=i$:
\beq P^* & = &
 \left[ \begin{array}{ccccc}\displaystyle
 0 &\displaystyle \pi(2)\,{f_0 \over s_1}\ \
   &\displaystyle \pi(3)\,{f_0 \over s_1}\ \
   &\displaystyle \pi(4)\,{f_0 \over s_1}\ \
   &\displaystyle \pi(5)\,{f_0 \over s_1}\\[14pt]\displaystyle
 \displaystyle \pi(1)\,{f_0 \over s_1}\ \ & 0
   &\displaystyle \pi(3)\,{f_1 \over s_2}\ \ 
   &\displaystyle \pi(4)\,{f_1 \over s_2}\ \ 
   &\displaystyle \pi(5)\,{f_1 \over s_2}\\[14pt]\displaystyle
 \displaystyle \pi(1)\,{f_0 \over s_1}\ \ 
   &\displaystyle \pi(2)\,{f_1 \over s_2}\ \ & 0
   &\displaystyle \pi(4)\,{f_2 \over s_3}\ \ 
   &\displaystyle \pi(5)\,{f_2 \over s_3}\\[14pt]\displaystyle
 \displaystyle \pi(1)\,{f_0 \over s_1}\ \ 
   &\displaystyle \pi(2)\,{f_1 \over s_2}\ \ 
   &\displaystyle \pi(3)\,{f_2 \over s_3}\ \ & 0
   &\displaystyle \pi(5)\,{f_3 \over s_4}\\[14pt]\displaystyle
 \displaystyle \pi(1)\,{f_0 \over s_1}\ \ 
   &\displaystyle \pi(2)\,{f_1 \over s_2}\ \ 
   &\displaystyle \pi(3)\,{f_2 \over s_3}\ \ 
   &\displaystyle \pi(4)\,{f_3 \over s_4}\ \ 
   & \displaystyle f_4
 \end{array}\right]\label{NAM-vis1}
\eeq
Notice that under the diagonal the values in a column are all the same,
and that above the diagonal the values in a row equal the 
probabilities from $\pi$ times a common factor.

When $\pi(\sigma(i))\ge s_i$ at some point, the transition matrix is the
same as above for rows before $i$ and for columns before $i$, but then
is different for the submatrix of rows and columns from $i$ and later, as 
is illustrated below, when $m=5$, $\sigma(i)=i$, and
$\pi(3)\ge s_3 = \pi(4)+\pi(5)$:
\beq P^* & = &
 \left[ \begin{array}{ccccc}\displaystyle
 0 &\displaystyle \pi(2)\,{f_0 \over s_1}\ \
   &\displaystyle \pi(3)\,{f_0 \over s_1}\ \
   &\displaystyle \pi(4)\,{f_0 \over s_1}\ \
   &\displaystyle \pi(5)\,{f_0 \over s_1}\\[14pt]\displaystyle
 \displaystyle \pi(1)\,{f_0 \over s_1}\ \ & 0
   &\displaystyle \pi(3)\,{f_1 \over s_2}\ \ 
   &\displaystyle \pi(4)\,{f_1 \over s_2}\ \ 
   &\displaystyle \pi(5)\,{f_1 \over s_2}\\[14pt]\displaystyle
 \displaystyle \pi(1)\,{f_0 \over s_1}\ \ 
   &\displaystyle \pi(2)\,{f_1 \over s_2}\ \ 
   &\displaystyle {\pi(3)\!-\!s_3 \over \pi(3)}\,f_2\ \
   &\displaystyle {\pi(4)\over\pi(3)}\,f_2\ \ 
   &\displaystyle {\pi(5)\over\pi(3)}\,f_2\,\\[14pt]\displaystyle
 \displaystyle \pi(1)\,{f_0 \over s_1}\ \ 
   &\displaystyle \pi(2)\,{f_1 \over s_2}\ \
   & f_2
   & 0 & 0 \\[14pt]\displaystyle
 \displaystyle \pi(1)\,{f_0 \over s_1}\ \ 
   &\displaystyle \pi(2)\,{f_1 \over s_2}\ \ 
   & f_2
   & 0 & 0
 \end{array}\right]\label{NAM-vis2}
\eeq

The eigenvalues and eigenvectors of these transition matrices are of
some interest.  For a transition matrix for the entire state, the
eigenvalues determine the rate of convergence of the Markov chain.
However, this connection does not hold for partial transitions that
update a single variable, rather than the entire state, as for Gibbs
sampling and its modifications.  Nevertheless, the eigenvalues provide
some insight.  An eigenvalue of one always exists, with right
eigenvector of all ones, since each row of transition probabilities
sums to one.  When the rows are all equal to $\pi$ (as for a Gibbs
sampling update of a single variable, seen in isolation from others),
all the remaining eigenvalues are zero, reflecting immediate
convergence to $\pi$ after one transition.  Negative eigenvalues
correspond to ``antithetic'' aspects of the transition, which reduce
asymptotic variance, even compared to when the eigenvalues.

\pagebreak

For notational simplicity, suppose that $\sigma(i)=i$ for all $i$.
Then at each NAM step, $i$, prior to any at which $\pi(i)\ge s_i$, we can
identify an eigenvalue of $\lambda_i = -\pi(i)f_{i-1}/s_i$. An associated
right eigenvector is\vspace{-6pt}
\beq
  v_i & = & [\, 0,\, \ldots,\, s_{i-1}-\pi(i),\, -\pi(i),\,\ldots,-\pi(i)\, ]^T
  \\[2pt]
  & = & [\, 0,\ldots,s_{i-1},0,\ldots, 0\, ]^T\, -\, 
        [\, 0,\ldots,\pi(i),\pi(i),\ldots, \pi(i)\, ]^T
  \label{reigenvec} \\[-23pt] \nonumber
\eeq
where there are $i\!-\!1$ leading zero elements in the vector.  These
eigenvectors (along with $v_0=[1,\ldots,1]^T$ with eigenvalue 1) are
orthogonal with respect to an inner product based on $\pi$, with
$v_i^T D v_j^T=0$ for $i \ne j$, where $D$ is the diagonal matrix with
$\pi$ on the diagonal.  If at some step, $i<m$, we find that
$\pi(i)\ge s_i$, we can identify an eigenvalue of
$\lambda_i=-f_{i-1}s_i/\pi(i)$, with the same eigenvector $v_i$ as above.
Since the submatrix of rows and columns after $i$ will be zero, all
remaining eigenvalues (to $\lambda_{m-1}$) are zero.\footnote{\rule{0pt}{8pt}%
Proof that $v_i$ is an eigenvector of $P^*$, with eigenvalue as
given above: First, $[P^* v_i]_j$ is zero for $j<i$ since it equals\vspace{-6pt}
\beq
 s_{i-1}P^*(j \rightarrow i) \ -\ \sum_{k=i}^m \pi(i) P^*(j \rightarrow k)
 \ =\ 
 s_{i-1}\pi(i){f_{j-1} \over s_j} 
   \ -\ \pi(i) \sum_{k=i}^m \pi(k){f_{j-1} \over s_j}
 \ =\ 
 s_{i-1}\pi(i){f_{j-1} \over s_j} \ -\ \pi(i)s_{i-1}{f_{j-1}\over s_j}\ =\ 0
 \nonumber\\[-11pt]\nonumber
\eeq
When $i$ is less than any $k$ for which $\pi(k)\ge s_k$, then for any $j>i$,
$[P^* v_i]_j$ equals\vspace{-6pt}
\beq
 s_{i-1}P^*(j \rightarrow i) \ -\ \sum_{k=i}^m \pi(i) P^*(j \rightarrow k)
 \ =\ 
 s_{i-1}\pi(i){f_{i-1} \over s_i} \ -\ \pi(i) f_{i-1}
 \ =\ 
 \pi(i){f_{i-1} \over s_i} (s_{i-1} - s_i)
 \ =\ 
 \left[-\pi(i){f_{i-1} \over s_i}\right] \big[\!-\pi(i)\big]
 \nonumber\\[-11pt]\nonumber
\eeq
consistent with an eigenvalue of $-\pi(i)f_{i-1}/s_i$. Finally, when 
$\pi(i)<s_i$, $[P^* v_i]_i$ equals\vspace{-6pt}
\beq
  - \sum_{k=i+1}^m\! \pi(i) P^*(i \rightarrow k)
  \ =\ -\ \pi(i) \sum_{k=i+1}^m\! \pi(k){f_{i-1} \over s_i}
  \ =\ -\pi(i) f_{i-1}
  \ =\ \left[-\pi(i){f_{i-1} \over s_i}\right] \big[s_{i-1}-\pi(i)\big]
 \nonumber\\[-11pt]\nonumber
\eeq
again consistent with the eigenvalue $-\pi(i)f_{i-1}/s_i$.
When $\pi(i)\ge s_i$, the eigenvalue is $-f_{i-1}s_i/\pi(i)$, since
$[P^* v_i]_i$ equals\vspace{-6pt}
\beq\lefteqn{
  (s_{i-1}-\pi(i))\, P^*(i \rightarrow i)\, -\! 
     \sum_{k=i+1}^m\! \pi(i) P^*(i \rightarrow k)
  \ \ =\ \ 
  (s_{i-1}-\pi(i))\,{\pi(i)-s_i\over\pi(i)}f_{i-1}\, -\!
     \sum_{k=i+1}^m\! \pi(i) {\pi(k)\over\pi(i)} f_{i-1}}
  & & \mbox{\hspace*{6in}} \nonumber\\[-3pt]
  & \!\!=\! & {f_{i-1}\over\pi(i)}\! \big[ s_{i-1}(\pi(i)\!-\!s_i)+\pi(i)s_i
         - \pi(i) \sum_{k=i}^m \pi(k) \big]
  \ =\ {f_{i-1}\over\pi(i)}\, \big[ s_{i-1}\pi(i)-s_{i-1}s_i+\pi(i)s_i
           - \pi(i)s_{i-1} \big]
  \ =\, \left[-{f_{i-1}s_i\over\pi(i)}\right]\! \big[s_{i-1}-\pi(i)\big]
 \nonumber\\[-11pt]\nonumber
\eeq
and when $j>i$, $[P^*v_i]_j$ equals
$\displaystyle (s_{i-1}-\pi(i))\,P^*(j\rightarrow i)
\ =\ (s_{i-1}-\pi(i))\,f_{i-1}
\ =\ s_i f_{i-1}
\ =\ \left[-{f_{i-1}s_i\over\pi(i)}\right] \big[\!-\pi(i)\big]$.
}

These eigenvalues (apart from the single eigenvalue of one) are all
negative, except that some are zero when $\pi(\sigma(i)) \ge s_i$ for
some $i$.  This provides an alternative proof, via Corollary~15 of (Neal
and Rosenthal 2023), that NAM methods efficiency-dominate Gibbs
sampling, in addition to the general proof of this for AM methods
given in Section~\ref{sec-AM}.

In isolation, the negative eigenvalues of the modified NAM transition
matrix introduce an element of ``antithetic'' sampling, reducing
asymptotic variance of estimates, while slowing convergence to $\pi$,
since the absolute values of the eigenvalues (other than the single
one) are greater than for Gibbs sampling.  However, when the
transitions are used to update single variables, rather than the
entire state, the modification will not necessarily lead to slower
convergence than Gibbs sampling with random selection of variable to
update --- that will depend on the eigenvalues of the full transition
matrix for an update of a randomly selected variable, which are not
zero for Gibbs sampling.

Different orders of focal values for NAM may produce different
transition probabilities, so different ways of choosing an order
produce different methods for modifying Gibbs sampling.  I will use
``NAM'' without a prefix to refer to a method in which the order of
focal values is fixed.  I next discuss a method in which focal values
are chosen to have non-decreasing probability, as in
Figure~\ref{NAM-mod1}.  This order may be different for each Gibbs
sampling update, as changes to other variables change the conditional
distribution of the variable updated.  This will be followed by
discussion of the opposite strategy, of focusing on values in
non-increasing order of probability, which can have rather different
properties.

\section{\hspace*{-8pt}
  The Upward Nested Antithetic Modification (UNAM) 
  method}\label{sec-UNAM}\vspace{-11pt}

In the example of Figure~\ref{NAM-mod1}, the focal values used (1, 2,
and 3) are in increasing order of probability: $\pi(1)\!=\!1/10\ <\
\pi(2)\!=\!2/10\ <\ \pi(3)\!=\!3/10$, with the final value having the
largest probability, $\pi(4)\!=\!4/10$.  I will refer to the NAM
method in which focal values are chosen in non-decreasing order of
probability as the Upwards Nested Antithetic Modification (UNAM)
method.

This method is not new.  It is equivalent to a method discussed by
Frigessi, Hwang, and Younes (1992), and later devised independently by
Tjelmeland (2004).  As these authors note, and I will discuss below,
the UNAM method will always produce a modified transition probability
matrix that Peskun-dominates the original matrix, and hence
efficiency-dominates it --- i.e., produces estimates with lower
asymptotic variance.  As discussed in Sections~\ref{sec-peskun}, this
Peskun-dominance and efficiency-dominance for updates of individual
variables will carry over to an overall method that randomly selects a
variable to update.  NAM methods do not in general produce transitions
that Peskun-dominate Gibbs sampling, as can be seen for the example of
Figure~\ref{NAM-mod2}, but as discussed for antithetic modifications
in general in Section~\ref{sec-AM}, they always efficiency-dominate
Gibbs sampling, and this also carries over to an overall method that
randomly selects a variable to update.

Unless an ordering by probability is already known, the UNAM method
will start by finding a permutation, $\sigma$, of the possible values
that orders them in non-decreasing probability, so that
\mbox{$\pi(\sigma(i)) \le \pi(\sigma(j))$ when $i \le j$}.  (In the
example of Figure~\ref{NAM-mod1}, values are already ordered by
probability, so $\sigma(i)=i$.)  Various sorting algorithms could be
used to find this ordering.  With $m$ possible values, this can be
done in time proportional to $m \log m$ using a comparison sort, or
in time linear in $m$ if a radix sort is used.

\begin{algorithm}[t]

\begin{tabbing}

\hspace{0.7in}\=
\bf Input:\ \ \ \ \= Gibbs sampling probabilities, 
                   $\pi^{\rule{0pt}{1pt}}(i)$, for $i=1,\ldots,m$
\\ \>
                \> The current state value, $k$, in $\{1,\ldots,m\}$
\\[5pt] \>
\bf Output:    \> UNAM transition probabilities from $k$, as $p(i)$, 
                  for $i=1,\ldots,m$
\\
\> \hspace*{20pt} \= \hspace*{20pt} \= \hspace*{20pt} \= \hspace*{20pt} \= 
   \hspace*{120pt} \= \\[-3pt]
\> Set $\sigma$ to some permutation on 
   $\{1,\ldots,m\}$ for which $\pi(\sigma(i)) \le \pi(\sigma(j))$
   when $i \le j$\\[4pt]
\> Set $s$ to 1
   \>\>\>\textit{The sum of probabilities for values that have not yet
                 been focal} \\[4pt]
\> Set $f$ to 1 
   \>\>\>\textit{The sum of transition probabilities from k to values that 
                 have not yet been focal} \\[7pt]
\> \textit{Find modified transition probabilities from the current value to
           successive focal values,} \\
\> \textit{until the focal value is the current value}\\[4pt]
\> Set $i$ to 1 \\[4pt]
\> While $\sigma(i) \ne k$: \\[4pt]
\>\> \textit{Let $q$ be the probability of the focal value; 
             update $s$ to be the sum}\\
\>\> \textit{of probabilities for remaining non-focal values} \\[4pt]
\>\> Set $q$ to $\pi({\sigma(i)})$ \\
\>\> Subtract $q$ from $s$ \>\>\>\> \textit{Sets variable s to $s_i$} \\[4pt]
\>\> \textit{Compute the transition probability from current value, $k$, to 
             the focal value,}\\
\>\> \textit{and find the new total probability for transitions to remaining 
             values}
     \\[4pt]
\>\> Set $p({\sigma(i)})$ to $\min (f,\, (q / s) f)$ 
     \>\>\>\> \textit{Min with $f$ done in case $q>s$ due to rounding} \\
\>\> Subtract $p({\sigma(i)})$ from $f$
    \>\>\>\> \textit{Sets variable f to $f_i$, was previously $f_{i-1}$} \\[4pt]
\>\> Add 1 to $i$ \\[4pt]
\>\textit{Compute modified transition probabilities from the current value, k,
          which is now focal, to} \\
\>\textit{values that have not previously been focal, as well as the self 
          transition probability for $k$} \\[4pt]
\> If $i = m$: \\
\>\> Set $p(k)$ to $f$ \\[4pt]
\> Else: \\
\>\> Subtract $\pi(k)$ from s \\
\>\> Set $p(k)$ to 0 \\
\>\> For $j=i\!+\!1,\ldots,m$:\ \ Set $p({\sigma(j)})$ 
     to $\min(f,\,(\pi({\sigma(j)})/s) f)$\ \ \
        \textit{Min guards against rounding}\\[-20pt]
\end{tabbing}\vspace{8pt}

\caption{Computation of UNAM transition probabilities, a simplification
         of Algorithm~\ref{alg-NAM} when focal values have non-decreasing 
         probability.}\label{alg-UNAM}

\end{algorithm}

Once a suitable sorted order, $\sigma$, has been found, UNAM can be
implemented by just applying Algorithm~\ref{alg-NAM} with that
$\sigma$.  However, when $\sigma$ puts values in non-decreasing order
of probability, this algorithm can be simplified, as shown in
Algorithm~\ref{alg-UNAM}.  In particular, within the loop, it is never
possible for $\pi(a_i)$, which is $q$ in the program, to be greater
than or equal to $s_i$.  As discussed above regarding Algorithm~\ref{alg-NAM},
it is possible to modify Algorithm~\ref{alg-UNAM} to return
a sampled value rather than transition probabilities.

It is useful to see, from looking at the update to $f$ in the
loop of Algorithm~\ref{alg-UNAM}, that for $i<k$, 
\beq
  f_i & = & f_{i-1}\ -\ f_{i-1}\,{\pi(\sigma(i)) \over s_i}
  \ \ =\ \ f_{i-1}\left(1 \, -\, {\pi(\sigma(i)) \over s_i}\right)
  \ \ =\ \ f_{i-1}\,{s_i\, -\, \pi(\sigma(i)) \over s_i}
  \label{eq-useful}
  \\[-16pt]\nonumber
\eeq

To show that the UNAM method never decreases non-self transition
probabilities, it suffices to show that the transition probability
from $\sigma(j)$ to $\sigma(i)$, with $i<j$, never decreases, since
reversibility then guarantees the same for the transition probability
from $\sigma(i)$ to $\sigma(j)$.  When Algorithm~\ref{alg-UNAM} is
applied with $k=\sigma(j)$, this will be so if $f_{i-1} \ge s_i$ for 
all $i<j$ (see the entries below the diagonal in (\ref{NAM-vis1}) above).  
Using $s_0=1$, $f_0=1$, and the update of equation~(\ref{eq-s-update}),
we can first see that $f_0 \ge s_1$, and then, using~(\ref{eq-useful}), 
that
if $f_{i-2} \ge s_{i-1}$,
\beq
   f_{i-1} 
   & = & f_{i-2}\, {s_{i-1} \,-\, \pi(\sigma(i\!-\!1)) \over s_{i-1}} 
         \label{eq-tmp} \\[2pt]
   & \ge & s_{i-1}\, {s_{i-1} \,-\, \pi(\sigma(i\!-\!1)) \over s_{i-1}} \\[2pt]
   & = & s_{i-1}-\pi(\sigma(i\!-\!1)) \\[2pt]
   & \ge & s_{i-1}-\pi(\sigma(i))\ \ =\ \ s_i
\eeq
where the second inequality is because $\sigma$ orders values by non-decreasing
probability. It follows that\pagebreak{}\linebreak{}%
$f_{i-1} \ge s_i$ 
for all $i<j$, and hence UNAM never decreases non-self transition probabilities.
Peskun's theorem therefore guarantees that estimates using UNAM
have lower asymptotic variance than estimates using Gibbs sampling, when
the variable to be updated is randomly selected. 

UNAM transitions also Peskun-dominate those produced by MHGS.  Again,
we need only look at transition probabilities from $\sigma(j)=k$ to
$\sigma(i)$ with $i<j$, for which $\pi(\sigma(i)) \le \pi(k)$.  For
such a transition, the MHGS transition probability, from
equation~(\ref{MHGS-nonself}), is $\pi(\sigma(i))\, /\,
(1\!-\!\pi(\sigma(i)))$.  From equation~(\ref{eq-fstrans}), we see
that the UNAM transition probabilities will be at least as large as
the MHGS transition probabilities if $\pi(\sigma(i)) f_{i-1} / s_i
\,\ge\, \pi(\sigma(i))\, /\, (1\!-\!\pi(\sigma(i)))$, as will be the
case if $f_{i-1} \,\ge\, s_i\,/\,(1\!-\!\pi(\sigma(i)))$, for all
$i<j$.  This holds for $i=1$, since $f_0=1$, and from~(\ref{eq-s-update}),
$s_1=s_0-\pi(\sigma(1))=1-\pi(\sigma(1))$.
Furthermore, if $f_{i-2} \,\ge\, s_{i-1}\,/\,(1\!-\!\pi(\sigma(i\!-\!1)))$,
then again using equation~(\ref{eq-useful}, we have
\beq
   f_{i-1} & = & f_{i-2}\, {s_{i-1} \,-\, \pi(\sigma(i\!-\!1)) \over s_{i-1}} 
     \\[2pt]
   & \ge & {s_{i-1} \over 1\,-\,\pi(\sigma(i\!-\!1))}
           \, {s_{i-1} \,-\, \pi(\sigma(i\!-\!1)) \over s_{i-1}} \\[2pt]
   & = & {s_{i-1} \,-\, \pi(\sigma(i\!-\!1)) \over 1\,-\,\pi(\sigma(i\!-\!1))}
   \\[2pt]
   & \ge & {s_{i-1} \,-\, \pi(\sigma(i)) \over 1\,-\,\pi(\sigma(i))}
   \ \ =\ \ {s_i \over 1\,-\,\pi(\sigma(i))}
\eeq
The second inequality follows from $\pi(\sigma(i)) \ge 
\pi(\sigma(i\!-\!1))$ and the fact that if $0 \le \delta \le A \le B$
then $A/B\ge(A\!-\!\delta)/(B\!-\!\delta)$.
So the UNAM non-self transition probabilities are at least as great as
those using MHGS, and hence Peskun's theorem implies that with random
selection of variable to update, UNAM leads to lower asymptotic
variance than MHGS.

If $\pi(\sigma(i))=\pi(\sigma(i\!+\!1))$, then for any $k=\sigma(j)$
with $j>i+1$, Algorithm~\ref{alg-UNAM} will produce the same
transition probabilities from $k$ to $\sigma(i)$ and from $k$ to
$\sigma(i\!+\!1)$.  To see this, note that in iteration $i$ of the
loop, $p(\sigma(i))$ will be set to $\pi(\sigma(i)) f_{i-1} / s_i$, and
in the next iteration, $p(\sigma(i\!+\!1))$ will be set to
\beq
  \pi(\sigma(i\!+\!1))\, {f_i \over s_{i+1}} & = & 
  \pi(\sigma(i))\, {f_i \over s_{i+1}} \ \ =\ \ 
  \pi(\sigma(i))\, f_{i-1}\, {s_i - \pi(\sigma(i)) \over s_i}
    {1 \over s_i-\pi(\sigma(i\!+\!1))}\ \ =\ \ 
  \pi(\sigma(i))\, {f_{i-1} \over s_i}\ \ \ \
\eeq
which is the same.  Furthermore, as for all NAM methods, the
transition probabilities to any $\sigma(i)$ from all the $\sigma(j)$
with $j>i$ are the same.  Accordingly, when two or more values have
equal probability, it makes no difference in what order $\sigma$
places them.

Algorithm~\ref{alg-UNAM} sets all UNAM self transition probabilities
to zero, except possibly that for $\sigma(m)$, which will be zero if
$\pi(\sigma(m\!-\!1))=\pi(\sigma(m))$ but not otherwise.  To see this,
note that this self transition probability will be $f_{m-1}$, which
from equation~(\ref{eq-useful}) has the factor
$(s_{m-1}-\pi(\sigma(m\!-\!1)))\,/\,s_{m-1}$, and since
$s_{m-1}=\pi(\sigma(m))$, this is zero when
$\pi(\sigma(m\!-\!1))=\pi(\sigma(m))$.

As discussed earlier, the MHGS method can be applied when the number
of possible values is countably infinite, provided these have a
tractable form.  Doing this seems much harder for the UNAM method, since
Algorithm~3 looks at the possible values starting from the least
probable, and so would take an infinite number of steps.  
Some hope for using UNAM with a countably infinite (or very large)
number of possible values comes from reversing the recursions in
equations~(\ref{eq-s-update}) and~(\ref{eq-useful}):\vspace{-4pt}
\beq
  f_{m-1}\ \ =\ P^*(\sigma(m)\rightarrow\sigma(m)),
    &&\ f_{i-1}\ =\ f_{i}\,{s_i \over s_i - \pi(\sigma(i))} 
  \\[4pt]
  s_{m-1}\ \ =\ \pi(\sigma(m)),
    &&\ s_{i-1} \ =\ s_{i} + \pi(\sigma(i))
\eeq
After defining these recursions for a finite $m$, one might find the
limiting form as $m$ goes to infinity.  One could then sample 
from the transition distribution computing only finitely many of the $f_i$,
as necessary.  Unfortunately, it is not clear how 
to compute $f_m = P^*(\sigma(m)\rightarrow\sigma(m))$ without looking at all
$m$ values, but perhaps this is tractable for some 
distributions.\footnote{\rule{0pt}{10pt}%
If it happens that $\pi(\sigma(m\!-\!1))=\pi(\sigma(m))$, we
know that $P^*(\sigma(m)\rightarrow\sigma(m))=0$, but then the recursion
from $f_{m-1}$ to $f_{m-2}$ is undefined, and we have a problem computing
$f_{m-2}$.} If we can sample from $\pi$ (i.e., the Gibbs sampling conditional
probabilities), we could apply rejection sampling, using our knowledge
of the \textit{relative} transition probabilities from the current state 
to the other states, which we can get from these recursions, \textit{except}
when the current state is the most probable, for which we would need
to know the self transition probability.

The UNAM method gives the same transition probabilities as the
method that is implicit in the statement and proof of Theorem~1 of
Frigessi, Hwang, and Younes (1992).  They note that their method can
be applied to the probabilities for a Gibbs sampling update of a
randomly-selected state variable, and that the Peskun-dominance of the
individual updates extends to this scenario.  They also found
eigenvalues and eigenvectors of their transition matrices, which
I presented above for the more general class of NAM methods.

Like my description of UNAM here, the procedure of Frigessi,
\textit{et al.}, illustrated in Figure~\ref{UNAM-frigessi},
focuses on values in order of non-decreasing
probability, alters transition probabilities to and from each such
focal value in turn, and then proceeds to apply the procedure to the
sub-matrix of remaining values.  However, in their description, the
values in the sub-matrix are not rescaled by a common factor in order
to keep the row sums equal to one, as happens in the UNAM procedure
--- instead, self transition probabilities in the sub-matrix are
reduced to keep the sum of transition probabilities equal to one,
which is always possible when the focal values are in non-decreasing
order of probability. The final result is the same as for UNAM.

\begin{figure}[t]

\vspace*{-20pt}

{\small
\[
\left[\begin{array}{cccc} 
{ 1 \over 10 } & { 2 \over 10 } & { 3 \over 10 } & { 4 \over 10 } \\[4pt]
{ 1 \over 10 } & { 2 \over 10 } & { 3 \over 10 } & { 4 \over 10 } \\[4pt]
{ 1 \over 10 } & { 2 \over 10 } & { 3 \over 10 } & { 4 \over 10 } \\[4pt]
{ 1 \over 10 } & { 2 \over 10 } & { 3 \over 10 } & { 4 \over 10 }
\end{array}\right]
\begin{array}{cc}\\[0.4in] \rightarrow \\[0.4in] \times {10 \over 9}\end{array}
\left[\begin{array}{cccc} 
\mathbf{ 0 } &
\mathbf{ 2 \over 9 } &
\mathbf{ 3 \over 9 } &
\mathbf{ 4 \over 9 } \\[4pt]
\mathbf{ 1 \over 9 } &
\mathbf{ 17 \over 90 } &
{ 3 \over 10 } &
{ 4 \over 10 } \\[4pt]
\mathbf{ 1 \over 9 } &
{ 2 \over 10 } &
\mathbf{ 26 \over 90 } &
{ 4 \over 10 } \\[4pt]
\mathbf{ 1 \over 9 } &
{ 2 \over 10 } &
{ 3 \over 10 } &
\mathbf{ 35 \over 90 }
\end{array}\right]
\begin{array}{cc}\\[0.4in] \rightarrow \\[0.4in] \times {80 \over 63}\end{array}
\left[\begin{array}{cccc} 
{ 0 } &
{ 2 \over 9 } &
{ 3 \over 9 } &
{ 4 \over 9 } \\[4pt]
{ 1 \over 9 } &
\mathbf{ 0 } &
\mathbf{ 24 \over 63 } &
\mathbf{ 32 \over 63 } \\[4pt]
{ 1 \over 9 } &
\mathbf{ 16 \over 63 } &
\mathbf{ 148 \over 630 } &
{ 4 \over 10 } \\[4pt]
{ 1 \over 9 } &
\mathbf{ 16 \over 63 } &
{ 3 \over 10 } &
\mathbf{ 211 \over 630 }
\end{array}\right]
\begin{array}{cc}\\[0.4in]\rightarrow \\[0.4in] \times {100\over\,63}\end{array}
\left[\begin{array}{cccc} 
{ 0 } &
{ 2 \over 9 } &
{ 3 \over 9 } &
{ 4 \over 9 } \\[4pt]
{ 1 \over 9 } &
{ 0 } &
{ 24 \over 63 } &
{ 32 \over 63 } \\[4pt]
{ 1 \over 9 } &
{ 16 \over 63 } &
\mathbf{ 0 } &
\mathbf{ 40 \over 63 } \\[4pt]
{ 1 \over 9 } &
{ 16 \over 63 } &
\mathbf{ 30 \over 63 } &
\mathbf{ 10 \over 63 }
\end{array}\right]
\]
\vspace{-20pt}
}

\caption{Modification of Gibbs sampling 
         transition probabilities using the procedure of
         Frigessi, Hwang, and Younes (1992), equivalent to UNAM.
         The probabilities altered at 
         each stage are shown in bold.  The factors by which
         non-self transition probabilities in the current row and column
         are multiplied are below the arrows.  At each stage, self
         transition probabilities are altered so that the probabilities in 
         a row sum to one.  Note that the final result is the
         same as obtained with UNAM, as shown in Figure~\ref{NAM-mod1}.
}\label{UNAM-frigessi}

\end{figure}

One feature of this procedure is that modified non-self transition
probabilities at every stage (not just the final stage) are at least
as large as the Gibbs sampling probabilities, and hence these
intermediate transition probabilities Peskun-dominate Gibbs sampling.
Indeed, Frigessi, \textit{et al.}\ consider in detail (on pages~624
and~626--627) only a simplified form of their method, in which only
the first stage of modifications is performed, involving the
least-probable state (except that when several states have the
smallest probability, they modify the transition probabilities for all
of them). In this regard, they remark (page 627),\vspace{-8pt}
\begin{quotation}\noindent
In the definition of the modified Gibbs sampler, we did not complete
all the procedure described in part (b) of Theorem 1, for two reasons:
The first is that we are not sure that, from any configuration which is
not a local minimum of the energy, this new Markov chain would reach
a bottom with positive probability. Our proof cannot be extended to
show that this new stochastic matrix has no eigenvalue $-1$ at temperature
0. The second reason is practical: Each new step of the procedure of 
Theorem 1(b) would involve more and more computational cost. We therefore
restrict ourselves to only one step, which is easy to
implement.\vspace{-8pt}
\end{quotation}
Their first reason is particular to applications that aim essentially at
optimization rather than sampling. Their second reason has some validity,
since if only one stage of the procedure is done, one needn't sort the possible
values by probability, but only 
find the state(s) of lowest probability.  But the $m \log m$ sorting cost
is not prohibitive in the typical case where all $m$ probabilities must
be computed in any case.  Perhaps they did not realize that only the
$m$ probabilities for transitions from the current state need be computed,
as in Algorithm~\ref{alg-UNAM}, rather than all $m^2$ transition probabilities.
They give a recursive formula for the eigenvalues (page~617, Remark 4), 
which might have led them to an efficient simulation procedure, but they 
do not exploit its computational possibilities.\footnote{\rule{0pt}{10pt}%
Note that for $i<k$,
$P^*(\sigma(k)\rightarrow\sigma(i))=\pi(\sigma(i))f_{i-1}/s_i
=-\lambda_i$, and
$P^*(\sigma(i)\rightarrow\sigma(k))=-\lambda_i\pi(\sigma(k))/\pi(\sigma(i))$,
so knowing the eigenvalues allows efficient computation of transition 
probabilities.}

Frigessi, \textit{et al.}\ also show their method, when when applied
to the entire state (not necessarily when used to sample individual
variables as in Gibbs sampling), minimizes the maximum asymptotic
variance of the estimated expectation of a function, maximizing over
all functions with variance one (under $\pi$).  This is of limited
practical relevance, however, since the worst-case function will be
proportional to the indicator function of the least likely state,
which is seldom of interest.

A method equivalent to UNAM is also described by Tjelmeland (2004).
Tjelmeland was apparently unaware of the work of Frigessi, {\em et
al.}, perhaps since the title and abstract of the paper by Frigessi,
{\em et al.}\ give little indication that it describes a general
method for improving Gibbs sampling, and as noted above, Frigessi,
{\em et al.} are dismissive of the utility of the full method.  The
title and abstract of Tjelmeland's paper also do not mention that it
contains a general-purpose improvement to Gibbs sampling, focusing
instead on a particular context involving multiple proposals.  The
method is described as ``Transition alternative~2'' on page~5.

The presentation of Tjelmeland's method is somewhat similar to
that of Frigessi, \textit{et al.}, but differs in several respects.
As described mathematically, it alters the entire sub-matrix at each
stage, rather than only the row and column involving the focal value,
and the diagonal.  The end result is the same, 
however.\footnote{\rule{0pt}{10pt}%
The equivalence is easier to see after simplifying Tjelmeland's equation (14),
that defines a factor for multiplying transition probabilities:\vspace{-12pt}
\beq
  u^t & = & \min_{k \in A^t} \left( {1 - \sum_{l \notin A^t} P^t_{k,l}(y)
    \over \sum_{l\in A^t \setminus \{k\}} P^t_{k,l}(y)} \right)\nonumber
\eeq
where $A_t$ is the set of states with non-zero self transition probabilities.
The numerator in the fraction here is the same for all~$k$, from which it 
follows that the minimum is for the $k$ with minimum value for 
$P^t_{k,k}$.}

Following the presentation of the method, Tjelmeland remarks that\vspace{-8pt}
\begin{quotation}\noindent
  The above process defines all elements in $\mathbf{P(y)}$. When 
  simulating the Markov chain one of course only needs the elements in
  row $\kappa$. These can easily be computed without computing the
  whole matrix $\mathbf{P(y)}$. This is computationally important if
  $m$ is large.\vspace{-7pt}
\end{quotation}
Tjelmeland gives no details, however.  Avoiding such unnecessary
computation of the full transition matrix is the point of
Algorithm~\ref{alg-UNAM} for UNAM, as well as the more general NAM
method of Algorithm~\ref{alg-NAM}.

Yet another path to a method equivalent to UNAM is mentioned by
Pollet, Rombouts, Van Houcke, and Heyde (2004).  The
Metropolis-Hastings modification of Gibbs sampling probabilities that
define the MHGS method can be generalized to modify any
set of reversible transition probabilities, $P(u \rightarrow v)$.  
We use a proposal distribution, $Q$, that gives zero probability to the 
current state, rescaling $P(u \rightarrow v)$ for $v \ne u$ to
sum to one:\vspace{-11pt}
\beq
  Q(u \rightarrow v) & = & {P(u \rightarrow v) \over 1-P(u \rightarrow u)}
\eeq
The acceptance probability for such an update will be
\beq
 \min\!\left(\!1,\ {\pi(v)\,Q(v \rightarrow u) \over \pi(u)\,Q(u \rightarrow v)}
       \right)
 & \!=\! & 
 \min\!\left(\!1,\ {\pi(v)\,P(v \rightarrow u)\,(1\!-\!P(u \rightarrow u))
  \over \pi(u)\,P(u \rightarrow v)\,(1\!-\!P(v \rightarrow v))} \right)
 \ \ \!=\!\ \
 \min\!\left(\!1,\ {1\!-\!P(u \rightarrow u) \over 1\!-\!P(v \rightarrow v)}
 \right)\ \ \ \ \
\eeq
using the fact that $\pi(v)\,P(v \rightarrow u)\ =\ \pi(u)\,P(u \rightarrow v)$
due to the reversibility of $P$.

The modified non-self transition probabilities will therefore be
\beq
  \mbox{when $u \ne v$},\ \ \ P^*(u \rightarrow v) & = & 
    \min\left({P(u \rightarrow v) \over 1\!-\!P(u \rightarrow u)},\
                {P(u \rightarrow v) \over 1\!-\!P(v \rightarrow v)}
    \right)\label{MH-mod-prob}
\eeq
with the self transition probabilities determined by probabilities summing to 
one.\footnote{\rule{0pt}{10pt}%
Pollet, \textit{et al.}\ (2004) give an incorrect expression
for $P^*(u \rightarrow v)$ (in their notation, $T'_{ij}$) on the bottom left of 
page 2, but this appears to be what they intended.\vspace*{2pt}}  

Since these modified transition probabilities are themselves
reversible (as for any Metropolis-Hastings method), the procedure can
be repeated as many times as desired.  $P^*(u \rightarrow v)$ will
equal $P(u \rightarrow v)$ if the self transition probability of
either $u$ or $v$ is zero, while otherwise $P^*(u \rightarrow v)$
will be greater than $P(u \rightarrow v)$.  Hence repetition of this
procedure asymptotically converges to a transition matrix with at
most one non-zero self transition probability.

The same result is obtained with at most $m$ repetitions if only the
sub-matrix with non-zero self transition probabilities is updated
(scaling it to have rows that sum to one, applying the
Metropolis-Hastings procedure, and then scaling it back).  The value
with the smallest self transition probability will have zero self
transition probability after this modification, so all but at most one
self transition probability will be zero after $m\!-\!1$ applications
of the procedure.

The results of these Metropolis-Hasting procedures, and of the methods
of Frigessi, \textit{et al.}\ and of Tjelmeland, are the same as the
result obtained by the UNAM method.  This is a consequence of three
characteristics that they share.  First, all these methods produce
transition probabilities that are reversible with respect to $\pi$.
Second, they ultimately set all self transition probabilities to zero,
except perhaps for the most probable value.  Third, for all methods,
the modified transition probabilities $P^*(\sigma(i) \rightarrow \sigma(j))$
with $j>i$ are equal to $\pi(\sigma(j))$ times a factor that depends
only on $i$, not on $j$, which due to reversibility implies also that
the modified transition probabilities $P^*(\sigma(i)\rightarrow\sigma(j))$
with $j<i$ are equal to $\pi(\sigma(i))$ times a factor that depends only on 
$j$, not on $i$ (so elements in a column below the diagonal are all the
same, as seen in~(\ref{NAM-vis1}) for example). For UNAM, this can be
seen from the last line of Algorithm~\ref{alg-UNAM}. For the methods that 
repeatedly apply a Metropolis-Hasting modification, this is a consequence of
equation~(\ref{MH-mod-prob}), along with the fact that at each stage
values ordered by $\sigma$ have non-decreasing self transition
probability, which is true for the initial GS transition probabilities, 
and is maintained by each MH update.\footnote{\rule{0pt}{10pt}%
Let $P$ be transition probabilities before the MH update of~(\ref{MH-mod-prob}),
and $P^*$ the transition probabilities after this update.  Let $\sigma(i)$
and $\sigma(i\!+\!1)$ be consecutive focal values, with $\pi(\sigma(i))\le
\pi(\sigma(i\!+\!1))$.  Let $R=\pi(\sigma(i))/\pi(\sigma(i\!+\!1))$,
and define
$s_0=P(\sigma(i)\rightarrow\sigma(i))$, 
$s_1=P(\sigma(i\!+\!1)\rightarrow\sigma(i\!+\!1))$,
$A=\sum_{k<i} P(\sigma(i)\rightarrow\sigma(k))
=\sum_{k<i} P(\sigma(i\!+\!1)\rightarrow\sigma(k))$, 
$b_0=P(\sigma(i)\rightarrow\sigma(i\!+\!1))$, 
$b_1=P(\sigma(i\!+\!1)\rightarrow\sigma(i))=Rb_0$, 
$C_0=\sum_{k>i+1} P(\sigma(i)\rightarrow\sigma(k))$,
$C_1=\sum_{k>i+1} P(\sigma(i\!+\!1)\rightarrow\sigma(k))$,
and define $s_0^*$, $s_1^*$, $A^*$, $b_0^*$, $b_1^*$, $C_0^*$, and $C_1^*$
analogously for $P^*$ rather than $P$.  We wish to show that if $s_0\le s_1$,
then $s^*_0\le s^*_1$. We have that $C_0=1-A-b_0-s_0$,
$C_1=1-A-b_1-s_1$, $s^*_0=1-A^*-b_0^*-C_0^*$ and
$s^*_1=1-A^*-b_1^*-C_1^*$. Since $b_0^*=b_0/(1\!-\!s_0)$,
$b_1^*=Rb_0^*=Rb_0/(1\!-\!s_0)$, $C_0^*=C_0/(1\!-\!s_0)$, and
$C_1^*=C_1/(1\!-\!s_1)$, we have\vspace{-3pt}
\beq
  s_0^* & = & 1\ -\ A^*\ -\ {b_0\over1\!-\!s_0}\ -\ {1-A-b_0-s_0\over1\!-\!s_0}
  \ \ =\ \ -A^*\ +\ {A \over 1\!-\!s_0} \nonumber\\[4pt]
  s_1^* & = & 1\ -\ A^*\ -\ {Rb_0\over1\!-\!s_0}-{1-A-Rb_0-s_1\over1\!-\!s_1}
  \ \ =\ \ -A^*\ +\ Rb_0\left({1\over1\!-\!s_1}-{1\over1\!-\!s_0}\right)
           \ +\ {A \over 1\!-\!s_1}
  \nonumber
\eeq
Since $s_1\ge s_0$, we see that the middle term in the expression for $s^*_1$
is non-negative, and the final term is at least as large as the final term 
in the expression for $s^*_0$, and hence $s^*_1 \ge s^*_0$.
}
These characteristics determine a unique final result,
once all self transition probabilities, apart perhaps for $\sigma(m)$, 
are zero.\footnote{\rule{0pt}{10pt}%
To see this, let $h_i$ for 
$i=1,\ldots,m\!-\!1$ be the factors that
are used to multiply $\pi(\sigma(j))$ to get
$P^*(\sigma(i)\rightarrow\sigma(j))$ for $j>i$, which due to reversibility
also determine $P^*(\sigma(j)\rightarrow\sigma(i))$, and let $g$ be the
self transition probability for $\sigma(m)$.  The requirement that
transition probabilities sum to one leads to $m$ linear equations in
$g$ and the $h_i$, which uniquely determine them.
}
All these methods must therefore produce the the same final result as
the UNAM method.



\section{\hspace*{-8pt}
  The Downward Nested Antithetic Modification (DNAM) 
  method}\label{sec-DNAM}\vspace{-11pt}

The Nested Antithetic Modification approach can also be applied with
values ordered by non-increasing probability, giving the Downward
Nested Antithetic Modification (DNAM) method.  DNAM sometimes leads
to smaller self transition probabilities than UNAM.  With order
reversed from UNAM, there is no guarantee that all non-self transition
probabilities with DNAM are at least as large as with Gibbs sampling,
so Peskun's theorem does not apply, but as discussed in Section~\ref{sec-AM},
the transition probabilities produced with DNAM nevertheless 
efficiency-dominate Gibbs sampling.

DNAM can be implemented by simply applying the NAM procedure of
Algorithm~\ref{alg-NAM}, passing it a $\sigma$ that orders values by
non-increasing probability.  However, finding this order by sorting
values according to probability can be avoided when the current value
has probability of $1/2$ or more, as shown in
Algorithm~\ref{alg-DNAM}.  DNAM sometimes produces transition
probabilities that are zero past some point in the downward ordering.
The algorithm could be modified to efficiently skip these zero
probabilities, as discussed in Section~\ref{sec-NAM}.

\begin{algorithm}[p]

\begin{tabbing}

\hspace{0.7in}\=
\bf Input:\ \ \ \ \= Gibbs sampling probabilities, 
                   $\pi^{\rule{0pt}{1pt}}(i)$, for $i=1,\ldots,m$
\\ \>
                \> The current state value, $k$, in $\{1,\ldots,m\}$
\\[5pt] \>
\bf Output:    \> DNAM transition probabilities, $p(i)$, for $i=1,\ldots,m$
\\
\> \hspace*{20pt} \= \hspace*{20pt} \= \hspace*{20pt} \= \hspace*{20pt} \= 
   \\[-3pt]
\> If $\pi(k) \ge 1/2$: \\[4pt]
\>\> \textit{Quickly handle the case where the current value has probability 
             half or more,} \\
\>\> \textit{without needing to order values by probability} \\[4pt]
\>\> For $i = 1,\ldots,m$: \\
\>\>\> If $i \ne k$: \\
\>\>\>\> Set $p(i)$ to $\min(1,\,\pi(i)/\pi(k))$\ \ \ \
           \textit{Min guards against round-off error} \\[4pt]
\>\> Set $p(k)$ to $(2\pi(k)-1)\,/\,\pi(k)$ \\[4pt]

\> Else: \\[4pt]
\>\> Set $\sigma$ to some permutation on 
   $\{1,\ldots,m\}$ for which $\pi(\sigma(i)) \ge \pi(\sigma(j))$
   when $i \le j$\\[4pt]

\>\> Set $p$ to the output of the NAM procedure of Algorithm~\ref{alg-NAM} with 
     inputs $\pi$, $\sigma$, and $k$

\end{tabbing}\vspace{0pt}

\caption{Computation of DNAM transition probabilities.}\label{alg-DNAM}

\end{algorithm}

Some examples of transition matrices obtained using UNAM are shown in
Figure~\ref{cmp-GS-UNAM-DNAM}, with comparison to UNAM and Gibbs sampling.

In example (a), both UNAM and DNAM have a single non-zero
self transition probability --- the value with largest probability for
UNAM, one of those with second-smallest probability for DNAM.  Unlike
UNAM, DNAM can treat values with the same probability (here,
$\pi(2)=\pi(3)$) in substantively different ways, so it matters how
the sorting algorithm used to produce $\sigma$ handles ties.  Note
that in this example, the non-zero self transition probability is
smaller for DNAM than for UNAM, but the reverse is also possible.

\begin{figure}[p]\vspace*{-1pt}

\begin{center}

\makebox[45pt]{GS}$\displaystyle
\left[\begin{array}{cccc} 
{ 1 \over 12 } & { 3 \over 12 } & { 3 \over 12 } & { 5 \over 12 } \\[4pt]
{ 1 \over 12 } & { 3 \over 12 } & { 3 \over 12 } & { 5 \over 12 } \\[4pt]
{ 1 \over 12 } & { 3 \over 12 } & { 3 \over 12 } & { 5 \over 12 } \\[4pt]
{ 1 \over 12 } & { 3 \over 12 } & { 3 \over 12 } & { 5 \over 12 }
\end{array}\right]
$\hspace{22pt}$\displaystyle
\left[\begin{array}{cccc} 
{ 2 \over 10 } & { 2 \over 10 } & { 3 \over 10 } & { 3 \over 10 } \\[4pt]
{ 2 \over 10 } & { 2 \over 10 } & { 3 \over 10 } & { 3 \over 10 } \\[4pt]
{ 2 \over 10 } & { 2 \over 10 } & { 3 \over 10 } & { 3 \over 10 } \\[4pt]
{ 2 \over 10 } & { 2 \over 10 } & { 3 \over 10 } & { 3 \over 10 }
\end{array}\right]
$\hspace{22pt}$\displaystyle
\left[\begin{array}{cccc} 
{ 1 \over 10 } & { 3 \over 10 } & { 3 \over 10 } & { 3 \over 10 } \\[4pt]
{ 1 \over 10 } & { 3 \over 10 } & { 3 \over 10 } & { 3 \over 10 } \\[4pt]
{ 1 \over 10 } & { 3 \over 10 } & { 3 \over 10 } & { 3 \over 10 } \\[4pt]
{ 1 \over 10 } & { 3 \over 10 } & { 3 \over 10 } & { 3 \over 10 } 
\end{array}\right]
$\hspace{22pt}$\displaystyle
\left[\begin{array}{cccc} 
{ 1 \over 10 } & { 1 \over 10 } & { 3 \over 10 } & { 5 \over 10 } \\[4pt]
{ 1 \over 10 } & { 1 \over 10 } & { 3 \over 10 } & { 5 \over 10 } \\[4pt]
{ 1 \over 10 } & { 1 \over 10 } & { 3 \over 10 } & { 5 \over 10 } \\[4pt]
{ 1 \over 10 } & { 1 \over 10 } & { 3 \over 10 } & { 5 \over 10 }
\end{array}\right]
$
\\\vspace{17pt}

\makebox[45pt]{UNAM}$\displaystyle
\left[\begin{array}{cccc} 
{     0      } & { 3 \over 11 } & { 3 \over 11 } & { 5 \over 11 } \\[4pt]
{ 1 \over 11 } & {     0      } & { 15\over 44 } & { 25\over 44 } \\[4pt]
{ 1 \over 11 } & { 15\over 44 } & {     0      } & { 25\over 44 } \\[4pt]
{ 1 \over 11 } & { 15\over 44 } & { 15\over 44 } & { 10\over 44 }
\end{array}\right]
$\hspace{22pt}$\displaystyle
\left[\begin{array}{cccc} 
\ {     0      } &\ { 2 \over  8 } &\ { 3 \over  8 } &\ { 3 \over  8 }\ \\[4pt]
\ { 2 \over  8 } &\ {     0      } &\ { 3 \over  8 } &\ { 3 \over  8 }\ \\[4pt]
\ { 2 \over  8 } &\ { 2 \over  8 } &\ {     0      } &\ { 1 \over  2 }\ \\[4pt]
\ { 2 \over  8 } &\ { 2 \over  8 } &\ { 1 \over  2 } &\ {     0      }\
\end{array}\right]
$\hspace{22pt}$\displaystyle
\left[\begin{array}{cccc} 
\ {     0      } &\ { 3 \over  9 } &\ { 3 \over  9 } &\ { 3 \over  9 }\ \\[4pt]
\ { 1 \over  9 } &\ {     0      } &\ { 4 \over  9 } &\ { 4 \over  9 }\ \\[4pt]
\ { 1 \over  9 } &\ { 4 \over  9 } &\ {     0      } &\ { 4 \over  9 }\ \\[4pt]
\ { 1 \over  9 } &\ { 4 \over  9 } &\ { 4 \over  9 } &\ {     0      }\
\end{array}\right]
$\hspace{22pt}$\displaystyle
\left[\begin{array}{cccc} 
\ {     0      } & { 1 \over 9 } & { 3 \over 9 } & { 5\over 9 }\ \\[4pt]
\ { 1 \over 9 } & {     0      } & { 3 \over 9 } & { 5\over 9 }\ \\[4pt]
\ { 1 \over 9 } & { 1 \over 9 } & {     0      } & { 7\over 9 }\ \\[4pt]
\ { 1 \over 9 } & { 1 \over 9 } & { 21 \over 45 } & { 14\over 45 }\
\end{array}\right]
$
\\\vspace{17pt}

\makebox[45pt]{DNAM}$\displaystyle
\left[\begin{array}{cccc} 
{     0      } & { 3 \over 42 } & { 3 \over 14 } & { 5 \over  7 }\ \\[4pt]
{ 1 \over 42 } & { 2 \over 42 } & { 3 \over 14 } & { 5 \over  7 }\ \\[4pt]
{ 1 \over 14 } & { 3 \over 14 } & {     0      } & { 5 \over  7 }\ \\[4pt]
{ 1 \over  7 } & { 3 \over  7 } & { 3 \over  7 } & {     0      }\
\end{array}\right]
$\hspace{22pt}$\displaystyle
\left[\begin{array}{cccc} 
\ {     0     } &\ { 1 \over 7 } &\ { 3 \over 7 } &\ { 3 \over 7 }\ \\[4pt]
\ { 1 \over 7 } &\ {     0     } &\ { 3 \over 7 } &\ { 3 \over 7 }\ \\[4pt]
\ { 2 \over 7 } &\ { 2 \over 7 } &\ {     0     } &\ { 3 \over 7 }\ \\[4pt]
\ { 2 \over 7 } &\ { 2 \over 7 } &\ { 3 \over 7 } &\ {     0     }\
\end{array}\right]
$\hspace{22pt}$\displaystyle
\left[\begin{array}{cccc} 
\ {     0     } & { 3 \over 21} & { 3 \over 7 } &\ { 3 \over 7 }\ \\[4pt]
\ { 1 \over 21} & { 2 \over 21} & { 3 \over 7 } &\ { 3 \over 7 }\ \\[4pt]
\ { 1 \over 7 } & { 3 \over 7 } & {     0     } &\ { 3 \over 7 }\ \\[4pt]
\ { 1 \over 7 } & { 3 \over 7 } & { 3 \over 7 } &\ {     0     }\
\end{array}\right]
$\hspace{22pt}$\displaystyle
\left[\begin{array}{cccc} 
\ {     0      } &\ {     0      } &\ {     0      } &\ {     1      }\,\\[4pt]
\ {     0      } &\ {     0      } &\ {     0      } &\ {     1      }\,\\[4pt]
\ {     0      } &\ {     0      } &\ {     0      } &\ {     1      }\,\\[4pt]
\ { 1 \over 5  } &\ { 1 \over 5  } &\ { 3 \over 5  } &\ {     0  }\,
\end{array}\right]
$

\vspace{12pt}

\end{center}

\hspace*{20pt}$\pi$\hspace{32pt}%
              ${1\over12}\ \ \, {3\over12}\ \ \, {3\over12}\ \ \, {5\over12}$
\hspace*{46pt}%
              ${2\over10}\ \ \, {2\over10}\ \ \, {3\over10}\ \ \, {3\over10}$
\hspace*{49pt}%
              ${1\over10}\ \ \, {3\over10}\ \ \, {3\over10}\ \ \, {3\over10}$
\hspace*{46pt}%
              ${1\over10}\ \ \, {1\over10}\ \ \, {3\over10}\ \ \, {5\over10}$

\vspace{9pt}

\hspace*{89pt}(a)\hspace*{108pt}(b)\hspace*{110pt}(c)\hspace*{110pt}(d)

\vspace{-3pt}

\caption{Some comparisons of transitions probabilities for Gibbs Sampling (GS),
         UNAM, and DNAM.  For all examples, values are ordered by non-decreasing
         probability, so UNAM focuses on values as ordered, and DNAM 
         focuses on values in the reverse order.\vspace*{-7pt}
        }\label{cmp-GS-UNAM-DNAM}

\end{figure}

In example (b), both UNAM and DNAM produce self transition
probabilities that are all zero.  This happens with UNAM when the two
largest probabilities under $\pi$ are equal.  It happens with DNAM
when some value has a probability under $\pi$ equal to the sum of
probabilities of values later in the order $\sigma$.  In this
example, this happens because the second-last value in the order
$\sigma$ has the same probability as the last value.  Note that
although both UNAM and DNAM produce zero self transition probabilities,
the other transition probabilities differ for the two methods.

Example (c) shows that UNAM can produce all zero self transition
probabilities while DNAM does not.  Example (d) shows the reverse, and
also shows that with DNAM a large sub-matrix of transition
probabilities may be all zero, a property that can sometimes be
exploited to reduce computational cost.

Since neither UNAM nor DNAM is clearly superior to the other in all
situations, one might consider randomly choosing between them, with
equal probabilities, hoping to obtain the advantages of both.  I call
this method UDNAM.  The transition probabilities for this method are
simply the averages of those for UNAM and those for DNAM.  Theorem~11
of (Neal and Rosenthal 2023) can be applied (twice) to show that since
UNAM and DNAM both efficiency-dominate Gibbs sampling, UDNAM must also
efficiency-dominate Gibbs sampling --- UDNAM, as the random
combination of UNAM and DNAM, must efficiency-dominate the random
combination of UNAM and GS, which must efficiency-dominate the random
combination of GS and GS, which is simply GS.

As noted earlier, Algorithm~\ref{alg-NAM} used in DNAM can be modified
to sample a value from the transition distribution, taking time
proportional only to the index of this sampled value in the order
$\sigma$, rather than computing all probabilities.  Since DNAM looks
at probabilities in decreasing order, this permits its use when the
number of possible values is countably infinite, provided a formula
for probabilities of values is available, and a non-increasing ordering
can be determined.

For the the geometric($\theta$) distribution of
equation~(\ref{eq-geom-cdf}), used as an example for MHGS, when $m$
goes to infinity, $\pi(i)=\theta(1-\theta)^{i-1}$ and $s_i=\sum_{j>i} \pi(j)
= (1-\theta)^i$.  The decreasing ordering is $\sigma(i)=i)$.
When $\theta \ge 1/2$, we will have $\pi(1)\ge 1/2$, so the DNAM
transition probabilities computed by Algorithm~\ref{alg-DNAM} will be
\beq
  P^*(1 \rightarrow 1) & = & (2\pi(1)-1)\,/\,\pi(1) 
                       \ \ =\ \ (2\theta-1)\,/\,\theta \\[4pt]
  P^*(1 \rightarrow j) & = & (1-\theta)^{j-1}\!\!\!\!\!\!,
                       \ \ \ \mbox{for $j>1$} \\[4pt]
  P^*(i \rightarrow 1) & = & 1, \ \ \ \mbox{for $i>1$} \\[4pt]
  P^*(i \rightarrow j) & = & 0, \ \ \ \mbox{for $i,j>1$}
\eeq
When $\theta < 1/2$, we will never have $\pi(i) \ge s_i$, so the
DNAM transition probabilities will follow the pattern 
of~(\ref{NAM-vis1}), giving:\footnote{\rule{0pt}{8pt}%
For $j<i$, we will have\vspace{-3pt}
\beq
P^*(i \rightarrow j)\ =\ \pi(j) {f_{j-1} \over s_j}
\ =\ {\theta (1-\theta)^{j-1} \over (1-\theta)^j} f_{j-1}\,=\,
{\theta \over 1-\theta} f_{j-1}
\eeq
So then, $f_j\, =\,
f_{j-1} - P^*(i \rightarrow j)\,=\, f_{j-1} - {\theta \over 1-\theta} f_{j-1}
\,=\, {1-2\theta \over 1-\theta}f_{j-1}$, from which it follows that $f_j\,=\,
\left(1-2\theta \over 1-\theta\right)^j\!\!\!,$ and hence that
$P^*(i \rightarrow j)\,=\,{\theta \over 1-\theta} 
\left(1-2\theta \over 1-\theta\right)^{j-1}\!\!\!.$\vspace{4pt}

For $j>i$,\vspace{-6pt}
\beq
P^*(i\rightarrow j)\,=\,\pi(j){f_{i-1} \over s_i}\ =\ 
{\theta(1-\theta)^{j-1}\over(1-\theta)^i}
\left(1-2\theta\over1-\theta\right)^{i-1}\!=\
{\theta (1-2\theta)^{i-1} \over (1-\theta)^{2i-1}}(1-\theta)^{j-1}
\eeq
}
\beq
  P^*(i \rightarrow j) & = & {\theta\over1-\theta}
                \left(1-2\theta\over1-\theta\right)^{j-1}\!\!\!\!\!\!\!\!,
                \ \ \ \ \ \mbox{for $j<i$} \\[4pt]
  P^*(i \rightarrow i) & = & 0 \\[4pt]
  P^*(i \rightarrow j) & = & 
    {\theta (1-2\theta)^{i-1} \over (1-\theta)^{2i-1}}
    (1-\theta)^{j-1}\!\!\!\!\!\!,
    \ \ \ \ \mbox{for $j>i$}
\eeq
So for this geometric distribution, the DNAM method produces the minimum
possible self transition probability.  The transition distributions are
piecewise geometric, and so are easily sampled from.

Finally, note that self transition probabilities that are all zero can
sometimes be obtained with NAM using an ordering that is neither
upward (as in UNAM) nor downward (as in DNAM).  For example~(a) of
Figure~\ref{cmp-GS-UNAM-DNAM}, we can obtain the transition
probability matrices below by using the ordering 1,4,2,3 (or 1,4,3,2),
shown on the left, and by using the ordering 4,1,2,3 (or 4,1,3,2),
shown on the right:
\beq
\left[\begin{array}{cccc} 
{     0      } & { 3 \over 11 } & { 3 \over 11 } & { 5 \over 11 } \\[4pt]
{ 1 \over 11 } & {     0      } & { 5 \over 33 } & { 25\over 33 } \\[4pt]
{ 1 \over 11 } & { 5 \over 33 } & {     0      } & { 25\over 33 } \\[4pt]
{ 1 \over 11 } & { 15\over 33 } & { 15\over 33 } & {     0      }
\end{array}\right]
& \ \ \ \ \ \ &
\left[\begin{array}{cccc} 
{     0      } & { 3 \over 21 } & { 3 \over 21 } & { 5\over 7 } \\[4pt]
{ 1 \over 21 } & {     0      } & { 5 \over 21 } & { 5\over 7 } \\[4pt]
{ 1 \over 21 } & { 5 \over 21 } & {     0      } & { 5\over 7 } \\[4pt]
{ 1 \over  7 } & { 3 \over  7 } & { 3 \over  7 } & {     0      }
\end{array}\right] 
\eeq

However, although a transition matrix having self transition
probabilities that are all zero always exists when no value has
probability greater than $1/2$, such a transition matrix cannot
generally be obtained using NAM with some ordering --- this is possible
only when there is exact equality between the probability of some
value and a sum of probabilities of some other values.  In the next
three sections, I will discuss methods that do always minimize
self transition probabilities.

\section{\hspace*{-8pt}
  The Zero-self DNAM (ZDNAM) method}\label{sec-ZDNAM}\vspace{-11pt}

A non-zero self transition probability is necessary only for a value
whose probability under $\pi$ is more than one half.  But DNAM will
produce a non-zero self transition probability for a value with
probability less than one half if this probability is greater than
the sum of the probabilities of values with lower probability, as is
the case in examples (a) and (c) of Figure~\ref{cmp-GS-UNAM-DNAM}.
Note that when this happens the transition probabilities among the
remaining values are all zero, so the DNAM procedure ends at this point.

Here, I describe a modified procedure, the Zero-self DNAM method
(ZDNAM), which modifies the DNAM procedure to operate differently at
the step just before the one where DNAM would produce a non-zero
self transition probability, substituting transition probabilities
that avoid this.  As for DNAM, the remaining transition probabilities
are all zero, so no further steps are necessary.

The idea can be illustrated by an example with $m=5$ and $\sigma(i)=i$, 
with
$\pi(1)=6/18$, $\pi(2)=5/18$, $\pi(3)=4/18$, $\pi(4)=2/18$, and $\pi(5)=1/18$.
DNAM modifies the original transitions as follows:\vspace{5pt}
\beq
\left[\begin{array}{ccccc} 
{ 6\over18 } & { 5\over18 } & { 4\over18 } & { 2\over18 } & { 1\over18 } \\[4pt]
{ 6\over18 } & { 5\over18 } & { 4\over18 } & { 2\over18 } & { 1\over18 } \\[4pt]
{ 6\over18 } & { 5\over18 } & { 4\over18 } & { 2\over18 } & { 1\over18 } \\[4pt]
{ 6\over18 } & { 5\over18 } & { 4\over18 } & { 2\over18 } & { 1\over18 } \\[4pt]
{ 6\over18 } & { 5\over18 } & { 4\over18 } & { 2\over18 } & { 1\over18 }
\end{array}\right]
\!\!\rightarrow\!\!
\left[\begin{array}{ccccc} 
{    0     } & { 5\over12 } & { 4\over12 } & { 2\over12 } & { 1\over12 } \\[4pt]
{ 6\over12 } & { 5\over36 } & { 4\over36 } & { 2\over36 } & { 1\over36 } \\[4pt]
{ 6\over12 } & { 5\over36 } & { 4\over36 } & { 2\over36 } & { 1\over36 } \\[4pt]
{ 6\over12 } & { 5\over36 } & { 4\over36 } & { 2\over36 } & { 1\over36 } \\[4pt]
{ 6\over12 } & { 5\over36 } & { 4\over36 } & { 2\over36 } & { 1\over36 }
\end{array}\right]
\!\!\rightarrow\!\!
\left[\begin{array}{ccccc} 
{    0     } & { 5\over12 } & { 4\over12 } & { 2\over12 } & { 1\over12 } \\[4pt]
{ 6\over12 } & {    0     } & { 4\over14 } & { 2\over14 } & { 1\over14 } \\[4pt]
{ 6\over12 } & { 5\over14 } & { 4\over49 } & { 2\over49 } & { 1\over49 } \\[4pt]
{ 6\over12 } & { 5\over14 } & { 4\over49 } & { 2\over49 } & { 1\over49 } \\[4pt]
{ 6\over12 } & { 5\over14 } & { 4\over49 } & { 2\over49 } & { 1\over49 }
\end{array}\right]
\!\!\rightarrow\!\!
\left[\begin{array}{ccccc} 
{    0     } & { 5\over12 } & { 4\over12 } & { 2\over12 } & { 1\over12 } \\[4pt]
{ 6\over12 } & {    0     } & { 4\over14 } & { 2\over14 } & { 1\over14 } \\[4pt]
{ 6\over12 } & { 5\over14 } & { 1\over28 } & { 2\over28 } & { 1\over28 } \\[4pt]
{ 6\over12 } & { 5\over14 } & { 4\over28 } & {     0    } & {    0     } \\[4pt]
{ 6\over12 } & { 5\over14 } & { 4\over28 } & {     0    } & {    0     }
\end{array}\right]\ \ \ \ \ \nonumber 
\eeq

\noindent
The non-zero self transition probability of $P^*(3\rightarrow3)=1/28$ 
results from $\pi(3)=4/18$ being greater than the sum of probabilities
for later values, which in this example is $s_3 = \pi(4)+\pi(5)=3/18$.

For this example, the ZDNAM method operates the same as DNAM for the
first step, but at step $i=2$, the ZDNAM algorithm
recognizes that $\pi(\sigma(i+1))>s_{i+1}=\sum_{j>i+1} \pi(\sigma(j))$
--- in this example, that $\pi(3)=4/18\ >\ \pi(4)+\pi(5)=3/18$ --- and
employs a special construction to avoid a non-zero self transition
probability for $\sigma(i+1)$ --- in this example, for the value $3$. The
result is as follows:\vspace{3pt}
\beq
\left[\begin{array}{ccccc} 
{ 6\over18 } & { 5\over18 } & { 4\over18 } & { 2\over18 } & { 1\over18 } \\[4pt]
{ 6\over18 } & { 5\over18 } & { 4\over18 } & { 2\over18 } & { 1\over18 } \\[4pt]
{ 6\over18 } & { 5\over18 } & { 4\over18 } & { 2\over18 } & { 1\over18 } \\[4pt]
{ 6\over18 } & { 5\over18 } & { 4\over18 } & { 2\over18 } & { 1\over18 } \\[4pt]
{ 6\over18 } & { 5\over18 } & { 4\over18 } & { 2\over18 } & { 1\over18 }
\end{array}\right]
\!\rightarrow\!
\left[\begin{array}{ccccc} 
{    0     } & { 5\over12 } & { 4\over12 } & { 2\over12 } & { 1\over12 } \\[4pt]
{ 6\over12 } & { 5\over36 } & { 4\over36 } & { 2\over36 } & { 1\over36 } \\[4pt]
{ 6\over12 } & { 5\over36 } & { 4\over36 } & { 2\over36 } & { 1\over36 } \\[4pt]
{ 6\over12 } & { 5\over36 } & { 4\over36 } & { 2\over36 } & { 1\over36 } \\[4pt]
{ 6\over12 } & { 5\over36 } & { 4\over36 } & { 2\over36 } & { 1\over36 }
\end{array}\right]
\!\rightarrow\!
\left[\begin{array}{ccccc} 
{    0     } & { 5\over12 } & { 4\over12 } & { 2\over12 } & { 1\over12 } \\[4pt]
{ 6\over12 } & {    0     } & { 12\over40} & { 4\over30 } & { 2\over30 } \\[4pt]
{ 6\over12 } & { 15\over40} & {    0     } & { 2\over24 } & { 1\over24 } \\[4pt]
{ 6\over12 } & { 10\over30} & { 4\over24 } & {     0    } & {    0     } \\[4pt]
{ 6\over12 } & { 10\over30} & { 4\over24 } & {     0    } & {    0     }
\end{array}\right]
\eeq

This special operation is uniquely determined by the requirements that
the result be reversible with respect to $\pi$, that it not alter
transition probabilities to or from $\sigma(j)$ for $j<i$ that were
found in previous steps, that transition probabilities among the
$\sigma(j)$ with $j>i+1$ be zero, and that transition probabilities
to $\sigma(j)$ for $j>i+1$ from both $\sigma(i)$ and $\sigma(i\!+\!1)$
be proportional to $\pi(\sigma(j))$.

The derivation of the general scheme can be illustrated with reference
to the NAM transition matrix shown in~(\ref{NAM-vis2}), which represents
the result of DNAM when $m=5$, $\sigma(i)=i$, and a non-zero self transition 
probability is produced at step $3$.  At step $2$, the ZDNAM 
method will alter the matrix produced so that it instead has the following form:
\beq P^* & = &
 \left[ \begin{array}{ccccc}\displaystyle
 0 &\displaystyle \pi(2)\,{f_0 \over s_1}\ \
   &\displaystyle \pi(3)\,{f_0 \over s_1}\ \
   &\displaystyle \pi(4)\,{f_0 \over s_1}\ \
   &\displaystyle \pi(5)\,{f_0 \over s_1}\\[14pt]\displaystyle
 \displaystyle \pi(1)\,{f_0 \over s_1}\ \ & 0
   &\displaystyle {1\over\pi(2)}Af_1\ \ 
   &\displaystyle {\pi(4)\over\pi(2)}Bf_1\ \ 
   &\displaystyle {\pi(5)\over\pi(2)}Bf_1\\[14pt]\displaystyle
 \displaystyle \pi(1)\,{f_0 \over s_1}\ \ 
   &\displaystyle {1\over\pi(3)}Af_1\ \ 
   &\displaystyle 0\ \
   &\displaystyle {\pi(4)\over\pi(3)}Cf_1\ \ 
   &\displaystyle {\pi(5)\over\pi(3)}Cf_1\,\\[14pt]\displaystyle
 \displaystyle \pi(1)\,{f_0 \over s_1}\ \ 
   &\displaystyle Bf_1\ \
   & Cf_1
   & 0 & 0 \\[14pt]\displaystyle
 \displaystyle \pi(1)\,{f_0 \over s_1}\ \ 
   &\displaystyle Bf_1\ \ 
   & Cf_1
   & 0 & 0
 \end{array}\right]\label{ZDNAM-vis}
\eeq
This transition matrix is reversible with respect to $\pi$ by construction.
$A$, $B$, and $C$ can be found from the requirement that the rows sum to one.

I will now switch to using a general notation, with $i$ being the step at
which ZDNAM recognizes that \mbox{$\pi(\sigma(i\!+\!1))>s_{i+1}$,} and hence the
special construction is needed.  The example above has $i=2$ and
$\sigma(i)=i$.  Recall that $s_i$ is the sum of $\pi(\sigma(j))$ for
all $j>i$, and that for any $k>i$, $f_i$ is the sum of transition
probabilities from $\sigma(k)$ to $\sigma(j)$ for all $j>i$.

When finding $A$, $B$, and $C$, the requirement that rows of the
matrix sum to one is equivalent to requiring that for
$k\ge i$, the sum of $P^*(\sigma(k)\rightarrow\sigma(j))$ for $j\ge i$
must be $f_{i-1}$.  This gives the following equations:
\beq
   A\ +\ Bs_{i+1} \ = \ \pi(\sigma(i)),\ \ \ \
   A\ +\ Cs_{i+1} \ = \ \pi(\sigma(i\!+\!1)),\ \ \ \
   B\ +\ C \ = \ 1
\eeq
Solving this system of equations, we get
\beq
   A = {\pi(\sigma(i))+\pi(\sigma(i\!+\!1))-s_{i+1} 
           \over 2},\ 
   B = {\pi(\sigma(i))-\pi(\sigma(i\!+\!1))+s_{i+1} 
           \over 2\,s_{i+1}},\ 
   C = {s_{i+1}+\pi(\sigma(i\!+\!1))-\pi(\sigma(i))
           \over 2\,s_{i+1}} \ \label{eq-ABC}
\eeq

Algorithm~\ref{alg-ZDNAM} implements this procedure.  As in
Algorithm~\ref{alg-DNAM} for DNAM, it starts by handling the case
where the current value has probability $1/2$ or more specially, which
avoids the need to sort by probability.  The case where the
most-probable value has probability $1/2$ or more is also handled
specially. Otherwise, the DNAM procedure is applied for $i$ from 1 on
up, until the current value is reached in the ordering found, while
also checking whether the \text{next} step, $i\!+\!1$, will be
one in which $\pi(\sigma(i\!+\!1))\ \ge\ s_{i+1}$, and hence the
special construction will be used. Because of this forward check, no
check for whether $\pi(\sigma(i))\ \ge\ s_i$ is needed within the
loop.

\begin{algorithm}[p]

\begin{tabbing}

\hspace{0.5in}\=
\bf Input:\ \ \ \ \ \ \ \ \ \ \ \ \ \ \ \ \ \ \ \= Gibbs sampling probabilities,
                   $\pi^{\rule{0pt}{1pt}}(i)$, for $i=1,\ldots,m$
\\ \>
                \> The current state value, $k$, in $\{1,\ldots,m\}$
\\[5pt] \>
\bf Output:     \> ZDNAM transition probabilities, $p(i)$, for $i=1,\ldots,m$
\\[-1pt]
\> \hspace*{20pt} \= \hspace*{20pt} \= \hspace*{20pt} \= \hspace*{20pt} \= 
   \hspace*{20pt} \= \hspace*{20pt} \= \hspace*{90pt} \= \\[-3pt]
\> If $\pi(k) \ge 1/2$: \\[4pt]
\>\> \textit{Quickly handle the case where the current value has probability 
             half or more,} \\
\>\> \textit{without needing to order values by probability} \\[4pt]
\>\> For $i = 1,\ldots,m$: \\
\>\>\> If $i \ne k$: \\
\>\>\>\> Set $p(i)$ to $\min(1,\,\pi(i)/\pi(k))$ \>\>\>\>
           \textit{Min guards against round-off error} \\[4pt]
\>\> Set $p(k)$ to $(2\pi(k)-1)\,/\,\pi(k)$ \\[4pt]
\> Else: \\[4pt]
\>\> Set $\sigma$ to some permutation on 
   $\{1,\ldots,m\}$ for which $\pi(\sigma(i)) \ge \pi(\sigma(j))$
   when $i \le j$\\[4pt]
\>\> If $\pi(\sigma(1)) \ge 1/2$: \\[4pt]
\>\>\> \textit{Handle the case where a value has probability of 1/2 or more. 
        Won't be the}\\
\>\>\> \textit{current value, since that's handled above.} \\[4pt]
\>\>\> Set $p(\sigma(1))$ to $1$ \\[2pt]
\>\>\> For $i = 2,\ldots,m$:\\
\>\>\>\> Set $p(\sigma(i))$ to 0 \\[4pt]
\>\> Else: \\[4pt]
\>\>\> Set $s$ to 1
   \>\>\>\textit{The sum of probabilities for values that have not yet
                 been focal} \\[4pt]
\>\>\> Set $f$ to 1 
   \>\>\>\textit{The sum of transition probabilities from k to values
                 not yet focal} \\[7pt]
\>\>\> \textit{Find modified transition probabilities from the current value to
           successive focal values,} \\
\>\>\> \textit{until the focal value is the current value, or special handling
              to avoid a non-zero}\\
\>\>\> \textit{self transition probability is needed.}\\[4pt]
\>\>\> Set $i$ to 1 \\[4pt]
\>\>\> While $f>0$ and $\sigma(i) \ne k$ and $\pi(\sigma(i\!+\!1))
                                < s-\pi(\sigma(i))-\pi(\sigma(i\!+\!1)$: \\[4pt]
\>\>\>\> \textit{Let $q$ be the probability of the focal value; 
             update $s$ to be the sum}\\
\>\>\>\> \textit{of probabilities for remaining non-focal values} \\[4pt]
\>\>\>\> Set $q$ to $\pi({\sigma(i)})$ \\
\>\>\>\> Subtract $q$ from $s$ \>\>\>\> 
         \textit{Sets variable s to $s_i$, guaranteed positive}\\[4pt]
\>\>\>\> \textit{Compute the transition probability from the current value, k, 
                 to the focal value,}\\
\>\>\>\>\textit{and find the new total probability for transitions to remaining 
                values} \\[4pt]
\>\>\>\> Set $p({\sigma(i)})$ to $(q / s) f$ 
  \>\>\>\> \textit{Guaranteed $p(\sigma(i))\le f\le 1$, even with rounding} \\
\>\>\>\> Subtract $p({\sigma(i)})$ from $f$ 
  \>\>\>\> \textit{Sets variable f to $f_i$, was previously $f_{i-1}$} \\[4pt]
\>\>\>\> Add 1 to $i$ \\[4pt]

\>\>\> Set $q$ to $\pi(\sigma(i))$ \\
\>\>\> Subtract $q$ from $s$ \\[4pt]
\>\>\> Continue with the procedure of Algorithm~\ref{alg-ZDNAM}: Part 2.

\end{tabbing}\vspace{6pt}

\caption{Part 1. Procedure for computing ZDNAM transition 
probabilities.}\label{alg-ZDNAM}

\end{algorithm}

\addtocounter{algorithm}{-1}

\begin{algorithm}[p]

\begin{tabbing}
\hspace{0.5in}\=
\hspace*{20pt} \= \hspace*{20pt} \= \hspace*{20pt} \= \hspace*{20pt} \= 
   \hspace*{20pt} \= \hspace*{20pt} \= \hspace*{20pt} \= 
   \hspace*{20pt} \= \hspace*{40pt} \= \\[-3pt]

\>\>\> Continuation of Algorithm~\ref{alg-ZDNAM}: Part 1.\\[6pt]

\>\>\>If $f > 0$ and $s > 0$ and $i<m$:\\[4pt]
\>\>\>\> Set $q_2$ to $\pi(\sigma(i\!+\!1))$ \\
\>\>\>\> Set $s_2$ to 
 $\max\,(0,\,s-q_2)$
   \>\>\>\>\>\>\textit{max guards against round-off error}\\[4pt]
\>\>\>\> If $q_2 \ge s_2$: \\[4pt]
\>\>\>\>\> \textit{Use the special construction to avoid a non-zero 
                   self transition probability.} \\[4pt]
\>\>\>\>\> Set $A$ to $(q+q_2-s_2)\,/\,2$ \\
\>\>\>\>\> If $k=\sigma(i)$: \\
\>\>\>\>\>\> Set $p(\sigma(i))$ to 0 \\
\>\>\>\>\>\> Set $p(\sigma(i\!+\!1)$ to $fA/q$ \\
\>\>\>\>\> Else If $k=\sigma(i\!+\!1)$: \\
\>\>\>\>\>\> Set $p(\sigma(i))$ to $fA/q_2$ \\
\>\>\>\>\>\> Set $p(\sigma(i\!+\!1)$ to 0 \\[4pt]
\>\>\>\>\> If $s_2 \le 0$: \\
\>\>\>\>\>\> Add 2 to $i$ \\
\>\>\>\>\> Else: \\
\>\>\>\>\>\> Set $B$ to $(q-q_2+s_2)\,/\,(2s_2)$ \\
\>\>\>\>\>\> Set $C$ to $(s_2+q_2-q)\,/\,(2s_2)$ \\
\>\>\>\>\>\> If $k=\sigma(i)$: \\
\>\>\>\>\>\>\> Add 2 to $i$ \\
\>\>\>\>\>\>\> While $i \le m$: \\
\>\>\>\>\>\>\>\> Set $p(\sigma(i))$ to $fB\pi(\sigma(i))/q$ \\
\>\>\>\>\>\>\>\> Add 1 to $i$ \\
\>\>\>\>\>\> Else If $k=\sigma(i\!+\!1)$: \\
\>\>\>\>\>\>\> Add 2 to $i$ \\
\>\>\>\>\>\>\> While $i \le m$: \\
\>\>\>\>\>\>\>\> Set $p(\sigma(i))$ to $fC\pi(\sigma(i))/q_2$ \\
\>\>\>\>\>\>\>\> Add 1 to $i$ \\
\>\>\>\>\>\> Else: \\
\>\>\>\>\>\>\> Set $p(\sigma(i))$ to $fB$ \\
\>\>\>\>\>\>\> Set $p(\sigma(i\!+\!1))$ to $fC$ \\
\>\>\>\>\>\>\> Add 2 to $i$ \\
\>\>\>\> Else: \\[2pt]
\>\>\>\>\> \textit{Compute modified transition probabilities from the 
           current value, k,} \\
\>\>\>\>\> \textit{which is now focal, to values that have not 
           previously been focal.} \\[4pt]
\>\>\>\>\> Set $p(\sigma(i))$ to 0 \\
\>\>\>\>\> Add 1 to $i$ \\
\>\>\>\>\> While $i \le m$: \\
\>\>\>\>\>\> Set $p(\sigma(i))$ to $(\pi(\sigma(i))\, /\, s)\, f$ \\
\>\>\>\>\>\> Add 1 to $i$ \\[4pt]
\>\>\> \textit{Set any remaining transition probabilities to zero.}\\[4pt]
\>\>\> While $i \le m$: \\
\>\>\>\> Set $p(\sigma(i))$ to 0 \\
\>\>\>\> Add 1 to $i$

\end{tabbing}\vspace{1pt}

\caption{Part 2. Continuation of procedure for computing 
                 ZDNAM transition probabilities.}

\end{algorithm}

If the special construction is needed, the values $A$, $B$, and $C$
of~(\ref{eq-ABC}) are computed and used, taking care to avoid division
by zero.

As is the case for other NAM methods, the ZDNAM algorithm computes
transition probabilities sequentially, and hence can easily be
modified to sample a value from the transition distribution based on a
uniform random variate, terminating once the cumulative probability
exceeds the uniform variate. The possibilities for handling distributions
with a countably infinite number of values are similar to DNAM.

The reduction in self transition probability for ZDNAM compared to
DNAM is not uniformly beneficial --- it is not always the case that
the ZDNAM transition matrix efficiency-dominates the DNAM transition
matrix.  This can be seen, for example, when $m=3$ and $\pi(1)=4/9$,
$\pi(2)=3/9$, and $\pi(3)=2/9$, for which
\beq
  P^*_{\mbox{\tiny DNAM}}\ \ =\ \ \left[\begin{array}{ccc}
    0         & {9\over15}& {6\over15}\\[4pt]
  {12\over15} & {1\over15}& {2\over15}\\[4pt]
  {12\over15} & {3\over15}& 0
  \end{array}\right],\ \ \ \ \
  P^*_{\mbox{\tiny ZDNAM}}\ \ =\ \ \left[\begin{array}{ccc}
    0         & {15\over24}& {9\over24}\\[4pt]
  {20\over24} & 0          & {4\over24}\\[4pt]
  {18\over24} & {6\over24}& 0
  \end{array}\right]
\eeq
Numerical calculation finds that the eigenvalues of 
$P^*_{\mbox{\tiny DNAM}} - P^*_{\mbox{\tiny ZDNAM}}$ are $0.10306$,
$-0.03639$, and zero. Since their signs are mixed, Theorem~9 of
(Neal and Rosenthal 2023) shows that neither $P^*_{\mbox{\tiny DNAM}}$
nor $P^*_{\mbox{\tiny ZDNAM}}$ efficiency-dominates the other.

A ZDNAM transition probability matrix has eigenvalues and eigenvectors
that can be associated with each step followed when constructing it.
Until the special construction is used, when $\pi(\sigma(i+1))\ge
s_{i+1}$, these are the same as for any NAM procedure, as described in
Section~\ref{sec-NAM} (e.g., equation~(\ref{reigenvec})).  Two 
eigenvalues and eigenvectors are associated with steps $i$ and $i+1$
when the special construction is applied at step $i$.  Assuming for
notational simplicity that the non-increasing ordering is $\sigma(i)=i$,
these two eigenvalues are given by
\beq \lambda & = & 
-{f_{i-1} \over 2} \left[ \,1\ \pm\
  \sqrt{1\ -\ \big(\pi(i)-\pi(i\!+\!1)+s_{i+1}\big)\,
              \big(\pi(i\!+\!1)^2-(\pi(i)-s_{i+1})^2\big)\,/\,
              \big(\pi(i)\pi(i\!+\!1)s_{i+1}\big)
     \rule{0pt}{10pt}}\rule{0pt}{12pt}\
  \right]\ \ \ \ \
\label{eq-zdnam-eig}\eeq
An associated right eigenvector for such a $\lambda$ is
\beq v & = &
  [\, 0,\, \ldots,\ \
   Cf_{i-1}s_{i+1}+\lambda\pi(i\!+\!1),\ \
   -Bf_{i-1}s_{i+1}-\lambda\pi(i),\ \
   Bf_{i-1}\pi(i\!+\!1)-Cf_{i-1}\pi(i),\ \ \ldots\,]^T\ \ \
\label{eq-zdnam-eigv}
\eeq
where there are $i\!-\!1$ leading zero elements in the vector, and the
elements after position $i+1$ are all the same.\footnote{\rule{0pt}{8pt}%
Here is the proof that either one of the
$\lambda$ of~(\ref{eq-zdnam-eig}), which can be written as 
\mbox{$\lambda\,=\,-(f_{i-1}/2)\big[1\pm\sqrt{D}\,\big]$}, in which
$D\ =\ 1\ -\ \big(\pi(i)-\pi(i\!+\!1)+s_{i+1}\big)\,
              \big(\pi(i\!+\!1)^2-(\pi(i)-s_{i+1})^2\big)\,/\,
              \big(\pi(i)\pi(i\!+\!1)s_{i+1}\big)$,
is an eigenvalue of the ZDNAM transitions $P^*$ visualized in~(\ref{ZDNAM-vis})
with the corresponding $v$ from~(\ref{eq-zdnam-eigv}) as an associated
eigenvector.\vspace{4pt}

\noindent We first show that 
$[P^*v]_j=0$ for $j<i$:\vspace{-8pt}
\beq [P^*v]_j\! & \!\!\!=\!\!\! &
 (Cf_{i-1}s_{i+1}+\lambda\pi(i\!+\!1))P^*(j\rightarrow i) 
 - (Bf_{i-1}s_{i+1}+\lambda\pi(i))P^*(j\rightarrow i\!+\!1)
 +\!\! \sum_{k=i+2}^m\! (Bf_{i-1}\pi(i\!+\!1)-Cf_{i-1}\pi(i))P^*(j\rightarrow k)
 \nonumber \\[-6pt]
 & \!\!\!=\!\!\! & {f_{j-1}\over s_j}\,\Big[
 (Cf_{i-1}s_{i+1}+\lambda\pi(i\!+\!1))\,\pi(i)
 \,-\, (Bf_{i-1}s_{i+1}+\lambda\pi(i))\,\pi(i\!+\!1)
 \,+\, (Bf_{i-1}\pi(i\!+\!1)-Cf_{i-1}\pi(i))\,s_{i+1}
 \Big]\ =\ 0
\nonumber\eeq

\noindent Next, we see that\vspace{-8pt}
\beq [P^*v]_i\! & \!\!\!=\!\!\! &
 -(Bf_{i-1}s_{i+1}+\lambda\pi(i))P^*(i\rightarrow i\!+\!1)
 \,+\! \sum_{k=i+2}^m\! (Bf_{i-1}\pi(i\!+\!1)-Cf_{i-1}\pi(i))P^*(i\rightarrow k)
 \nonumber\\[-1pt]
 & \!\!=\!\! & \big(f_{i-1}/\pi(i)\big)\,\big(
 -(Bf_{i-1}s_{i+1}+\lambda\pi(i))\,A
 \,+\, (Bf_{i-1}\pi(i\!+\!1)-Cf_{i-1}\pi(i))\,Bs_{i+1}
 \,\big) 
 \nonumber\\[2pt]
 & \!\!=\!\! & -\big(f_{i-1}^2/(2\pi(i))\big)\,\big(
   \big((\pi(i)\!-\!\pi(i\!+\!1)\!+\!s_{i+1})\,-\,
     \big[1\pm\sqrt{D}\big]\pi(i)\big)A
     \, -\, (B\pi(i\!+\!1)\!-\!C\pi(i))\,(\pi(i)\!-\!\pi(i\!+\!1)\!+\!s_{i+1})
 \,\big)\ \ \ \ \
 \nonumber\\[2pt]
 & \!\!=\!\! & -\big(f_{i-1}^2/4\big)\,\big(
 (-\!\pi(i\!+\!1)\!+\!s_{i+1})\,(\pi(i)\!+\!\pi(i\!+\!1)\!-\!s_{i+1})\,/\,\pi(i)
     \ \mp\ (\pi(i)+\pi(i\!+\!1)-s_{i+1})\sqrt{D}
 \nonumber\\
 && \ \ \ \ \ \ \ \ \ \ \ \ \ \ \ \ \
  -\,((\pi(i)-\pi(i\!+\!1)+s_{i+1})\pi(i\!+\!1)
        - (s_{i+1}+\pi(i\!+\!1)-\pi(i))\pi(i))
          \,(\pi(i)\!-\!\pi(i\!+\!1)\!+\!s_{i+1})
     \,/\,(\pi(i)s_{i+1})
 \,\big)
 \nonumber\\[2pt]
 & \!\!=\!\! & -\big(f_{i-1}^2/4\big)
       \,\big(\pm(s_{i+1}\!-\!\pi(i)\!-\!\pi(i\!+\!1))\sqrt{D} 
       \ +\ (\pi(i)\pi(i\!+\!1)^2\!+\!\pi(i)^2\pi(i\!+\!1)\!+\!
             \pi(i\!+\!1)s_{i+1}^2\!+\!\pi(i\!+\!1)^2s_{i+1}
    \ \ \ \nonumber \\
 && \ \ \ \ \ \ \ \ \ \ \ \ \ \ \ \ \ \ \ \ \ \ \ \ \ \ \ \ \ \ \ \ \ \ \ \ \
    \ \ \ \ \ \ \ \ \ \ \ \ \ \ \ \ \ \ \ \ \ \
     +\,2\pi(i)s_{i+1}^2\!-\!\pi(i)^3\!-\!\pi(i\!+\!1)^3
       \!-\!s_{i+1}^3\!-3\pi(i)\pi(i\!+\!1)s_{i+1})
       \,/\,(\pi(i)s_{i+1})\,\big)
 \nonumber\\[2pt]
 & \!\!=\!\! & -\big(f_{i-1}^2/4\big)\,\big[1\pm\sqrt{D}\,\big]\,
      \big(s_{i+1}+\pi(i\!+\!1)-\pi(i)-\big[1\pm\sqrt{D}\,\big]\pi(i\!+\!1)\big)
 \nonumber\\[2pt]
 & \!\!=\!\! & \lambda\,(Cf_{i-1}s_{i+1}+\lambda\pi(i\!+\!1))
\nonumber\eeq

\noindent In similar fashion, we have:\vspace{-8pt}
\beq [P^*v]_{i+1}\!\!\! & \!\!\!=\!\!\! &
 (Cf_{i-1}s_{i+1}+\lambda\pi(i\!+\!1))P^*(i\!+\!1\rightarrow i)
 \,+\! \sum_{k=i+2}^m\! (Bf_{i-1}\pi(i\!+\!1)-Cf_{i-1}\pi(i))P^*(i\rightarrow k)
 \nonumber\\[-1pt]
 & \!\!=\!\! & \big(f_{i-1}/\pi(i\!+\!1)\big)\,\big(
 (Cf_{i-1}s_{i+1}+\lambda\pi(i\!+\!1))A
 \,+\, (Bf_{i-1}\pi(i\!+\!1)-Cf_{i-1}\pi(i))\,Cs_{i+1}
 \,\big) 
 \nonumber\\[2pt]
 & \!\!=\!\! & -\big(f_{i-1}^2/(2\pi(i\!+\!1))\big)\,\big(
 (-(s_{i+1}\!+\!\pi(i\!+\!1)\!-\!\pi(i))+\big[1\pm\sqrt{D}\,\big]\pi(i\!+\!1))A
 \,-\, (B\pi(i\!+\!1)\!-\!C\pi(i))\,(s_{i+1}\!+\!\pi(i\!+\!1)\!-\!\pi(i))
 \,\big) 
 \nonumber\\[2pt]
 & \!\!=\!\! & -\big(f_{i-1}^2/4)\big)\,\big(
 (-s_{i+1}\!+\!\pi(i))\,(\pi(i)\!+\!\pi(i\!+\!1)-s_{i+1})\,/\,\pi(i\!+\!1)
  \ \pm\ (\pi(i)\!+\!\pi(i\!+\!1)-s_{i+1})\sqrt{D} 
 \nonumber\\
 && \ \ \ \ \ \ \ \ \ \ \ \ \ \ \ \ \
  -\,((\pi(i)-\pi(i\!+\!1)+s_{i+1})\pi(i\!+\!1)
        - (s_{i+1}+\pi(i\!+\!1)-\pi(i))\pi(i))
      \,(s_{i+1}\!+\!\pi(i\!+\!1)\!-\!\pi(i))
     \,/\,(\pi(i\!+\!1)s_{i+1})
 \,\big)
 \nonumber\\[2pt]
 & \!\!=\!\! & -\big(f_{i-1}^2/4\big)\,\big(
   \pm\ (\pi(i)\!+\!\pi(i\!+\!1)-s_{i+1})\sqrt{D}\ +\ 
   (\pi(i)\pi(i\!+\!1)^2-\pi(i)^2\pi(i\!+\!1)-\pi(i)s_{i+1}^2-\pi(i)^2s_{i+1}
   \nonumber\\
 && \ \ \ \ \ \ \ \ \ \ \ \ \ \ \ \ \ \ \ \ \ \ \ \ \ \ \ \ \ \ \ \ \ \ \ \ \
    \ \ \ \ \ \ \ \ \ \ \ \ \ 
    -\,2\pi(i\!+\!1)s_{i+1}^2\!+\!\pi(i)^3\!+\!\pi(i\!+\!1)^3\!+\!s_{i+1}^3
    \!+3\pi(i)\pi(i\!+\!1)s_{i+1})
      \,/\,(\pi(i\!+\!1)s_{i+1})
 \,\big) 
 \nonumber\\[2pt]
 & \!\!=\!\! & -\big(f_{i-1}^2/4\big)\,\big[1\pm\sqrt{D}\,\big]\,
      \big(-\pi(i)+\pi(i\!+\!1)-s_{i+1}+\big[1\pm\sqrt{D}\,\big]\pi(i)\big)
 \nonumber\\[2pt]
 & \!\!=\!\! & \lambda\,(-Bf_{i-1}s_{i+1}-\lambda\pi(i))
\nonumber\eeq

\noindent Finally, for $j>i+1$, we have that\vspace{-3pt}
\beq [P^*v]_j\! & \!\!=\!\! &
 (Cf_{i-1}s_{i+1}+\lambda\pi(i\!+\!1))P^*(j\rightarrow i) 
 \,-\, (Bf_{i-1}s_{i+1}+\lambda\pi(i))P^*(j\rightarrow i\!+\!1)
 \nonumber \\[2pt]
 & \!\!=\!\! &
 (Cf_{i-1}s_{i+1}+\lambda\pi(i\!+\!1))Bf_{i-1}
 \,-\, (Bf_{i-1}s_{i+1}+\lambda\pi(i))Cf_{i-1}
 \nonumber \\[2pt]
 & \!\!=\!\! & \lambda\, (Bf_{i-1}\pi(i\!+\!1)-Cf_{i-1}\pi(i))
\nonumber\eeq\vspace*{-10pt}
}
The eigenvalues after those associated with steps $i$ and $i+1$ are
all zero.

The eigenvalues of a ZDNAM transition matrix are all zero or negative,
apart from the one eigenvalue of 1 with eigenvector $[1,\ldots,1]^T$.
This is so for the eigenvalues associated with the NAM steps before
step $i$, where the special construction is needed, as demonstrated in
Section~\ref{sec-NAM}.  The two eigenvalues given
by~(\ref{eq-zdnam-eig}) are also negative --- the value of the square
root is less than one, hence the quantity in square brackets is
positive, and the eigenvalue is negative. To see that the square root
is less than one, note first that $\pi(i)\ge\pi(i\!+\!1)$, since the
ordering is non-increasing, and hence the factor
$(\pi(i)-\pi(i\!+\!1)+s_{i+1})$ is positive.  Also,
$\pi(i)<\pi(i\!+\!1)+s_{i+1}$, since otherwise the special
construction would have been used before step $i$, and
$\pi(i\!+\!1)\ge s_{i+1}$, since the special construction was used at
step $i$, and hence $\pi(i)\ge s_{i+1}$.  It follows that
$0\le\pi(i)-s_{i+1}<\pi(i\!+\!1)$, and hence the factor
$(\pi(i\!+\!1)^2-(\pi(i)-s_{i+1})^2)$ is positive.

Since the eigenvalues (apart from the single 1) are all zero or
negative, Corollary~15 of (Neal and Rosenthal 2023) can then be applied
to show that ZDNAM transitions efficiency-dominate Gibbs sampling.  As
discussed for antithetic modifications in Section~\ref{sec-AM},
Theorem~12 of (Neal and Rosenthal 2023) allows us to then conclude
that using ZDNAM to update a randomly selected variable
efficiency-dominates using Gibbs sampling with such random updates.

\section{\hspace*{-8pt}
  The Shifted Tower (ST) and Half Shifted Tower (HST) 
  methods}\label{sec-ST}\vspace{-11pt}

Suwa and Todo (2010) and Suwa (2022) describe a class of methods for
defining transition probabilities that can be viewed in terms of
building a ``tower'' of probabilities for values, applying a circular
shift operation to produce a second tower, and then defining
transition probabilities by the alignment of the first and second
towers.  

The first method, of Suwa and Todo (2010), shifts by the probability
of the most probable value. I will refer to this as the Shifted
Tower (ST) method.  It always reduces self transitions to the minimum
possible.  Unlike all the methods considered previously in this paper,
it may produce non-reversible transitions (though
note that when there are only two possible values, transitions leaving $\pi$
invariant are always reversible with respect to $\pi$).  Suwa (2022)
generalized this method to an arbitrary shift, and in particular noted
that shifting by $1/2$ minimizes self transitions while also producing
transitions that are reversible.  I call this the Half Shifted Tower
(HST) method.

\begin{figure}[p]
\begin{center}

\setlength{\unitlength}{2.45in}
\begin{picture}(3.02,1.13)

  \put(0.0,1.05){\makebox(0.8,0.08){Shifted Tower}}

  \put(0.0,0.0){\framebox(0.3,0.4){1 (0.4)}}
  \put(0.0,0.4){\framebox(0.3,0.3){2 (0.3)}}
  \put(0.0,0.7){\framebox(0.3,0.1){3 (0.1)}}
  \put(0.0,0.8){\framebox(0.3,0.2){4 (0.2)}}

  \put(0.35,0.05){\vector(1,0){0.1}}
  \put(0.35,0.15){\vector(1,0){0.1}}
  \put(0.35,0.25){\vector(1,0){0.1}}
  \put(0.35,0.35){\vector(1,0){0.1}}
  \put(0.35,0.45){\vector(1,0){0.1}}
  \put(0.35,0.55){\vector(1,0){0.1}}
  \put(0.35,0.65){\vector(1,0){0.1}}
  \put(0.35,0.75){\vector(1,0){0.1}}
  \put(0.35,0.85){\vector(1,0){0.1}}
  \put(0.35,0.95){\vector(1,0){0.1}}

  \put(0.5,0.4){\framebox(0.3,0.4){1 (0.4)}}
  \put(0.5,0.8){\framebox(0.3,0.2){2 (0.2)}}
  \put(0.5,0.0){\framebox(0.3,0.1){2 (0.1)}}
  \put(0.5,0.1){\framebox(0.3,0.1){3 (0.1)}}
  \put(0.5,0.2){\framebox(0.3,0.2){4 (0.2)}}

  \put(0.77,0){\makebox(0.6,1){$\displaystyle
    \left[\begin{array}{cccc}
      0 & {1\over4} & {1\over4} & {1\over2} \\[4pt]
      1 & 0 & 0 & 0 \\[4pt]
      1 & 0 & 0 & 0 \\[4pt]
      0 & 1 & 0 & 0
    \end{array}\right]
  $}}

  \put(1.56,1.05){\makebox(0.8,0.08){Half Shifted Tower}}

  \put(1.56,0.0){\framebox(0.3,0.4){1 (0.4)}}
  \put(1.56,0.4){\framebox(0.3,0.3){2 (0.3)}}
  \put(1.56,0.7){\framebox(0.3,0.1){3 (0.1)}}
  \put(1.56,0.8){\framebox(0.3,0.2){4 (0.2)}}

  \put(1.91,0.05){\vector(1,0){0.1}}
  \put(1.91,0.15){\vector(1,0){0.1}}
  \put(1.91,0.25){\vector(1,0){0.1}}
  \put(1.91,0.35){\vector(1,0){0.1}}
  \put(1.91,0.45){\vector(1,0){0.1}}
  \put(1.91,0.55){\vector(1,0){0.1}}
  \put(1.91,0.65){\vector(1,0){0.1}}
  \put(1.91,0.75){\vector(1,0){0.1}}
  \put(1.91,0.85){\vector(1,0){0.1}}
  \put(1.91,0.95){\vector(1,0){0.1}}

  \put(2.06,0.5){\framebox(0.3,0.4){1 (0.4)}}
  \put(2.06,0.9){\framebox(0.3,0.1){2 (0.1)}}
  \put(2.06,0.0){\framebox(0.3,0.2){2 (0.2)}}
  \put(2.06,0.2){\framebox(0.3,0.1){3 (0.1)}}
  \put(2.06,0.3){\framebox(0.3,0.2){4 (0.2)}}

  \put(2.34,0){\makebox(0.6,1){$\displaystyle
    \left[\begin{array}{cccc}
      0 & {1\over2} & {1\over4} & {1\over4} \\[4pt]
      {2\over3} & 0 & 0 & {1\over3} \\[4pt]
      1 & 0 & 0 & 0 \\[4pt]
      {1\over2} & {1\over2} & 0 & 0
    \end{array}\right]
  $}}
\end{picture}

\end{center}

\caption{The Shifted Tower (ST) and Half Shifted Tower (HST)
methods. In this example, values 1, 2, 3, and 4 have probabilities
of 0.4, 0.3, 0.1, and 0.2.  On the left for each method is the tower
of regions for each value, with heights proportional to their
probabilities.  On the right of this tower is a shifted tower,
with regions that move out of the top moving into the bottom, which
may result in the region for a value being split between top and
bottom.  For the ST method, the shift is by the probability of the
most probable symbol.  For the HST method, the shift is always by
$1/2$. Transitions are defined by randomly sampling from the region of
the left tower corresponding to the current value, then following the
arrows right to a region of the shifted tower.  The resulting matrices
of transition probabilities are shown to the right.}\label{fig-ST}

\end{figure}
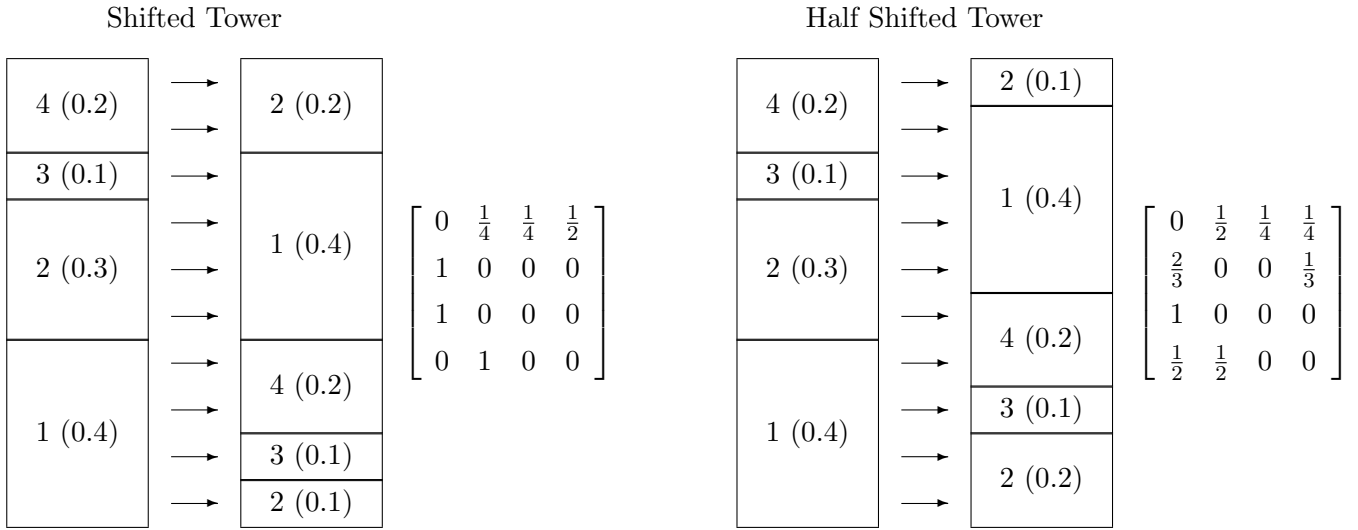

The ST and HST methods are illustrated in Figure~\ref{fig-ST}.
Algorithm~\ref{alg-ST} implements these methods, for any specified shift,
and any ordering of values, using a formula adapted from one given by 
Suwa (2022).\footnote{\rule{0pt}{10pt}%
Suwa's formula appears to erroneously treat the values as having 
the reverse of their specified order, comparing to Fig.~1 of Suwa (2022),
though this has no practical effect if the ordering was arbitrary anyway.
The formula used in Algorithm~\ref{alg-ST} corrects for this. Note that
$F_i$ in Suwa's formulas~(12) and~(13) corresponds to 
$C(i)+\pi(i)$ in the notation used here.}

For a given shift amount, $s \in [0,1]$, and ordering of values, $\sigma$,
the formula computes the ``flow'' from value $k$ to value $i$,
defined by $v_{ki}\, =\, \pi(k)\,P^*(k\rightarrow i)$, as
\beq
   v_{ki} & = & \max\,(0,\,\min\,(\Delta_1,\,\pi(k)+\pi(i)-\Delta_1,\,
                                   \pi(k),\,\pi(i))) \nonumber\\
   & & \ \ \ +\ \max\,(0,\,\min\,(\Delta_2,\,\pi(k)+\pi(i)-\Delta_2,\,
                                   \pi(k),\,\pi(i))) \label{ST-formula}\\[4pt]
   & & \mbox{where $\Delta_1 \ = \ \pi(k) - s + C_k - C_i$ and
                   $\Delta_2 \ = \ \Delta_1 + 1$} \nonumber
\eeq
Here, $C_k$ is the sum of probabilities for values before $k$ in the
ordering $\sigma$. We can compute these as follows:
\beq
   C_{\sigma(i)} & = & \sum_{j=1}^{i-1} \pi(\sigma(j))
\eeq
Once $v_{ki}$ has been computed, we can find the transition probability
from $k$ to $i$ as 
\beq 
  P^*(k \rightarrow i) & = & v_{ki}\, /\, \pi(k)
\eeq

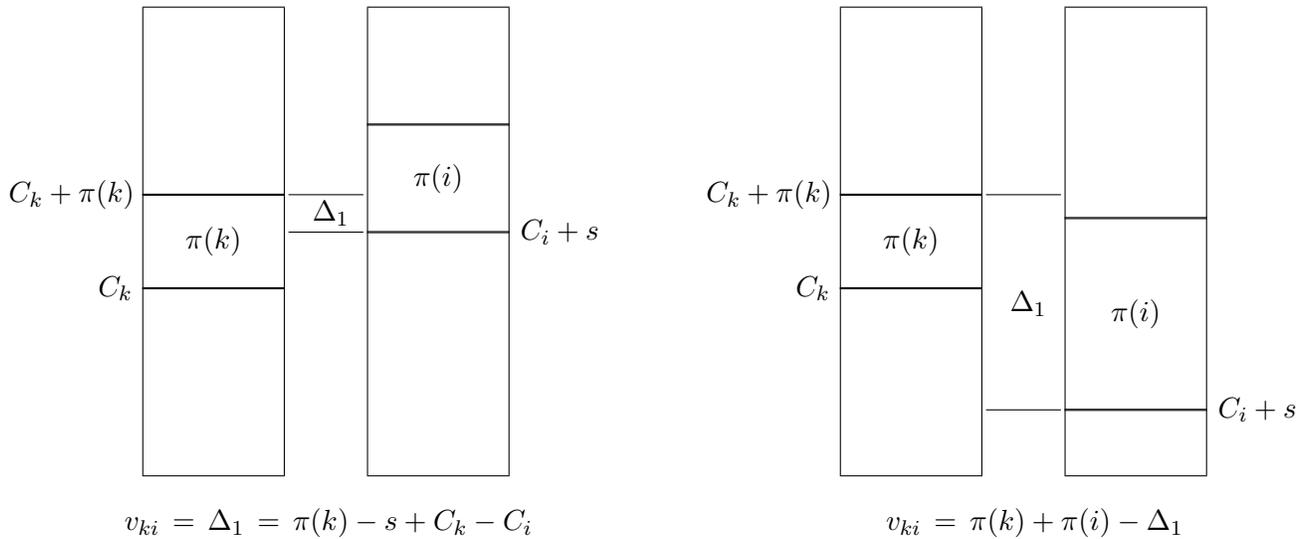
\begin{figure}[p]

\begin{center}

\setlength{\unitlength}{2.45in}
\hspace*{0.1in}
\begin{picture}(1.0,1.04)

  \put(0.065,0.35){\makebox(0.05,0.1){$C_k$}}
  \put(0.065,0.55){\makebox(0.05,0.1){\hspace*{-33pt}$C_k+\pi(k)$}}

  \put(0.15,0.0){\framebox(0.3,0.4){}}
  \put(0.15,0.4){\framebox(0.3,0.2){$\pi(k)$}}
  \put(0.15,0.6){\framebox(0.3,0.4){}}

  \put(0.47,0.544){$\ \ \Delta_1$}
  \put(0.46,0.52){\line(1,0){0.16}}
  \put(0.46,0.6){\line(1,0){0.16}}

  \put(0.63,0.0){\framebox(0.3,0.52){}}
  \put(0.63,0.52){\framebox(0.3,0.23){$\pi(i)$}}
  \put(0.63,0.75){\framebox(0.3,0.25){}}

  \put(0.94,0.47){\makebox(0.2,0.1){$C_i+s$}}

\end{picture}
\hspace{1.1in}
\begin{picture}(1.1,1.04)

  \put(0.065,0.35){\makebox(0.05,0.1){$C_k$}}
  \put(0.065,0.55){\makebox(0.05,0.1){\hspace*{-33pt}$C_k+\pi(k)$}}

  \put(0.15,0.0){\framebox(0.3,0.4){}}
  \put(0.15,0.4){\framebox(0.3,0.2){$\pi(k)$}}
  \put(0.15,0.6){\framebox(0.3,0.4){}}

  \put(0.47,0.35){$\ \ \Delta_1$}
  \put(0.46,0.14){\line(1,0){0.16}}
  \put(0.46,0.6){\line(1,0){0.16}}

  \put(0.63,0.0){\framebox(0.3,0.14){}}
  \put(0.63,0.14){\framebox(0.3,0.41){$\pi(i)$}}
  \put(0.63,0.55){\framebox(0.3,0.45){}}

  \put(0.94,0.09){\makebox(0.2,0.1){$C_i+s$}}

\end{picture}

\end{center}

\vspace{0.1in}

\hspace*{0.62in} $v_{ki} \,=\, \Delta_1 \,=\, \pi(k) - s + C_k - C_i$%
\hspace*{1.8in} $v_{ki} \,=\, \pi(k)+\pi(i)-\Delta_1$

\caption{Illustration of how $v_{ki}$ for the shifted tower method
can be found in two situations.}\label{fig-ST-formula}

\end{figure}

Figure~\ref{fig-ST-formula} illustrates how the formula for $v_{ki}$
of equation~(\ref{ST-formula}) is derived.  The left of the figure
shows a situation in which $v_{ki}=\Delta_1$, while the right shows a
situation in which $v_{ki}=\pi(k)+\pi(i)-\Delta_1$.  When the shifted
region for value $i$ completely encloses the original region for value
$k$, the flow will be $\pi(k)$, and in the opposite situation, the
flow will be $\pi(i)$.  Taking the minimum of all these possibilities,
and then replacing a negative value by zero, gives the flow in all
situations where wrap-around is not an issue.  To this, we need to
add the value for the flow that is found accounting for the possibility
that after shifting by $s$, the start of the region for value $i$ wraps
around from $1$ to $0$, which we do by replacing $\Delta_1$ by 
$\Delta_2=\pi(k)-s+C_k-(C_i-1)=\Delta_1+1$. The final result is 
given by equation~(\ref{ST-formula}).

In Algorithm~\ref{alg-ST}, this procedure is modified to avoid issues
with round-off error.  Rather than compute $\Delta_2$ as $\Delta_1+1$,
the program sets $\Delta_2$ to $\Delta_1+S$, where $S$ is the sum
of probabilities for all values. If the probabilities are normalized,
one would expect this to be 1, but it may not be due to round-off error.
Similarly, the transition probability $P^*(k\rightarrow i)$ is not
found as $v_{ki}/\pi(k)$, but rather as $v_{ki}\,/\sum_j v_{kj}$, which
guarantees that these transition probabilities are not greater than one
even if $\sum_j v_{kj}$ is not exactly $\pi(k)$.

\begin{algorithm}[p]

\begin{tabbing}

\hspace{0.5in}\=
\bf Input:\ \ \ \ \ \ \ \ \ \ \ \ \ \ \ \ \ \ \ \= Gibbs sampling probabilities,
                   $\pi^{\rule{0pt}{1pt}}(i)$, for $i=1,\ldots,m$
\\ \>
                \> The current state value, $k$, in $\{1,\ldots,m\}$
\\ \>
                \> Amount of shift, $s$, in $(0,1)$
\\ \>
                \> A permutation, $\sigma$, on $\{1,\ldots,m\}$, giving
                   an ordering of values
\\[5pt] \>
\bf Output:     \> ST transition probabilities, $p(i)$, for $i=1,\ldots,m$
\\[5pt] \>
\bf Temporary storage: \>Flows of probability, $v(i)$, from $k$ to each value,
                         for $i=1,\ldots,m$
\\ \>
                      \> Cumulative probabilities, $C(i)$, for $i=1,\ldots,m$,
                          with $C(i) = \sum_{j=1}^{i-1} \pi(i)$
\\[-8pt]
\> \hspace*{20pt} \= \hspace*{20pt} \= \hspace*{20pt} \= \hspace*{20pt} \= 
   \hspace*{145pt} \= \\[-5pt]
\> If $\pi(k) \ge 1/2$: \\[4pt]
\>\> \textit{Quickly handle the case where the current value has probability 
             half or more,} \\
\>\> \textit{without needing to compute cumulative probabilities.} \\[4pt]
\>\> For $i = 1,\ldots,m$: \\
\>\>\> If $i \ne k$: \\
\>\>\>\> Set $p(i)$ to $\min(1,\,\pi(i)/\pi(k))$
         \>\>\textit{Min guards against round-off error} \\[4pt]
\>\> Set $p(k)$ to $(2\pi(k)-1)\,/\,\pi(k)$ \\[4pt]

\> Else: \\[4pt]
\>\> \textit{Compute cumulative probabilities, in the order given by $\sigma,$
             but stored in the original order.}\\
\>\>\textit{Set $S$ to the sum of all
            probabilities, which should be one, but may differ due to rounding.}
     \\[4pt]
\>\> Set $S$ to 0 \\
\>\> For $i = 1,\ldots,m$: \\
\>\>\> Set $C(\sigma(i))$ to $S$ \\
\>\>\> Add $\pi(\sigma(i))$ to $S$ \\[8pt]
\>\> \textit{Find the flows from the current value to each value,
             and the total flow.} \\[4pt]
\>\> Set $t$ to 0 \\
\>\> For $i = 1,\ldots,m$: \\
\>\>\> Set $\Delta_1$ to $\pi(k)-s+C(k)-C(i)$ 
       \>\>\> \textit{Will be exactly zero if $i=k$ and $s=\pi(k)$} \\
\>\>\> Set $\Delta_2$ to $\Delta_1+S$ \\
\>\>\> Set $v(i)$ to $\max\,(0,\,\min\,(\Delta_1,\,\pi(k)+\pi(i)-\Delta_1,\,
                                   \pi(k),\,\pi(i)))$ \\
\>\>\>\ \hspace{53pt} $+\ \max\,(0,\,\min\,(\Delta_2,\,\pi(k)+\pi(i)-\Delta_2,\,
                                   \pi(k),\,\pi(i)))$ \\
\>\>\> Add $v(i)$ to $t$ \\[8pt]
\>\> If $t=0$: \\[4pt]
\>\>\> \textit{If the total flow is zero, return a result giving probability 1
             to the most probable value.} \\[4pt]
\>\>\> Set $j$ to 1 \\
\>\>\> For $i = 2,\ldots,m$: \\
\>\>\>\> If $\pi(i)>\pi(j)$:\\
\>\>\>\>\> Set $j$ to $i$ \\
\>\>\> For $i = 1,\ldots,m$:\\
\>\>\>\> Set $p(i)$ to 1 if $i=j$, otherwise to $0$ \\
\>\> Else:\\[4pt]
\>\>\> \textit{Find transition probabilities by normalizing flows by their 
               sum, which should be $\pi(k)$,} \\
\>\>\> \textit{but may differ due to rounding.}\\[4pt]
\>\>\> For $i = 1,\ldots,m$: \\
\>\>\>\> Set $p(i)$ to $v(i)/t$

\end{tabbing}\vspace{3pt}

\caption{Computation of ST transition probabilities.}\label{alg-ST}

\end{algorithm}

For both ST and HST, the ordering of values can matter.  I will use ST
and HST to refer to these methods with the original order retained.  I
use Ordered HST (OHST) to refer to HST with values ordered by
probability --- whether by non-increasing or non-decreasing
probability makes no difference.  I use Upward ST (UST) or Downward ST
(DST) to refer to the ST method in which the most probable value is
followed by the other values in non-decreasing or non-increasing
order.  For all these methods, how values with equal probability are
ordered may matter.

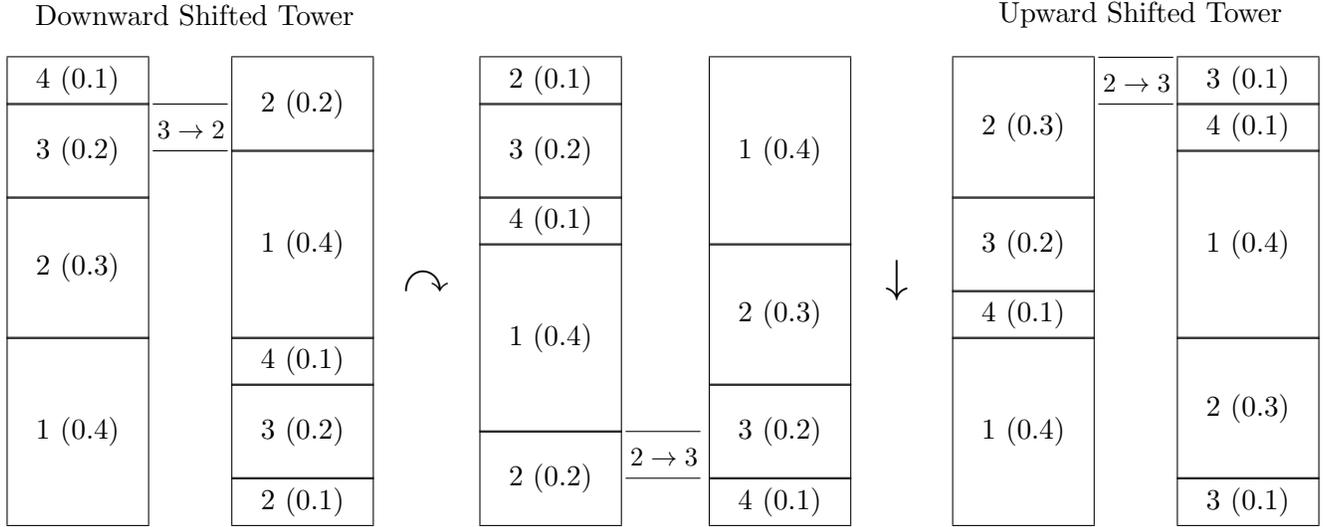
\begin{figure}[t]

\begin{center}

\setlength{\unitlength}{2.45in}
\begin{picture}(2.8,1.13)

  \put(0.0,1.05){\makebox(0.8,0.08){Downward Shifted Tower}}

  \put(0.0,0.0){\framebox(0.3,0.4){1 (0.4)}}
  \put(0.0,0.4){\framebox(0.3,0.3){2 (0.3)}}
  \put(0.0,0.7){\framebox(0.3,0.2){3 (0.2)}}
  \put(0.0,0.9){\framebox(0.3,0.1){4 (0.1)}}

  \put(0.32,0.827){\small$3 \rightarrow 2$}
  \put(0.31,0.8){\line(1,0){0.16}}
  \put(0.31,0.9){\line(1,0){0.16}}

  \put(0.48,0.4){\framebox(0.3,0.4){1 (0.4)}}
  \put(0.48,0.8){\framebox(0.3,0.2){2 (0.2)}}
  \put(0.48,0.0){\framebox(0.3,0.1){2 (0.1)}}
  \put(0.48,0.1){\framebox(0.3,0.2){3 (0.2)}}
  \put(0.48,0.3){\framebox(0.3,0.1){4 (0.1)}}

  \put(0.845,0.5){\LARGE $\curvearrowright$}

  \put(1.01,0.0){\framebox(0.3,0.2){2 (0.2)}}
  \put(1.01,0.2){\framebox(0.3,0.4){1 (0.4)}}
  \put(1.01,0.6){\framebox(0.3,0.1){4 (0.1)}}
  \put(1.01,0.7){\framebox(0.3,0.2){3 (0.2)}}
  \put(1.01,0.9){\framebox(0.3,0.1){2 (0.1)}}

  \put(1.33,0.127){\small$2 \rightarrow 3$}
  \put(1.32,0.1){\line(1,0){0.16}}
  \put(1.32,0.2){\line(1,0){0.16}}

  \put(1.50,0.0){\framebox(0.3,0.1){4 (0.1)}}
  \put(1.50,0.1){\framebox(0.3,0.2){3 (0.2)}}
  \put(1.50,0.3){\framebox(0.3,0.3){2 (0.3)}}
  \put(1.50,0.6){\framebox(0.3,0.4){1 (0.4)}}

  \put(1.875,0.5){\LARGE $\downarrow$}

  \put(2.02,1.05){\makebox(0.8,0.08){Upward Shifted Tower}}

  \put(2.02,0.0){\framebox(0.3,0.4){1 (0.4)}}
  \put(2.02,0.4){\framebox(0.3,0.1){4 (0.1)}}
  \put(2.02,0.5){\framebox(0.3,0.2){3 (0.2)}}
  \put(2.02,0.7){\framebox(0.3,0.3){2 (0.3)}}

  \put(2.34,0.927){\small$2 \rightarrow 3$}
  \put(2.33,0.9){\line(1,0){0.16}}
  \put(2.33,1.0){\line(1,0){0.16}}

  \put(2.5,0.0){\framebox(0.3,0.1){3 (0.1)}}
  \put(2.5,0.1){\framebox(0.3,0.3){2 (0.3)}}
  \put(2.5,0.4){\framebox(0.3,0.4){1 (0.4)}}
  \put(2.5,0.8){\framebox(0.3,0.1){4 (0.1)}}
  \put(2.5,0.9){\framebox(0.3,0.1){3 (0.1)}}

\end{picture}

\end{center}

\caption{Illustration of why UST and DST are reversals of each
other. On the left is an illustration of DST transition probabilities,
showing in particular that $\pi(3)\,P_{DST}(3 \rightarrow 2)=0.1$. In
the middle is the result of rotating the diagram on the left by 180
degrees, which produces reversed transition probabilities, where in
particular $\pi(2)\,P_{DST}(2 \rightarrow 3)=0.1$.  On the right is the
result of shifting the two towers in the middle down by 0.2 (wrapping
bottom to top).  This shift of both towers has no effect on the
transition probabilities, which are now seen to be those of
$P_{UST}$.}  \label{UST-DST-rev}

\end{figure}

Both UST and DST produce transition probabilities are (in general) non-reversible, but which are, however, reverses of each other --- that is,
\beq
   \pi(u)\,P_{UST}(u \rightarrow v) & = & \pi(v)\,P_{DST}(v \rightarrow u)
\eeq
This relationship is illustrated in Figure~\ref{UST-DST-rev}.  Averaging
the transition probabilities produced by UST and DST therefore gives a method,
which I will call UDST, that is reversible:
\beq
   \pi(u)\,P_{UDST}(u \rightarrow v) 
   & = & \pi(u)\,(P_{UST}(u \rightarrow v)+P_{DST}(u \rightarrow v))\,/\,2
   \\
   & = & \pi(v)\,(P_{DST}(v \rightarrow u)+P_{UST}(v \rightarrow u))\,/\,2
   \\
   & = & \pi(v)\,P_{UDST}(v \rightarrow u)
\eeq
Like UST and DST, UDST produces the minimum possible self transition
probabilities, so it will provide interesting information on the
effect of reversibility in the experimental comparisons.

Algorithm~\ref{alg-ST} also starts by checking whether the current
value has probability of $1/2$ or more, and if so, finds the
transition probabilities from this value quickly, without needing to
compute cumulative probabilities.  This check could be omitted, as
might be desirable if it is known that probabilities of a half or more
are unlikely. Also, when this check is omitted, it is not necessary for
the input probabilities, $\pi$, to be normalized to sum to one, given
the adjustments described in the previous paragraph, provided the
shift amount, $s$, is on the same scale as these unnormalized
probabilities. Indeed, the procedure described by Suwa (2022) does not
assume that probabilities are normalized.

The ST method can be implemented by applying Algorithm~\ref{alg-ST}
with $s$ set to the maximum value of $\pi$.  For the HST method,
Algorithm~\ref{alg-ST} is called with $s$ set to $1/2$.

In many contexts, computing transition probabilities is not necessary
--- all that is needed is a way of sampling from the transition
distribution given the current state value. For ST methods, sampling
directly may be significantly faster than first computing transition
probabilities and then sampling using them.
Algorithm~\ref{alg-ST-sample} implements such a direct sampling
method, based on randomly choosing a point within the region of the
``tower'' corresponding to the current value, then moving this point
down by the shift amount, with wrap-around (equivalent to moving the
tower up), and choosing a new value using this shifted point as if it
were a random $[0,1]$ variate.

\begin{algorithm}[t]

\begin{tabbing}

\hspace{0.8in}\=
\bf Input:\ \ \ \ \ \ \= Gibbs sampling probabilities,
                   $\pi^{\rule{0pt}{1pt}}(i)$, for $i=1,\ldots,m$
\\ \>
                \> The current state value, $k$, in $\{1,\ldots,m\}$
\\ \>
                \> Amount of shift, $s$, in $(0,1)$
\\ \>
                \> A permutation, $\sigma$, on $\{1,\ldots,m\}$, giving
                   an ordering of values
\\[5pt] \>
\bf Output:     \> A state value, $j$, sampled from the ST transition 
                   probabilities from state $k$, \\
              \>\> guaranteed not to be a value with transition probability zero
\\[-8pt]
\> \hspace*{20pt} \= \hspace*{20pt} \= \hspace*{20pt} \= \hspace*{20pt} \= 
   \hspace*{115pt} \= \\[4pt]
\> \textit{Find the sum, $u$, of probabilities of values before $k$ in the
           ordering used.} \\[4pt]
\> Set $i$ to 1 \\
\> Set $u$ to 0 \\
\> While $\sigma(i) \ne k$:\\
\>\> Set $u$ to $u + \pi(\sigma(i))$ \\
\>\> Add 1 to $i$ \\[4pt]
\> \textit{Add a random amount to $u$, while subtracting the shift, with
           wrap-around.}\\[4pt]
\> Set $r$ to a uniform random variate on $[0,1]$ \\
\> Add $r\pi(k) - s$ to $u$ \>\>\>\>\> \textit{Guaranteed not to increase $u$
                                               when $s=\max_j\pi(j)$}\\
\> If $u \le 0$: \\
\>\> Add 1 to $u$
       \>\>\>\> \textit{Guarantees that $u$ is greater than zero}\\[4pt]
\> \textit{Use this value of $u$ to pick a value, $j$, to transition to, picking
    an arbitrary value with}\\
\> \textit{non-zero probability if no value is chosen due to round-off 
           error.}\\[4pt]
\> Set $i$ to 0 \\
\> Set $s$ to 0 \\
\> While $i<m$ and $u>s$: \\
\>\> Add 1 to $i$ \\
\>\> If $\pi(\sigma(i))>0$: \\
\>\>\> Add $\pi(\sigma(i))$ to $s$ \\
\>\>\> Set $j$ to $\sigma(i)$\\[4pt]
\> \textit{The value $j$ is now a sample from the transition probabilities
           from the current state, $k$.}
\end{tabbing}

\caption{Sampling from ST transition probabilities.}\label{alg-ST-sample}

\end{algorithm}

This technique could be used when the state has a countably infinite
number of values, as long as the cumulative distribution function and
its inverse can be computed efficiently.

\section{\hspace*{-8pt}
Flattened slice sampling methods (FSS and ZFSS)}\label{sec-slice}\vspace{-11pt}

Modified Gibbs sampling methods can also be derived using the ``slice
sampling'' framework (Neal 2003).  For discrete distributions, slice
sampling can be visualized using bars associated with each possible
value, with the height of a bar equal to its value's probability.  A
vertical level within the bar for the current value is sampled
uniformly, and some update is then made that moves amongst the bars
that intersect the horizontal line drawn at this level, with the
property of leaving the uniform distribution on this horizontal
``slice'' invariant.

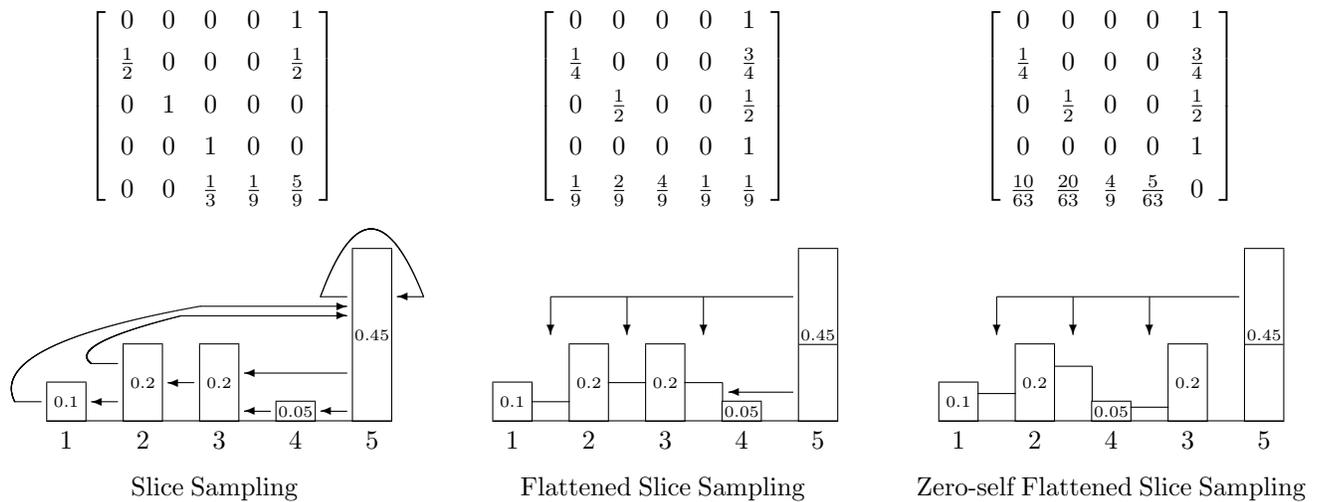
\begin{figure}[t]

{\small
\hspace{39pt}
$
\left[\begin{array}{ccccc}
  0   & 0  & 0  & 0  & 1 \\[4pt]
  1\over2 & 0  & 0  & 0  & 1\over2 \\[4pt]
  0   & 1  & 0  & 0  & 0 \\[4pt]
  0   & 0  & 1  & 0  & 0 \\[4pt]
  0   & 0  & 1\over3& 1\over9& 5\over9 
\end{array}\right]
$
\hspace{69pt}
$
\left[\begin{array}{ccccc}
  0   & 0  & 0  & 0  & 1 \\[4pt]
  1\over4 & 0  & 0  & 0  & 3\over4 \\[4pt]
  0   & 1\over2& 0  & 0  & 1\over2 \\[4pt]
  0   & 0  & 0  & 0  & 1 \\[4pt]
  1\over9 & 2\over9& 4\over9& 1\over9& 1\over9 
\end{array}\right]
$
\hspace{67pt}
$
\left[\begin{array}{ccccc}
  0   & 0  & 0  & 0  & 1 \\[4pt]
  1\over4 & 0  & 0  & 0  & 3\over4 \\[4pt]
  0   & 1\over2& 0  & 0  & 1\over2 \\[4pt]
  0   & 0  & 0  & 0  & 1 \\[4pt]
  \!{10\over63}\! & \!{20\over63}\! & \!{4\over9}\! & \!{5\over63}\! & 0 
\end{array}\right]
$
}

\vspace*{-50pt}

\begin{center}\tiny

\setlength{\unitlength}{2.0in}
\hspace{17pt}%
\begin{picture}(1.15,0.8)

\put(0.05,0.1){\line(1,0){0.9}}
\put(0.05,0.1){\framebox(0.1,0.1){0.1}}
\put(0.035,0.15){\line(-1,0){0.07}}
\put(0.45,0.4){\vector(1,0){0.385}}
\qbezier(-0.035,0.15)(-0.11,0.3)(0.45,0.4)
\put(0.05,0.0){\makebox(0.1,0.1){\small1}}
\put(0.25,0.1){\framebox(0.1,0.2){0.2}}
\put(0.235,0.15){\vector(-1,0){0.07}}
\put(0.235,0.25){\line(-1,0){0.07}}
\put(0.4,0.375){\vector(1,0){0.435}}
\qbezier(0.165,0.25)(0.1,0.3)(0.4,0.375)
\put(0.25,0.0){\makebox(0.1,0.1){\small2}}
\put(0.45,0.1){\framebox(0.1,0.2){0.2}}
\put(0.45,0.0){\makebox(0.1,0.1){\small3}}
\put(0.435,0.2){\vector(-1,0){0.07}}
\put(0.65,0.1){\framebox(0.1,0.05){0.05}}
\put(0.65,0.0){\makebox(0.1,0.1){\small4}}
\put(0.635,0.125){\vector(-1,0){0.07}}
\put(0.85,0.1){\framebox(0.1,0.45){0.45}}
\put(0.835,0.225){\vector(-1,0){0.27}}
\put(0.835,0.125){\vector(-1,0){0.07}}
\put(0.835,0.425){\line(-1,0){0.07}}
\put(1.035,0.425){\vector(-1,0){0.07}}
\qbezier(0.765,0.425)(0.895,0.78)(1.035,0.425)
\put(0.85,0.0){\makebox(0.1,0.1){\small5}}
\end{picture}
\begin{picture}(1.15,1)

\put(0.05,0.1){\line(1,0){0.9}}
\put(0.05,0.1){\framebox(0.1,0.1){0.1}}
\put(0.05,0.0){\makebox(0.1,0.1){\small1}}
\put(0.15,0.1){\makebox(0.1,0.05){}}
\put(0.15,0.15){\line(1,0){0.1}}
\put(0.25,0.1){\framebox(0.1,0.2){0.2}}
\put(0.25,0.0){\makebox(0.1,0.1){\small2}}
\put(0.35,0.1){\makebox(0.1,0.1){}}
\put(0.35,0.2){\line(1,0){0.1}}
\put(0.45,0.1){\framebox(0.1,0.2){0.2}}
\put(0.45,0.0){\makebox(0.1,0.1){\small3}}
\put(0.55,0.1){\makebox(0.1,0.1){}}
\put(0.55,0.2){\line(1,0){0.1}}
\put(0.65,0.15){\line(0,1){0.05}}
\put(0.65,0.1){\framebox(0.1,0.05){0.05}}
\put(0.65,0.0){\makebox(0.1,0.1){\small4}}
\put(0.85,0.1){\framebox(0.1,0.45){0.45}}
\put(0.85,0.3){\line(1,0){0.1}}
\put(0.85,0.0){\makebox(0.1,0.1){\small5}}
\put(0.835,0.425){\line(-1,0){0.635}}
\put(0.20,0.425){\vector(0,-1){0.1}}
\put(0.40,0.425){\vector(0,-1){0.1}}
\put(0.60,0.425){\vector(0,-1){0.1}}
\put(0.835,0.175){\vector(-1,0){0.17}}

\end{picture}
\begin{picture}(1.15,1)

\put(0.05,0.1){\line(1,0){0.9}}
\put(0.05,0.1){\framebox(0.1,0.1){0.1}}
\put(0.05,0.0){\makebox(0.1,0.1){\small1}}
\put(0.15,0.1){\makebox(0.1,0.0714){}}
\put(0.15,0.1714){\line(1,0){0.1}}
\put(0.25,0.1){\framebox(0.1,0.2){0.2}}
\put(0.25,0.0){\makebox(0.1,0.1){\small2}}
\put(0.35,0.1){\makebox(0.1,0.1429){}}
\put(0.35,0.2429){\line(1,0){0.1}}
\put(0.45,0.2429){\line(0,-1){0.0929}}
\put(0.45,0.1){\framebox(0.1,0.05){0.05}}
\put(0.45,0.0){\makebox(0.1,0.1){\small4}}
\put(0.55,0.1){\makebox(0.1,0.0357){}}
\put(0.55,0.1357){\line(1,0){0.1}}
\put(0.65,0.1){\framebox(0.1,0.2){0.2}}
\put(0.65,0.0){\makebox(0.1,0.1){\small3}}
\put(0.85,0.1){\framebox(0.1,0.45){0.45}}
\put(0.85,0.3){\line(1,0){0.1}}
\put(0.85,0.0){\makebox(0.1,0.1){\small5}}
\put(0.835,0.425){\line(-1,0){0.635}}
\put(0.20,0.425){\vector(0,-1){0.1}}
\put(0.40,0.425){\vector(0,-1){0.1}}
\put(0.60,0.425){\vector(0,-1){0.1}}

\end{picture}\hspace*{-20pt}

\end{center}

{ \small
\hspace{56pt} Slice Sampling
\hspace{79pt} Flattened Slice Sampling
\hspace{37pt} Zero-self Flattened Slice Sampling
}

\caption{Illustration of SS, FSS, and ZFSS methods. 
These diagrams portray transitions
that leave invariant the distribution on $\{1,2,3,4,5\}$ with probabilities
$0.1$, $0.2$, $0.2,$, $0.05$, and $0.45$.  The left diagram shows simple
slice sampling, in which a vertical position is randomly chosen from the bar 
for the current value, with height equal to its probability, and a 
movement to the left (with wrap-around) is then made until the next bar is
encountered.
In the middle diagram, the self transition probability for the most probable
value is reduced by distributing the excess of its probability over that of the
next-most probable value to new bars that follow the bars for values other
than the most probable value and the one before it.  Arrows showing 
the subsequent transitions are omitted, except for one left arrow showing that
there is still a non-zero self transition probability, going from the most
probable value to another bar also associated with this value.  In the
right diagram, value 3 is moved to just before the most-probable value, which
blocks such a self transition. The resulting transition probability matrices
are shown above the diagrams.}\label{fig-FSS}

\end{figure}

One update that seems promising for avoiding self transitions is to
move from the bar for the current value, at the sampled level, to the
next bar to the left that rises to that level, wrapping around to the
right side if the left end is reached.  This method is shown in the
left illustration of Figure~\ref{fig-FSS}.  Unfortunately, the method
will produce a non-zero self transition probability if one value has a
probability greater than all other values, as is the case for value 5
in Figure~\ref{fig-FSS}.  If the vertical level sampled when this is
the current value is greater than the probabilities of all other
values, the movement to the left will wrap around to the same value.
In this example, the resulting self transition probability from value
5 is $(0.45-0.2)/0.45 = 5/9$, but the minimum possible self transition
probability for this distribution is zero.

This self transition probability can be reduced by distributing the
portion of the probability of the most-probable symbol that is greater
than all other symbols amongst another set of bars, which follow the
bars for values other than the most probable value and the value to
its left (with wraparound).  This modification, called Flattened Slice
Sampling (FSS), is shown in the middle illustration of
Figure~\ref{fig-FSS}. The 0.25 excess probability for value 5 is moved
to bars to the right of values 1, 2, and 3, in proportion to their
probabilities.  When 5 is the current value, a bar is selected from
amongst these three new bars and the original bar with probabilities
0.05/0.45, 0.1/0.45, 0.1/0.45, and 0.2/0.45.  Movement to the left
then occurs as before.  If the bar moved to is any of those associated
with the most-probable value, that becomes the new state.

However, the self transition probability for FSS is still not zero
in this example.  The bar for value 4 is lower than the new bar to the
right of value 3. Consequently, a portion of the bar for value 5 
encounters this new bar, which is also associated with value 5, when
leftward movement occurs, resulting in a self transition probability
of 1/9 when value 5 is the current state.

The Zero-self Flattened Slice Sampling (ZFSS) method avoids such
unnecessary self transitions by re-ordering values to put a value that
blocks such movement immediately to the left of the most-probable
value, while leaving the order of values otherwise unchanged. The
value moved is the one closest on the left to the most-probable value
that will block any resulting movement from the original bar for the
most-probable value to one of the new bars also associated with this
value.  In the example of Figure~\ref{fig-FSS}, value 3 is moved to
the left of value 5.

This procedure assumes all values have probability less than
one half, which also implies that $m>2$.  Situations where a value has
probability one half or more are handled specially, in the same manner
as for ZDNAM and the ST methods, which, as will be discussed below in
Section~\ref{sec-lim}, is the only method that minimizes the
probability of a self transition in this situation. The FSS and ZFSS
methods are implemented in Algorithm~\ref{alg-FSS}, with an input flag
specifying whether the possible re-ordering for ZFSS is done.

As for the ST methods, an FSS transition can be simulated directly
more efficiently than it can be by first computing transition
probabilities from the current value and then sampling a new value
according to these probabilities.  Since the flow is computed in
Algorithm~\ref{alg-FSS} by adding portions (with the flow never
decreasing), one can keep track of the cumulative flow computed so
far, and make a transition to the value associated with the portion
just computed when this cumulative sum exceeds a random variate chosen
at the beginning.  Algorithm~\ref{alg-FSS-sample} implements this
approach.

\begin{algorithm}[p]

\begin{tabbing}

\hspace{0.5in}\=
\bf Input:\ \ \ \ \ \ \ \ \ \ \ \ \ \ \ \ \ \ \ \= Gibbs sampling probabilities,
                   $\pi^{\rule{0pt}{1pt}}(i)$, for $i=1,\ldots,m$
\\ \>
                \> The current state value, $k$, in $\{1,\ldots,m\}$
\\ \>
                \> A flag, ZERO, for whether ZFSS should be used
\\[5pt] \>
\bf Output:     \> FSS or ZFSS transition probabilities, $p(i)$, 
                   for $i=1,\ldots,m$
\\[5pt] \>
\bf Temporary storage: \>Flows of probability, $v(i)$, from $k$ to each value,
                         for $i=1,\ldots,m$
\\
\> \hspace*{20pt} \= \hspace*{20pt} \= \hspace*{20pt} \= \hspace*{20pt} \= 
   \hspace*{20pt} \= \hspace*{20pt} \= 
   \hspace*{50pt} \= \hspace*{55pt} \= \hspace*{30pt} \= \\[-5pt]
\> If $\pi(k) \le 0$: \\[4pt]
\>\> \textit{Handle transition from zero-probability value specially.} \\[4pt]
\>\> For $i=1,\ldots,m$: \\
\>\>\> Set $p(i)$ to $\pi(i)$ \\[4pt]
\> Else: \\[4pt]
\>\> \textit{Find the index, $x_1$, of the most probable value, and its
             probability, $\pi_1$.} \\[4pt]
\>\> Set $x_1$ to 1 \\
\>\> For $i=2,\ldots,m$: \\
\>\>\> If $\pi(i)>\pi(x_1)$: \\
\>\>\>\> Set $x_1$ to $i$ \\
\>\> Set $\pi_1$ to $\pi(x_1)$ \\[4pt]
\>\> If $\pi(x_1) \ge 1/2$ or $m \le 2$:
\>\>\>\>\>\> \textit{Checking for $m \le 2$ guards against round-off error}
             \\[4pt]
\>\>\> \textit{Handle the case where the current value has probability 
             half or more specially.} \\[4pt]
\>\>\> If $k = x_1$: \\
\>\>\>\> For $i=1,\ldots,m$: \\
\>\>\>\>\> Set $p(i)$ to $(2\pi_1-1)\,/\,\pi_1$ if $i=k$, otherwise to 
                         $\pi(i)/\pi_1$ \\
\>\>\> Else: \\
\>\>\>\> For $i=1,\ldots,m$: \\
\>\>\>\>\> Set $p(i)$ to 1 if $i=x_1$, otherwise to 0 \\
\>\> Else: \\[4pt]
\>\>\> \textit{Find the probability, $\pi_2$, of the second most probable 
               value.} \\[4pt]
\>\>\> Set $\pi_2$ to 0\\
\>\>\> for $i=1,\ldots,m$:\\
\>\>\>\> If $i \ne x_1$ and $\pi(i)>\pi_2$:\\
\>\>\>\>\> Set $\pi_2$ to $\pi(i)$ \\[4pt]
\>\>\> \textit{Find the index, $x_0$, of the value before the most probable
               value, or for ZFSS, the index} \\
\>\>\> \textit{of the first value before the most probable
  value which will block movement beyond it}\\
\>\>\> \textit{from encountering a piece of the most probable value.} \\[4pt]
\>\>\> Set $x_0$ to $x_1$ \\
\>\>\> Loop: \\
\>\>\>\> Set $x_0$ to $m$ if $x_0=1$, otherwise to $x_0-1$ \\
\>\>\>\> Set $\pi^*$ to $(0.5-\pi_1) + (0.5-\pi(x_0))$
\>\>\>\>\> \textit{Computing this way reduces round-off error} \\
\>\>\>\> Set $f$ to $(\pi_1-\pi_2)\,/\,\pi^*$
\>\>\>\>\> \textit{Guaranteed to be in $[0,1]$ even with rounding} \\
\>\>\> Repeat loop as long as ZERO and $\pi(x_0) < f \pi_2$ \\[4pt]

\>\>\> Continue with the procedure of Algorithm~\ref{alg-FSS}: Part 2.

\end{tabbing}\vspace{6pt}

\caption{Part 1. Procedure for computing FSS or ZFSS transition 
         probabilities.}\label{alg-FSS}

\end{algorithm}

\addtocounter{algorithm}{-1}

\begin{algorithm}[p]

\begin{tabbing}
\hspace*{0.5in} \=
   \hspace*{20pt} \= \hspace*{20pt} \= \hspace*{20pt} \= \hspace*{20pt} \= 
   \hspace*{20pt} \= \hspace*{20pt} \= 
   \hspace*{50pt} \= \hspace*{55pt} \= \hspace*{30pt} \= \\[-5pt]

\>\>\> Continuation of Algorithm~\ref{alg-FSS}: Part 1.\\[6pt]
\>\>\> \textit{Find the part of the flow due to distributing the difference in 
               probability between most} \\
\>\>\> \textit{probable and second-most probable 
               values among other values.  Here, $f$ is the factor to} \\
\>\>\> \textit{multiply probabilities of values besides $x_1$ 
               and $x_0$ by to get the part of $x_1$ flowing there.} \\[4pt]

\>\>\> For $i=1,\ldots,m$: \\
\>\>\>\> If $k=x_1$ and $i \ne x_1$ and $i \ne x_0$: \\
\>\>\>\>\> Set $v(i)$ to $f \pi(i)$ \\
\>\>\>\> Else: \\
\>\>\>\>\> Set $v(i)$ to 0 \\[4pt]

\>\>\> \textit{Find the flow due to slice movement.}\\[4pt]

\>\>\> Set $\ell$ to 0 
       \>\>\>\>\>\> \textit{Lower end of probability region to move}\\
\>\>\> Set $u$ to $\pi_2$ if $k=x_1$, otherwise to $\pi(k)$
       \>\>\>\>\>\> \textit{Upper end of probability region to move}\\
\>\>\> Set $i$ to $k$ \\[4pt]
\>\>\> While $\ell < u$: \\[4pt]
\>\>\>\> \textit{Move $i$ backwards, going from $x_1$ to $x_0$, from $x_0$
          to before $x_1$, and skipping $x_0$}\\
\>\>\>\> \textit{when otherwise going back.}\\[4pt]
\>\>\>\> If $i=x_1$: \\
\>\>\>\>\> Set $i$ to $x_0$ \\
\>\>\>\> Else: \\
\>\>\>\>\> If $i=x_0$: \\
\>\>\>\>\>\> Set $i$ to $m$ if $x_1=1$, otherwise to $x_1-1$ \\
\>\>\>\>\> Else: \\
\>\>\>\>\>\> Set $i$ to $m$ if $i=1$, otherwise to $i-1$ \\
\>\>\>\>\> If $i=x_0$: \\
\>\>\>\>\>\> Set $i$ to $m$ if $x_0=1$, otherwise to $x_0-1$ \\[4pt]
\>\>\>\> \textit{Add to flow from slice movement of $[\ell,u]$ region,
                 and update $\ell$ and $u$.}\\[4pt]
\>\>\>\> If $\ell < \pi(i)$: \\
\>\>\>\>\> If $i \ne x_1$ and $i \ne x_0$: \\
\>\>\>\>\>\> Set $t$ to $\min(u,f\pi(i))$ \\
\>\>\>\>\>\> If $\ell < t$: \\
\>\>\>\>\>\>\> Add $t-\ell$ to $v(x_1)$ \\
\>\>\>\>\>\>\> Set $\ell$ to $t$ \\
\>\>\>\>\> Set $t$ to $\min(u,\pi(i))$ \\
\>\>\>\>\> Add $t-\ell$ to $v(i)$ \\
\>\>\>\>\> Set $\ell$ to $t$ \\[4pt]
\>\>\>\textit{Return transition probabilities derived from flow.}\\[4pt]
\>\>\> For $i = 1,\ldots,m$: \\
\>\>\>\> Set $p(i)$ to $v(i)/\pi(k)$

\end{tabbing}\vspace{6pt}

\caption{Part 2. Continuation of procedure for computing FSS or ZFSS transition 
         probabilities.}

\end{algorithm}

\begin{algorithm}[p]

\begin{tabbing}

\hspace{0.5in}\=
\bf Input:\ \ \ \ \ \ \ \ \= Gibbs sampling probabilities,
                   $\pi^{\rule{0pt}{1pt}}(i)$, for $i=1,\ldots,m$
\\ \>
                \> The current state value, $k$, in $\{1,\ldots,m\}$
\\ \>
                \> A flag, ZERO, for whether ZFSS should be used
\\[3pt] \>
\bf Output:     \> A state value, $j$, sampled from the FSS/ZFSS transition
                   probabilities from state $k$,\\
              \>\> guaranteed not to be a value with transition probability zero
\\[-7pt]
\> \hspace*{20pt} \= \hspace*{20pt} \= \hspace*{20pt} \= \hspace*{20pt} \= 
   \hspace*{20pt} \= \hspace*{20pt} \= \hspace*{20pt} \= \hspace*{20pt} \= 
   \hspace*{20pt} \= \hspace*{20pt} \= \hspace*{20pt} \= 
   \hspace*{40pt} \= \\[-5pt]
\> If $\pi(k) \le 0$: \\[4pt]
\>\> \textit{Handle transition from zero-probability value specially.} \\[4pt]
\>\> Set $r$ to a uniform random variate on $[0,1]$ \\
\>\> For $i=1,\ldots,m$: \\
\>\>\> Set $s$ to $0$;\ \ \ Set $i$ to $0$ \\
\>\>\> While $i<m$ and $r \ge s$: \\
\>\>\>\> Add $1$ to $i$ \\
\>\>\>\> If $\pi(i)>0$:\\
\>\>\>\>\> Add $\pi(i)$ to $s$;\ \ \ Set $j$ to $i$ \\
\> Else: \\[2pt]
\>\> \textit{Find the index, $x_1$, of the most probable value, and its
             probability, $\pi_1$.} \\[4pt]
\>\> Set $x_1$ to 1 \\
\>\> For $i=2,\ldots,m$: \\
\>\>\> If $\pi(i)>\pi(x_1)$: \\
\>\>\>\> Set $x_1$ to $i$ \\
\>\> Set $\pi_1$ to $\pi(x_1)$ \\[4pt]
\>\> If $\pi(x_1) \ge 1/2$ or $m \le 2$:
\>\>\>\>\>\> \textit{Checking for $m \le 2$ guards against round-off error}
             \\[4pt]
\>\>\> \textit{Handle the case where the current value has probability 
             half or more specially.} \\[4pt]
\>\>\> If $k \ne x_1$: \\
\>\>\>\> Set $j$ to $x_1$ \\
\>\>\> Else: \\
\>\>\>\> Set $r$ to a uniform random variate on $[0,1]$ \\
\>\>\>\> Set $s$ to 0;\ \ \ Set $i$ to 0 \\
\>\>\>\> While $i<m$ and $r \ge s$:\\
\>\>\>\>\> Add 1 to $i$ \\
\>\>\>\>\> If $\pi(i)>0$:\\
\>\>\>\>\>\> If $i=k$: \\
\>\>\>\>\>\>\> Add $(2\pi_1-1)\,/\,\pi_1$ to $s$ \\
\>\>\>\>\>\> Else: \\
\>\>\>\>\>\>\> Add $\pi(i)\,/\,\pi_1$ to $s$ \\
\>\>\>\>\>\> Set $j$ to $i$ \\[2pt]
\>\> Else: \\[2pt]
\>\>\> \textit{Find the probability, $\pi_2$, of the second most probable 
               value, and set $j$ to its index.} \\[4pt]
\>\>\> Set $\pi_2$ to 0\\
\>\>\> for $i=1,\ldots,m$:\\
\>\>\>\> If $i \ne x_1$ and $\pi(i)>\pi_2$:\\
\>\>\>\>\> Set $\pi_2$ to $\pi(i)$;\ \ \ Set $j$ to $i$ \\[4pt]
\>\>\> \textit{Generate a uniform random variate from zero to
               height of bar for the current value.} \\[4pt]
\>\>\> Set $r$ to a uniform random variate on $[0,\pi(k)]$\\[4pt]
\>\>\> Continue with the procedure of Algorithm~\ref{alg-FSS-sample}: Part 2.

\end{tabbing}

\caption{Part 1. Procedure for sampling from FSS or ZFSS transition 
         probabilities.}\label{alg-FSS-sample}

\end{algorithm}

\addtocounter{algorithm}{-1}

\begin{algorithm}[p]
\begin{tabbing}
\hspace{0.5in}\=
   \hspace*{20pt} \= \hspace*{20pt} \= \hspace*{20pt} \= \hspace*{20pt} \= 
   \hspace*{20pt} \= \hspace*{20pt} \= \hspace*{20pt} \= \hspace*{20pt} \= 
   \hspace*{20pt} \= \hspace*{20pt} \= \hspace*{20pt} \= 
   \hspace*{40pt} \= \\[-5pt]

\>\>\> \textit{Find the index, $x_0$, of the value before the most probable
               value, or for ZFSS, the index} \\
\>\>\> \textit{of the first value before the most probable
  value which will block movement beyond it}\\
\>\>\> \textit{from encountering a piece of the most probable value.} \\[4pt]
\>\>\> Set $x_0$ to $x_1$ \\
\>\>\> Loop: \\
\>\>\>\> Set $x_0$ to $m$ if $x_0=1$, otherwise to $x_0-1$ \\
\>\>\>\> Set $\pi^*$ to $(0.5-\pi_1) + (0.5-\pi(x_0))$
 \>\>\>\>\>\>\>\> \textit{Computing this way reduces round-off error} \\
\>\>\>\> Set $f$ to $(\pi_1-\pi_2)\,/\,\pi^*$
 \>\>\>\>\>\>\>\> \textit{Guaranteed in $[0,1]$ even with rounding} \\
\>\>\> Repeat loop as long as ZERO and $\pi(x_0) < f \pi_2$ \\[4pt]
\>\>\> If $k=x_1$ and $r \ge \pi_2$: \\[4pt]
\>\>\>\> \textit{If the transition is from the most probable value, $x_1$,
                 and $r$ is in the region to be} \\
\>\>\>\> \textit{distributed among values other than 
         $x_1$ and $x_0$, then select such a value, $j$.} \\[4pt]
\>\>\>\> Subtract $\pi_2$ from $r$ \\
\>\>\>\> Set $s$ to 0;\ \ \ Set $i$ to 0 \\
\>\>\>\> While $i<m$ and $r \ge s$: \\
\>\>\>\>\> Add 1 to $i$ \\
\>\>\>\>\> If $i \ne x_1$ and $i \ne x_0$ and $\pi(i)>0$: \\
\>\>\>\>\>\> Add $f\pi(i)$ to $s$,\ \ \ Set $j$ to $i$ \\[2pt]
\>\>\> Else:\\[2pt]
\>\>\>\> \textit{Return a value that is transitioned to due to slice 
                 movement.}\\[4pt]
\>\>\>\> Set $\ell$ to $0$ 
          \>\>\>\>\>\>\>\>\> \textit{Lower end of region to move} \\
\>\>\>\> Set $u$ to $\pi_2$ if $k=x_1$, otherwise to $\pi(k)$
          \>\>\>\>\>\>\>\>\> \textit{Upper end of region to move} \\[4pt]
\>\>\>\> Set $i$ to $k$,\ \ \ Set $s$ to 0 \\[4pt]
\>\>\>\> While $\ell<u$ and $r \ge s$: \\[4pt]
\>\>\>\>\> \textit{Move $i$ backwards, going from $x_1$ to $x_0$, from
                   $x_0$ to before $x_1$, and skipping $x_0$}\\
\>\>\>\>\> \textit{when otherwise going back.} \\[4pt]
\>\>\>\>\> If $i=x_1$:\\
\>\>\>\>\>\> Set $i$ to $x_0$ \\
\>\>\>\>\> Else: \\
\>\>\>\>\>\> If $i=x_0$: \\
\>\>\>\>\>\>\> Set $i$ to $m$ if $x_1=1$, otherwise to $x_1-1$ \\
\>\>\>\>\>\> Else: \\
\>\>\>\>\>\>\> Set $i$ to $m$ if $i=1$, otherwise to $i-1$ \\
\>\>\>\>\>\> If $i=x_0$: \\
\>\>\>\>\>\>\> Set $i$ to $m$ if $x_0=1$, otherwise to $x_0-1$ \\[4pt]
\>\>\>\>\> \textit{Look at slice movement from
                   $[\ell,u]$ region, and update $\ell$ and $u$.} \\[4pt]
\>\>\>\>\> If $\ell<\pi(i)$: \\
\>\>\>\>\>\> If $i \ne x_1$ and $i \ne x_0$: \\
\>\>\>\>\>\>\> Set $t$ to $\min(u,f\pi(i))$ \\
\>\>\>\>\>\>\> if $\ell<t$:\\
\>\>\>\>\>\>\>\> Add $t-l$ to $s$;\ \ \ 
                 Set $j$ to $x_1$;\ \ \ Set $\ell$ to $t$\\
\>\>\>\>\>\> If $r \ge s$: \\ 
\>\>\>\>\>\>\> Set $t$ to $\min(u,\pi(i))$ \\
\>\>\>\>\>\>\>\> Add $t-l$ to $s$;\ \ \ 
                 Set $j$ to $i$;\ \ \ Set $\ell$ to $t$
\end{tabbing}

\caption{Part 2. Continuation of procedure for sampling from FSS or ZFSS 
         transition probabilities.}

\end{algorithm}

The FSS and ZFSS methods are non-reversible, whenever the maximum
probability is less than one half. For FSS, this non-reversibility
takes the form of consistent movement to the left (with wrap-around),
except for possible transitions to the most-probable value.  One might
speculate that such consistent movement improves efficiency.  The
re-ordering that may be done for ZFSS is designed to disturb this
leftward movement as little as possible. self transitions could
instead be avoided by ordering the values by non-decreasing
probability, but this would often disturb the original ordering more,
and could lead to the ordering changing from one update to another,
preventing consistent movement.

FSS and ZFSS are feasible for some distributions with a countably
infinite number of values.  Consider the geometric($\theta$)
distribution on $\{1,2,\ldots\}$ used previously as an example for
MHGS and DNAM, with $\theta<1/2$.  If we use a reverse order, so that
value 1 is rightmost, the excess in probability of the most probable
value (1) over the next-most probable (2) will be
$\theta-\theta(1-\theta)\,=\,\theta^2$, which will be distributed over
new bars that follow values 3, 4, 5, etc.\ to the right, in proportion to the
probabilities of these values.  The height of the new bar to the right of
value $i+1$ will be $\theta^2 \cdot \theta (1-\theta)^{i-2} \,=\, \theta^3
(1-\theta)^{i-2}$.  In comparison, the height of the bar for value
$i$ will be $\theta(1-\theta)^{i-1}$. The transition probability from
value $i$ (for $i>1$) to value~$1$ will be the sum of the ratio of 
these, $\theta^2\,/\,(1-\theta)$, plus the ratio of the excess of the
probability for value~$i$ over that for value $i+1$ to the probability for 
value $i$, which is $\theta$. This gives the transition
probabilities from value $i$ for $i>1$ as\vspace{-4pt}
\beq
P^*(i \rightarrow 1) & = & \theta^2\,/\,(1-\theta)\ +\ \theta
                     \ \ =\ \ \theta\,/\,(1-\theta) \\
P^*(i \rightarrow i+1) & = & 1\ -\ P^*(i \rightarrow 1)
                     \ \ =\ \ 1\ -\ \theta\,/\,(1-\theta) \\
P^*(i \rightarrow j) & = & 0,\ \ \ \ \mbox{for $j\ne1$ and $j\ne i+1$}
\eeq
For value $1$, we have\vspace{-4pt}
\beq
P^*(1 \rightarrow 1) & = & 0 \\
P^*(1 \rightarrow 2) & = & \theta(1-\theta)\,/\,\theta\ \ =\ \ 1-\theta \\
P^*(1 \rightarrow j) & = & \theta^3 (1-\theta)^{j-3}\,/\,\theta
                     \ \ =\ \ \theta^2 (1-\theta)^{j-3}\!\!\!\!\!,\ \ \ \ 
                     \mbox{for $j>2$}
\eeq

\section{\hspace*{-8pt}
  Non-domination of reversible methods minimizing 
  self transitions}\label{sec-nondom}\vspace{-11pt}

Proposition~13 of (Neal and Rosenthal 2023) provides a way of showing
that a method cannot be efficiency-dominated by another (see also
(Mira and Geyer 1999)). It states that for reversible, irreducible
transitions $P$ and $Q$, if $P$ efficiency-dominates $Q$, then $P$
eigen-dominates $Q$, where eigen-dominance of $P$ over $Q$ means that
if the eigenvalues of $P$ and $Q$ are ordered (retaining
multiplicity), all eigenvalues of $P$ are less than or equal to the
corresponding eigenvalue of $Q$.  Put in contrapositive form, this
proposition says that if $P$ does not eigen-dominate $Q$, it does not
efficiency-dominate $Q$.  Corollary~17 of (Neal and Rosenthal 2023)
shows that if $P$ and $Q$ are different, but have the same set of
eigenvalues, then neither efficiency-dominates the other.  

It was shown in Section~\ref{sec-ZDNAM} that ZDNAM always
efficiency-dominates Gibbs sampling, but this is not true for the
other reversible methods that minimize self transitions.  For example,
with $m=4$ and $\pi(1)=0.4$, $\pi(2)=0.3$, $\pi(3)=0.2$, and
$\pi(4)=0.1$, the UDST method produces a transition matrix with
eigenvalues of $-0.69246$, $-0.35046$, $0.04292$, and one. Gibbs
sampling transition matrices have all zero eigenvalues (apart from the
single eigenvalue of one).  So neither UDST nor GS eigen-dominates the
other (two of the eigenvalues of UDST are less than those of GS, but
one eigenvalue is greater), and hence neither can efficiency-dominate
the other. There are functions that are more efficiently estimated by
Gibbs sampling, and other functions that are more efficiently
estimated by UDST. HST and OHST also do not efficiency-dominate Gibbs
sampling for this example.

Several methods for modifying Gibbs sampling probabilities always
produce transitions with the minimum possible self transition
probability --- zero when $\pi_{\mbox{\tiny max}} =\max_i \pi(i) \le
1/2$, and $(2\pi_{\mbox{\tiny max}}\!-1)\,/\, \pi_{\mbox{\tiny max}}$
when $\pi_{\mbox{\tiny max}}\ge 1/2$ --- specifically, ZDNAM, all the ST
methods, and ZFSS.  The transitions produced by ZDNAM, UDST, HST, and
OHST are also reversible.  

Theorem 19 of (Neal and Rosenthal 2023) shows that an irreducible,
reversible transition matrix with minimum possible self transition
probabilities cannot be efficiency-dominated by any other reversible
transition matrix.  So, considered in isolation, transition matrices
produced by ZDNAM, UDST, HST, and OHST cannot be dominated by a
different reversible method.  

This can be extended to when any
reversible method minimizing self transitions is used to update a
randomly-chosen variable --- the resulting overall method cannot be
efficiency-dominated by any other reversible method that updates a
single variable chosen randomly in the same way.

The key fact to note is that the trace of a reversible transition
matrix is both the sum of its self transition probabilities (which are
on the diagonal) and the sum of its eigenvalues (Horn and Johnson
2013, p.~51).  Theorem~16 of (Neal and Rosenthal 2023) states (in
contrapositive form) that if $\mbox{trace}(P) \ge \mbox{trace}(Q)$,
and $P \ne Q$, then $P$ cannot efficiency-dominate $Q$.

The full transition matrix for a Gibbs sampling update of a particular
variable will (with a suitable ordering of values) be block diagonal,
with one block for each possible combination of values for other
variables, as was previously discussed in Section~\ref{sec-AM}.  If
the Gibbs sampling updates are modified to minimize self transitions,
the trace of the full transition matrix will be the sum of the traces
for each block, which will each have the minimum possible value.  If a
variable to update is chosen randomly with probabilities
$a_1,\ldots,a_n$ (for example, with each $a_k=1/n$), the combined
transition matrix can be written as \mbox{$P=\sum_k a_k P_k$}, where
$P_k$ is the transition matrix for an update of variable $k$.  The
trace of $P$ will $\sum_k a_k \mbox{trace}(P_k)$.

If each block of each $P_k$ minimizes self transition probabilities,
then $P$ will have the minimum possible trace of any such method.
That is, if $Q$ is any other method (not equal to $P$) that operates
by updating a variable chosen at random with probabilities
$a_1,\ldots,a_n$, then $\mbox{trace(Q)}\ge\mbox{trace(P)}$. It follows that
$Q$ cannot efficiency-dominate $P$ 


A stronger result applies when, for some particular problem, a method
produces self transition probabilities that are always zero (something
that is not always possible) --- random updating of variables using
this method cannot in this case be efficiency-dominated by any
reversible method at all, including methods that simultaneously change
the values of several (or all) variables, since no transition matrix
can have a trace (sum of self transition probabilities) less than
zero.

When variables are updated sequentially in some order that is randomly
chosen from a distribution in which an order and its reversal are
equally likely, an even stronger result is possible --- as long as it
is guaranteed that at least one of the variable updates has zero
self transition probability, the random order scan will have zero
probability of leaving the state unchanged, and hence the scan as a
whole cannot be efficiency-dominated by any other reversible
method.\footnote{\rule{0pt}{10pt}%
Note that this applies only to
estimates based on the states after each full scan (that is, on
``thinned'' estimates, as described below in Section~\ref{sec-emp}),
not necessarily to estimates that use the state after every variable
update within a scan.}

One should keep in mind, though, that such non-domination results are
rather weak justifications for using a method in practice. They say
only that for estimating the mean of \textit{some} function the method
is better than whatever alternative is being considered. But that does
not rule out the possibility that the method is much worse for the
functions of actual interest.

\section{\hspace*{-8pt}
  Comparisons on simple distributions}\label{sec-lim}\vspace{-11pt}

We can gain some insight into the differences between the various
methods by seeing how they behave on some simple distributions.  Note,
though, that in real applications, the distributions will generally be
more complex, and will change from one update to the next, as other
variables change (unless the variables are independent, which would be
an uninteresting case). So behaviour in these simple situations should
not be taken as a definite indication of how well the methods will
work in practice.

To begin, consider distributions in which all $m$ values have equal
probability --- that is, $\pi(i)=1/m$ for $i=1,\ldots,m$. (Similar
behaviour will occur for distributions that are approximately uniform
over some subset of values, with the total probability of other values
being small.) The probability of a self transition from a state chosen
from $\pi$ when using these probabilities directly as in Gibbs
Sampling will be
\beq
   p^{\mbox{\tiny self}}_{\rule{0pt}{9pt}GS} 
    & = & \sum_i \pi(i)\, P_{GS}(i \rightarrow i)
    \ \ =\ \ \sum_i \pi(i)\, \pi(i) \ \ =\ \ m\,(1/m)^2 \ \ =\ \ 1/m
\eeq
The minimum possible self transition probability for such a
distribution is zero, which will of course be achieved by the methods that
always produce minimum self transition probabilities --- namely,
ZDNAM, ST, UST, DST, UDST, HST, OHST, and ZFSS. It is easy to see
that, for this distribution, zero self transition probabilities will
also be produced by all the other methods besides Gibbs sampling ---
that is, by MHGS, UNAM, DNAM, UDNAM, and FSS.

However, these methods do not all produce the same transition
probabilities.  MHGS, UNAM, DNAM, UDNAM, and ZDNAM all produce
transitions in which $P^*(i\rightarrow j)\,=\,1/(m\!-\!1)$ for $i\ne
j$.  ST, UST, DST, FSS, and ZFSS produce transitions that are periodic
with period $m$ --- cycling through the $m$ states --- while UDST
produces transitions that have probability $1/2$ of moving to the
value before or after the current value, performing a random walk
around the cycle of values.  For even values of $m$, HST and OHST
produce transitions with period two that are not irreducible, while
for odd values of $m$, their transitions are irreducible and
aperiodic.  The effects of these differences when these transitions
are applied to multiple variables with changing distributions, using
various scan orders, are not obvious.

One intuitive measure of how much benefit we might expect from using a
method for avoiding self transitions is the ratio of the probabilities
of a non-self transition for such a method to that for Gibbs sampling.
For the uniform distribution, this ratio is
$1\,/\,(1-1/m)\,=\,m\,/\,(m-1)$ for all the non-GS methods.  If we see
self transitions as wasted effort, and non-self transitions as
useful, this ratio represents the factor by which we might (rather
naively) expect efficiency to be improved over Gibbs sampling.

Another simple case to look at is when one value has much larger
probability than any other value.  Specifically, let $\pi(1)=p$,
let $\pi(j)\,=\,(1-p)\,/\,(m-1)$ for $j=2,\ldots,m$, and look at the
limit as $m$ increases.

\begin{figure}

\begin{center}

\vspace*{-30pt}

\includegraphics[scale=0.6]{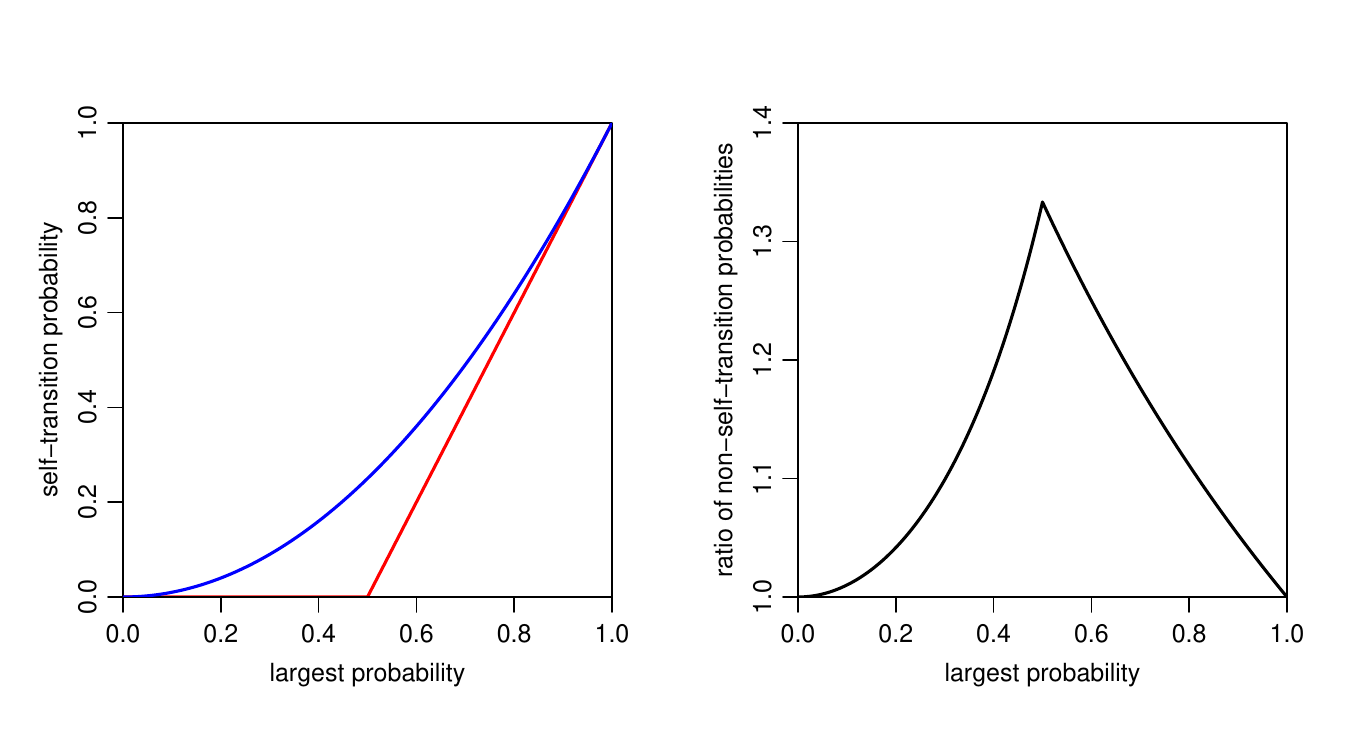}

\vspace*{-20pt}

\end{center}

\caption{Behaviour of self transition probabilities for distributions
with largest probability $p$ and other probabilities of
$(1-p)\,/\,(m-1)$, in the limit as $m$ goes to infinity.  
The plot on the left shows the self transition
probability as a function of $p$ for GS, MHGS, and UNAM in blue, and
all other methods (except UDNAM) in red.  The plot on the right shows
the ratio of the non-self transition probability for methods other
than GS, MHGS, UNAM, and UDNAM to the non-self transition probability
for GS, MHGS, and UNAM, as a function of $p$.}\label{fig-largep}

\end{figure}

In this scenario, the probability of a self transition using Gibbs
sampling from a value $j \ne 1$ is zero in the limit as $m$ increases,
while the probability of a self transition from value 1
is $p$, giving an overall self transition probability for GS of $p^2$.

One can easily compute that this is also the self transition
probability for MHGS and UNAM, in the limit as $m$ increases.  MHGS
and UNAM in fact produce exactly the same transition probabilities in
this situation (for any $m$).

For methods that produce minimum self transition probabilities, the
overall self transition probability is zero if $p \le 1/2$, and $2p-1$
if $p > 1/2$.  Also, in this situation DNAM and FSS produce the same
transition probabilities as ZDNAM and ZFSS. (UDNAM of course has a
self transition probability halfway between UNAM and DNAM.)

Figure~\ref{fig-largep} shows the self transition probabilities for
this scenario, as well as the ratio of the non-self transition
probability for all the methods minimizing self transitions to the
non-self transition probability for GS, MHGS, and UNAM.  This ratio
peaks at $4/3$ when $p=1/2$.

When $p<1/2$, the transition probabilities in this scenario produced
by ZDNAM, ST, HST, and ZFSS are all different (both for finite $m$ and
in the limit), even though they all have zero self transition
probability.  However, when $p \ge 1/2$, all these methods produce the
same transition probabilities.  

Indeed, for \textit{any} distribution with maximum probability one
half or more, \textit{all} methods that produce the minimum overall
self transition probability of $2p-1$ must produce the same transition
probabilities.  Specifically, if $\pi(1)=p\ge1/2$, then these 
transition probabilities must be as follows:
\beq
P^* & = &
 \left[ \begin{array}{cccc}
  \displaystyle {2p-1 \over p} 
    & \displaystyle {\pi(2)\over p} 
    & \displaystyle \ldots 
    & \displaystyle {\pi(m)\over p} \\[16pt]
  \displaystyle 1 
    & \displaystyle 0
    & \displaystyle \ldots 
    & \displaystyle 0 \\[4pt]
  \displaystyle \vdots & \vdots & & \vdots \\[4pt]
  \displaystyle 1 
    & \displaystyle 0
    & \displaystyle \ldots 
    & \displaystyle 0
 \end{array}\right]\label{eq-highpr}
\eeq
To see this, note that for $P^*$ to leave $\pi$ invariant, we must 
have\vspace{-4pt}
\beq
  \pi(1)\ \ =\ \ p \ \ =\ \ \sum_{i=1}^m \pi(i)\, P^*(i\rightarrow1)
     \ \ =\ \ p\, {2p-1 \over p} \ +\ \sum_{i=2}^m \pi(i)\, P^*(i\rightarrow1)
\eeq
and hence\vspace{-4pt}
\beq
  \sum_{i=2}^m \pi(i)\, P^*(i\rightarrow1) 
    \ \ =\ \ p\ -\ p\, {2p-1 \over p}\ \ =\ \ 1-p
\eeq
Since $\sum_{i=2}^m \pi(i)\,=\,1-p$, this is possible only if
$P^*(i\rightarrow1)=1$ for $i=2,\ldots,m$.  Note that this $P^*$ is
reversible with respect to $\pi$, so any method that produces minimal
self transition probabilities produces reversible transitions in this
context, even if the method is not generally reversible.

\section{\hspace*{-8pt}
  Framework for empirical comparisons}\label{sec-emp}\vspace{-11pt}

I will empirically compare the performance of the modified Gibbs
sampling procedures with each other and with standard Gibbs sampling
for three problems: the well-known Potts model used in statistical
physics and image processing, sampling of mixture indicators for a
Bayesian mixture model, and sampling of unobserved variables in a
belief network.  Of course, the results of these experiments are only
suggestive of performance in other applications, in which the
distributions sampled may have different characteristics.

I will evaluate all the fourteen methods discussed earlier, which
can be grouped as follows:\vspace{-6pt}
\begin{enumerate}
\item[1)] Gibbs sampling and methods that can be viewed
          as deriving from it: GS, MHGS, UNAM, DNAM, UDNAM, and ZDNAM.
\item[2)] Shifted tower methods: ST, DST, UST, UDST, HST, and OHST.
\item[3)] Slice sampling methods: FSS and ZFSS.\vspace{-6pt}
\end{enumerate}
Of these, ZDNAM, all the shifted tower methods, and ZFSS
always minimize self transition probability, and all the methods 
in group (1) plus UDST, HST, and OHST always produce reversible 
transitions.

Each method will be used in combination with several schemes for
choosing which of the $n$ variables are updated in each iteration.
For all schemes, $n$ variable updates are considered to constitute a 
\textit{scan}, which is sometimes viewed as a single iteration. 
The schemes used may include the following:\vspace{-6pt}
\begin{enumerate}
\item[1)] \textbf{Random.}  For each iteration, one of the $n$ variables
is randomly selected to be updated, independently of previous iterations.
\item[2)] \textbf{Sequential.} The variables are updated in a predefined
order from $1$ to $n$, which constitutes one scan.  Not done for
mixture models, for which there is no meaningful predefined order.
\item[3)] \textbf{Shuffled sequential.} The variables are randomly shuffled,
once, at the beginning of a run, the same way for all runs. They are then
repeatedly updated in this shuffled order, with each set of $n$ updates 
considered one scan.
\item[4)] \textbf{Checkerboard.} Only done for the Potts models, for which
the $n$ variables are arranged in a square array, on which one can imagine
a checkerboard pattern being placed.  A scan consists of updates for all the 
variables on black squares, followed by updates for all the variables on 
white squares.
\item[5)] \textbf{Random order.} For each scan, an order of the
$n$ variables is chosen at random, and the variables are then updated
in this order.  A new random order is chosen for the next scan.
\item[6)] \textbf{Random order times four.} Like the random order method, except
that the same random order is used for four scans in a row, before a new
random order is chosen.\vspace{-6pt}
\end{enumerate}

For each combination of method and scan order, the Markov chain is
simulated for a large number, $K$, of scans, starting with a random
state, producing a total of $nK$ states.  These states are then used
to form estimates for the expectation of several functions of the
state variables.  No iterations are discarded as ``burn-in'', since
the length of the runs and the speed of convergence make this
unnecessary.  Both \textit{thinned} and \textit{unthinned} estimates
are found.  The unthinned estimate for the expectation of a function
is the average value of the function at all iterations.  The thinned
estimate is the average over only the values after the last update of
a scan.  The unthinned estimates are therefore averages over $nK$
function values, whereas the thinned estimates are averages over $K$
function values.

For each distribution tested, three groups of methods are tested
using sets of four runs for each method,
with all runs being independent (using different
random number seeds).  The three groups compare the following selections
of methods:\vspace{-6pt}
\begin{itemize}
\item[1)] GS, MHGS, UNAM, DNAM, UDNAM, ZDNAM.
\item[2)] ST, DST, UST, UDST, HST, OHST.
\item[3)] UNAM, ZDNAM, ST, UDST, FSS, ZFSS.\vspace{-6pt}
\end{itemize}
The first group compares methods related to Gibbs sampling, the
second compares the shifted tower methods, and the third compares
what appear to be the best from the first two groups along with
the slice sampling methods.\footnote{\rule{0pt}{10pt}%
Note that runs in the third group
are independent of those in the first two groups for the same method.}
Summary graphs are produced for each group, for each of the distributions
tested.

The efficiency of a method and scan order for a particular function is
measured by an estimate of the asymptotic variance
(equation~(\ref{eq-asymvar})) for that function, found using the 
following formula:
\beq
  v(f,P) & = & \gamma_0 \ +\ \, 2\, \sum_{k=1}^{\infty} \gamma_k
\label{eq-estasymv}
\eeq
where $\gamma_k$ is the autocovariance of $f$ at lag $k$, defined by
\beq
 \gamma_k & = &  = E\Big[\Big(f(X^{(t)})-\mu\Big)\Big(f(X^{(t+k)})-\mu\Big)\Big]
\eeq
where the expectation is over realizations of the Markov chain with
transitions $P$, which leave $\pi$ invariant, started from a state
drawn from $\pi$, and $\mu$ is the expectation of $f$ with respect to $\pi$.
Since the realization will be stationary, the choice of $t$ in the above
formula makes no difference.
Note that $\gamma_0$ is the variance of $f$ with respect to $\pi$. 

This formula is proved for homogeneous reversible chains with a finite
state space in (Neal and Rosenthal 2023, Proposition~3)), but holds
more generally, including for chains with non-reversible transitions,
and those in which the transitions depend on the time index in a periodic
way (as for Gibbs sampling with a sequential scan), if we interpret
$\gamma_k$ as the average covariance between $f(X^{(t)})$ and $f(X^{(t+k)})$
as transitions at time $t$ vary periodically, provided that the 
distribution at time $t$ converges to $\pi$ as $t$ goes to infinity, and the
variance is finite (as is always the case for a finite state
space).\footnote{\rule{0pt}{10pt}%
Supposing that $\mu=0$ for simplicity,
this is a simple consequence of 
expanding $N$ times the variance of the mean estimate from a run of length $N$
as\vspace{-2pt}
\beq\textstyle
N E \Big[ \Big( {\textstyle 1 \over \textstyle N} 
    \sum\limits_{t=1}^N f(X^{(t)})\Big)^2 \Big]
\ =\ E\Big[ {\textstyle 1 \over \textstyle N} \sum\limits_{i=1}^N 
      \sum\limits_{j=1}^N f(X^{(i)}) f(X^{(j)}) \Big]
\ =\ E\Big[ {\textstyle 1 \over \textstyle N} \sum\limits_{t=1}^N f(X^{(t)})^2
     \ +\ 2 \sum\limits_{k=1}^{N-1} {\textstyle 1 \over \textstyle N}
        \sum\limits_{t=1}^{N-k} f(X^{(i)}) f(X^{(i+k)}) \Big] 
\\[-11pt] \nonumber
\eeq
which equals $E\big[\hat\gamma_0+2\!\sum\limits_{k=1}^{N-1} \hat\gamma_k\big]$, 
where $\hat\gamma_k$ is the estimate for $\gamma_k$ given in 
equation~(\ref{eq-hatgamma}) below.  Since the expectations of these
estimates converge to the true $\gamma_k$ as $N$ goes to infinity, 
equation~(\ref{eq-estasymv}) will hold generally.
}

From a realization of the chain of length $N$, the standard estimate of
$\gamma_k$ is
\beq
  \hat\gamma_k & = & {1 \over N}\, 
    \sum_{t=1}^{N-k} \Big(f(x^{(t)})-\mu\Big)\Big(f(x^{(t+k)})-\mu\Big)
  \label{eq-hatgamma}
  \\[-11pt] \nonumber 
\eeq
If $\mu$ is not known, it may be replaced by 
$\hat\mu=(1/N)\sum\limits_{t=1}^N f(x^t)$.\vspace{-2pt}

The asymptotic variance for $f$ is then estimated as\vspace{-5pt}
\beq
  \hat v(f,P) & = & \hat \gamma_0\ +\ \, 2\, \sum_{k=1}^M \hat\gamma_k 
  \label{eq-asymvar-est}
\eeq
where $M$ is selected such that $\hat\gamma_k$ is nearly zero for 
$k>M$.  Note that this estimate will be good only
if the length of the run, $N$, is much larger than a suitably chosen
value of $M$.\footnote{\rule{0pt}{10pt}%
Note that setting $M$ to the largest possible value 
of $N-1$ is not good, since the estimate will then have a large variance 
dominated by estimates $\hat\gamma_k$ whose means are close to zero.}

For unthinned estimates, the estimate for the asymptotic variance
based on a run of $K$ scans will use $N=nK$ in the above formulas.
For thinned estimates, there are only $N=K$ function values used, but
in the presentations of results, the asymptotic variance estimates
with thinning are multiplied by $n$ to account for each value used in
estimation requiring a factor of $n$ more computation time.

The practical motivation for thinning is usually to reduce memory
requirements and time for computing function values by a factor of
$n$, with the expectation that the efficiency of estimation will be
worse than if all values were used for estimates, though only slightly
worse for typical problems.  The belief that thinning gives worse
estimates is generally correct (provably so for reversible updates on
randomly chosen variables (Geyer 1992, Section 3.6)), but as will be
seen below, there is one context in which thinning actually improves
estimation efficiency.

\section{\hspace*{-8pt}
  Comparisons for Potts models}\label{sec-potts}\vspace{-11pt}

The Potts model originates in statistical physics (e.g., Landau and
Binder 2009, Section 4.3.2), but similar models are also used for
image analysis (e.g., Geman and Geman 1984) and other applications in
which some discrete aspect of a system exhibits local spatial
coherence.  

I will consider two-dimensional Potts models, which define a
distribution on the space of arrays of variables, $x_{r,c} \in
\{1,\ldots,m\}$, for $r=0,\ldots,R-1$ and $c=0,\ldots,C-1$, with the
total number of variables being $n=RC$.  Variables are ``neighbors''
if one is immediately above, below, left, or right of the other, with row or
column positions wrapping around from $R-1$ or $C-1$ to $0$.  The
distribution is defined by
\beq
  \pi(x) & = & {1 \over Z}\,
   \exp\Big(\, b\, \sum_{r=0}^{R-1}\, \sum_{c=0}^{C-1}\,
    \big( I(x_{r,c} = x_{r^+,c}) \,+\, I(x_{r,c} = x_{r,c^+}) \big)\Big)
\label{eq-potts}
\eeq
where $r^+ = r+1 \mod R$ and $c^+ = c+1 \mod C$.  Here, $Z$ is the 
normalizing constant needed to make these probabilities sum to one.
The parameter $b$ controls how strongly variables at neighboring
positions tend to be the same (if $b>0$) or different (if $b<0$).
In physical terms, minus the sum inside the exponential above is
proportional to the ``energy'' of the system, and $1/b$ is
proportional to the ``temperature''.

A Gibbs sampling update for this model will choose a new value for
$x_{r,c}$ from $\{1,\ldots,m\}$, with~$r$ and~$c$ chosen either
randomly or sequentially in some order.  The conditional distribution
for $x_{r,c}$ given the other variables depends only on the four
variables above, below, left, or right of $x_{r,c}$.  There are $m^4$
possible values of these four neighbors, so when $m$ is fairly small,
it is possible to precompute the Gibbs sampling probabilities for all
combinations of neighboring values.  Similarly, modified Gibbs
sampling probabilities, found with any of the methods discussed, could
be precomputed for all combinations of neighboring values and all
possibilities for the current value of the variable being updated.
All methods would then take close to the same computation time (though
for ordinary Gibbs sampling, the table of possible distributions would
be $m$ times smaller).

For my experiments, I used models with $m=4$, and either $R=C=8$
($n=64$) with $b=0.85$ or $R=C=5$ ($n=25$) with $b=-0.4$. In actual
applications, $R$ and $C$ are typically larger, but with these smaller
values, very long runs can be done to obtain accurate comparisons of
asymptotic variance.  For simplicity of implementation, I did not
precompute probabilities for these experiments.

All the scan orders described in Section~\ref{sec-emp} were tested.
The pre-defined sequential order was a raster scan over rows and
columns.  When $R$ and $C$ are even, note that the checkerboard scan
updates of black positions can be done in parallel, since there are no
interactions between the sites being updated, after which the updates
of the white positions can be done in parallel.  This will often be a
significant computational advantage of this scan.  However, when $R$
or $C$ are odd, the checkerboard scan will have adjacent sites of the
same colour at the point of wrapping around from $R-1$ or $C-1$ to
$0$, which inhibits parallel updates at these positions.  In these
experiments, I did not do updates in parallel for the checkerboard
scans, nor do the presentations of results account for any efficiency
advantage of using a checkerboard scan.

The expectations of three functions of state were estimated from
the runs done:\vspace{-8pt}
\begin{itemize}

\item[1)] \textbf{Count of 1s.}
The number of the $n=RC$ variables whose value is 1.  Since the
distribution is symmetrical with respect to the $m$ possible values,
this function has the same expectation as that of the number of
variables with any other value.  From symmetry, the expectation 
of this function must be $RC/m$, but its variance will vary with $b$.

\item[2)] \textbf{Sum of squared counts.}
The sum of the squares of the counts of how many variables have each
of the $m$ possible values. This has its largest possible value, of
$(RC)^2$, when all variables have the same value, so one value has
a count of $RC$ and the others have counts of zero.

\item[3)] \textbf{Number of neighbor pairs with equal values.}
The number of the $2RC$ pairs of neighboring variables that have
the same value.  If the variables took on the $m$ values uniformly
and independently, the expected value would be $2RC/m$ which is $RC/2$
when $m=4$ as in these experiments. Note that this function is
proportional to minus the ``energy'' in the physical 
interpretation; it is also proportional to the log of the
joint probability of all variables.\vspace{-6pt}

\end{itemize}

The $8\times8$ Potts models used a positive value for $b$ of $0.85$,
so there will be a tendency for neighboring sites to have the same
values.  Three arrays of values sampled from this distribution are
shown in Figure~\ref{fig-8x8}.  

\begin{figure}

\begin{verbatim}
      1  3  3  4  4  4  3  4  |  1  2  1  3  2  2  4  3  |  3  2  4  4  4  1  3  3
      4  1  2  2  3  3  3  4  |  4  4  1  4  4  4  4  4  |  3  1  4  4  2  1  1  3
      2  4  4  1  4  1  4  4  |  4  3  3  1  2  2  3  4  |  3  3  3  3  2  3  3  3
      2  3  2  2  2  2  4  2  |  3  3  1  1  1  1  3  3  |  3  3  3  3  2  1  3  3
      4  2  2  2  2  1  4  4  |  2  4  1  4  4  4  3  3  |  3  3  3  3  3  3  3  3
      4  2  4  4  3  2  2  4  |  4  3  4  4  4  3  1  1  |  2  2  3  4  4  3  3  3
      4  2  4  3  3  2  2  4  |  4  3  4  4  2  3  4  4  |  2  4  4  4  4  4  3  2
      4  1  2  4  1  1  1  1  |  2  2  2  2  2  2  4  4  |  3  2  4  4  1  1  3  3
\end{verbatim}

\caption{Three $8\times8$ arrays of values sampled from the Potts distribution 
         with $m=4$ and $b=0.85$.}\label{fig-8x8}

\end{figure}

For this distribution, the average count of 1 values is exactly 16,
from symmetry, with a variance of approximately 66.  The sum of
squared counts of the four possible values has an average of
approximately $1.29\times10^3$ and variance of approximately
$5.6\times10^4$. The average number of neighbor pairs with equal
values is approximately 61.9, more than 32, which it would be if
values for sites were drawn uniformly and independently, as expected
with a positive $b$.  The variance is approximately 67.

Each run for the $8\times8$ Potts model consisted of $K=200000$ scans,
each with $n=RC=64$ variable updates.  For each of the three groups of
methods, four independent runs of this length were done for each
method in the group.

\begin{figure}[p]
\begin{center}\vspace{-8pt}
\hspace*{-6pt}\includegraphics[scale=1]{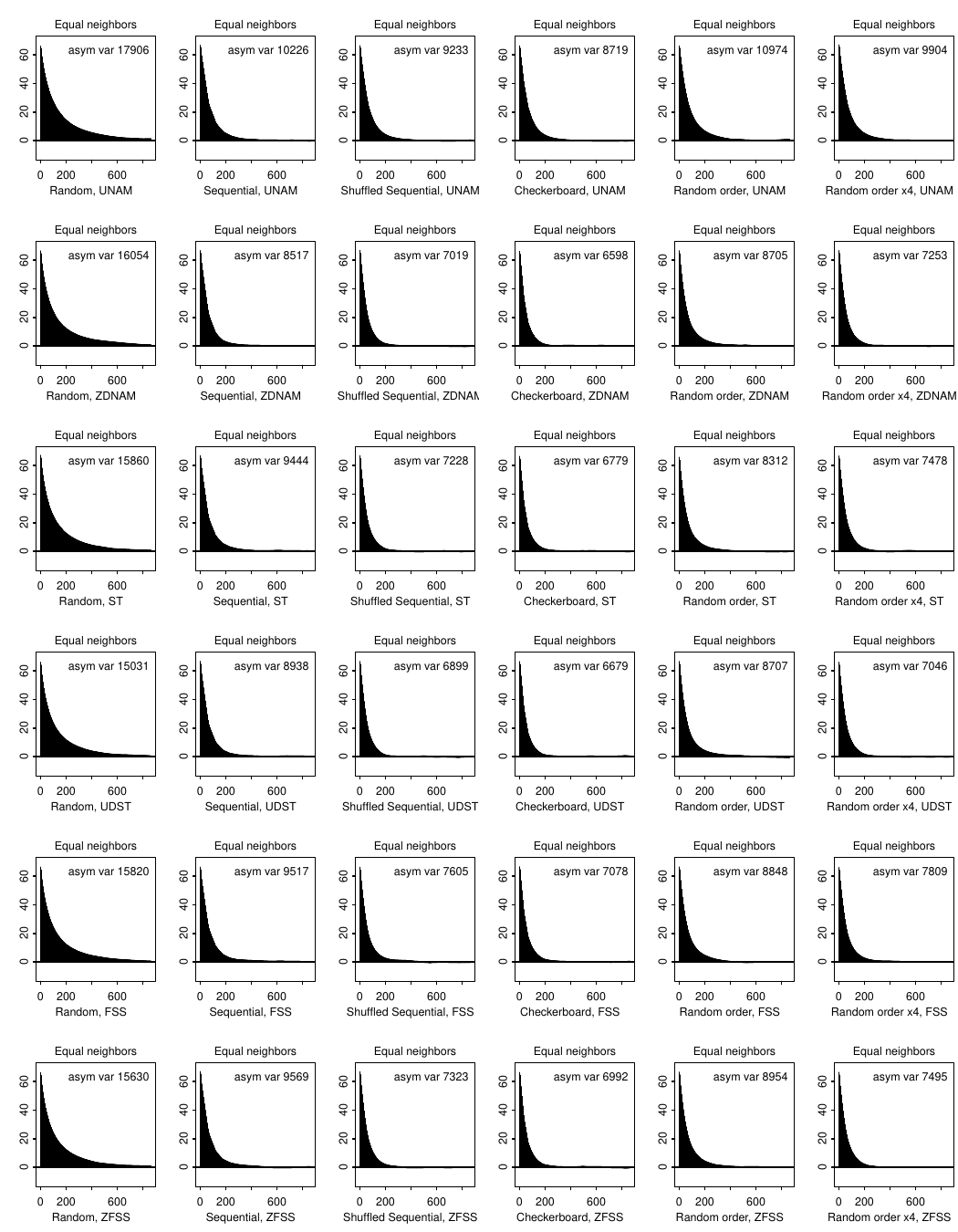}\vspace{-8pt}
\end{center}
\caption{Autocovariance function estimates for the number of equal
neighbors, from one set of runs for the $8\times8$ Potts model, 
for methods in the third group.}\label{fig-8x8-autocov}

\end{figure}

Estimates of autocovariance functions for the number of equal
neighbors (proportional to the energy) based on one of the four sets
of runs done for the third group of methods are shown in
Figure~\ref{fig-8x8-autocov}. These plots show that for the 
$8\times8$ Potts model autocovariances for all methods with all scan orders
are positive.  This is expected, since a fairly large positive value
for $b$ of $0.85$ leads to variables often having most neighbors the
same (as seen in Figure~\ref{fig-8x8}), and hence the conditional
distribution for that variable is concentrated on this dominant neighboring
value, leading to slow movement through the state space, and high
autocovariances for most functions.  

This can also be
seen from the frequencies of self transitions of the various methods
for the $8\times8$ Potts model
(which are the same for all scan orders):\vspace{-4pt}
\begin{quotation}\noindent
  GS: 0.46,\ \ \ MHGS: 0.33,\ \ \ UNAM: 0.31,\ \ \ DNAM: 0.24,\ \ \ 
  UDNAM: 0.28,\ \ \ FSS: 0.24 \\[4pt]
  ZDNAM, ST, DST, UST, UDST, HST, OHST, ZFSS: 0.23\ \ 
  (the minimum possible)\vspace{-6pt}
\end{quotation}
The maximum conditional probability for an update was half or more
40\% of the time.

\begin{figure}[p]
\begin{center}
\includegraphics[scale=0.65]{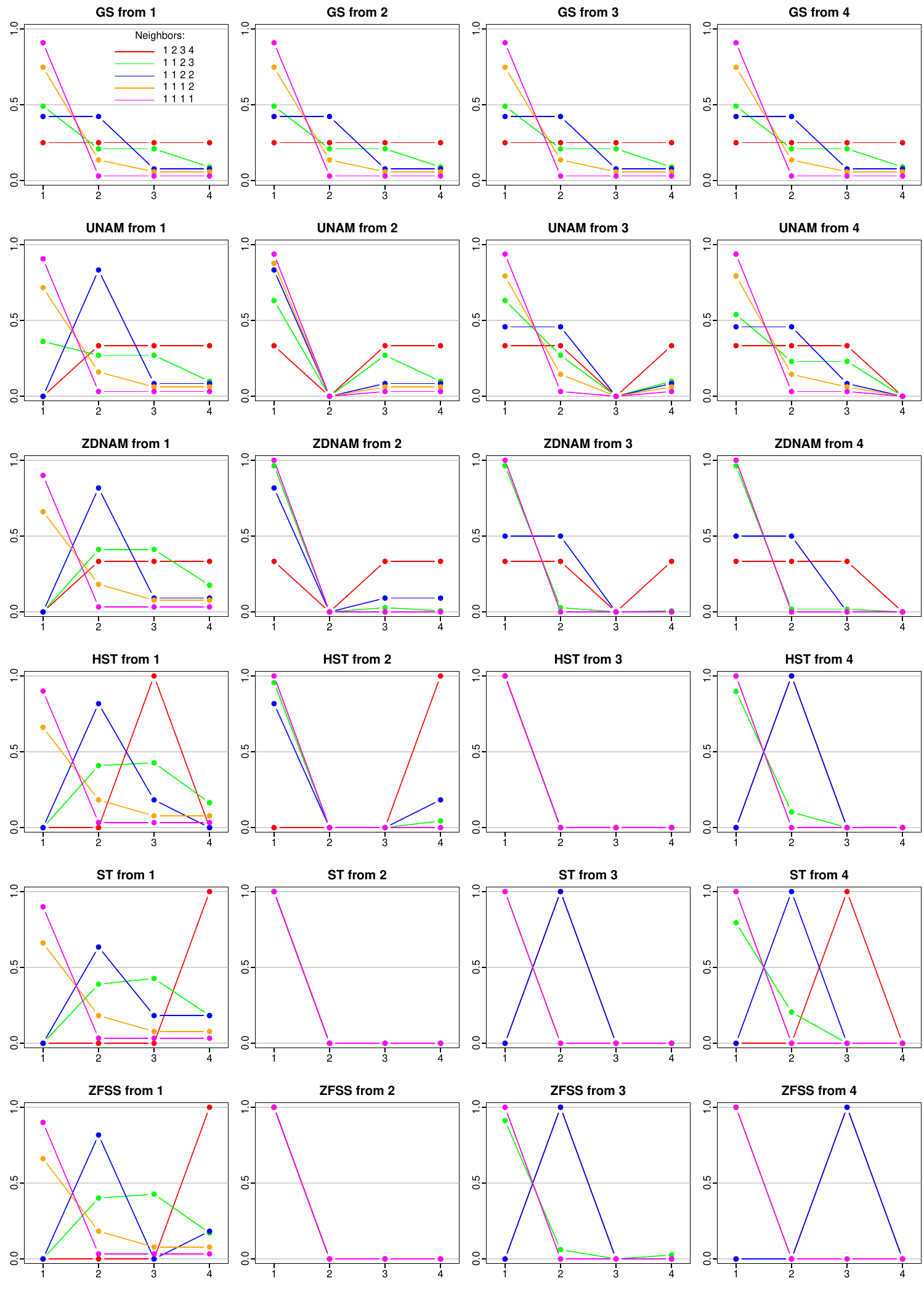}
\vspace*{-12pt}
\end{center}
\caption{Transition probabilities in different contexts for the $8\times8$
Potts model.}\label{fig-8x8-contexts}
\end{figure}

Although eight methods achieve the minimum self transition probability,
these methods have substantially different transition probabilities.
Figure~\ref{fig-8x8-contexts} shows, for each method, how the
transition probabilities vary depending on current value of the variable
being updated and the context of values for its neighbors.  The five
possible contexts are (1) all neighbors different, (2) two neighbors
the same, the others different, (3) two neighbors the same, other two
also the same, but different from the first two, (4) three neighbors the
same, the remaining one different, and (5) all
neighbors the same.  For each context, the transition probabilities
are shown for each current value of the variable (all the same
for Gibbs sampling, since it ignores the current value).

\begin{figure}[p]
\begin{center}
\includegraphics[scale=1]{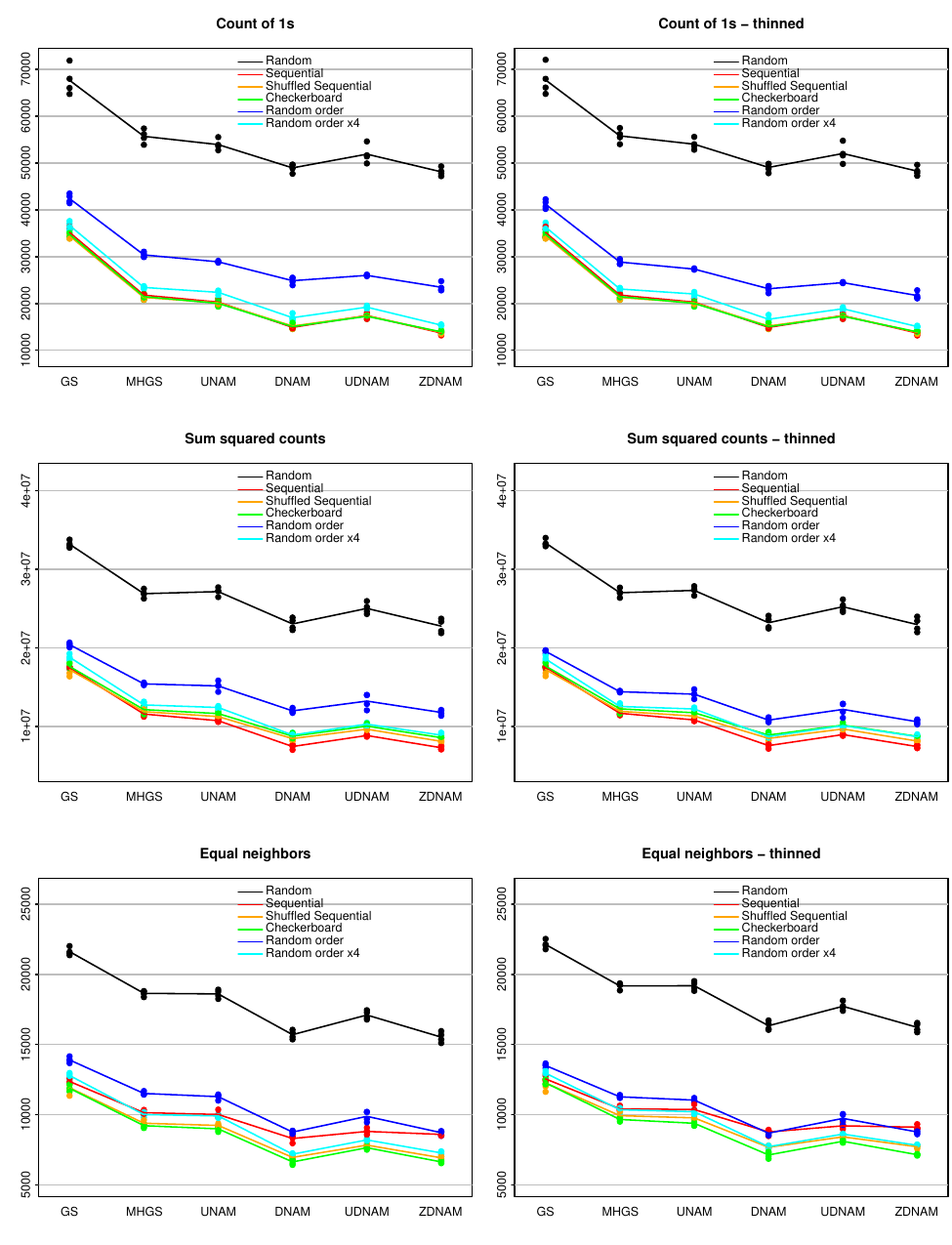}
\end{center}
\caption{Summaries of autocovariance function estimates for the $8\times8$
Potts model, for the first group of methods.}\label{fig-8x8-g1}
\end{figure}

\begin{figure}[p]
\begin{center}
\includegraphics[scale=1]{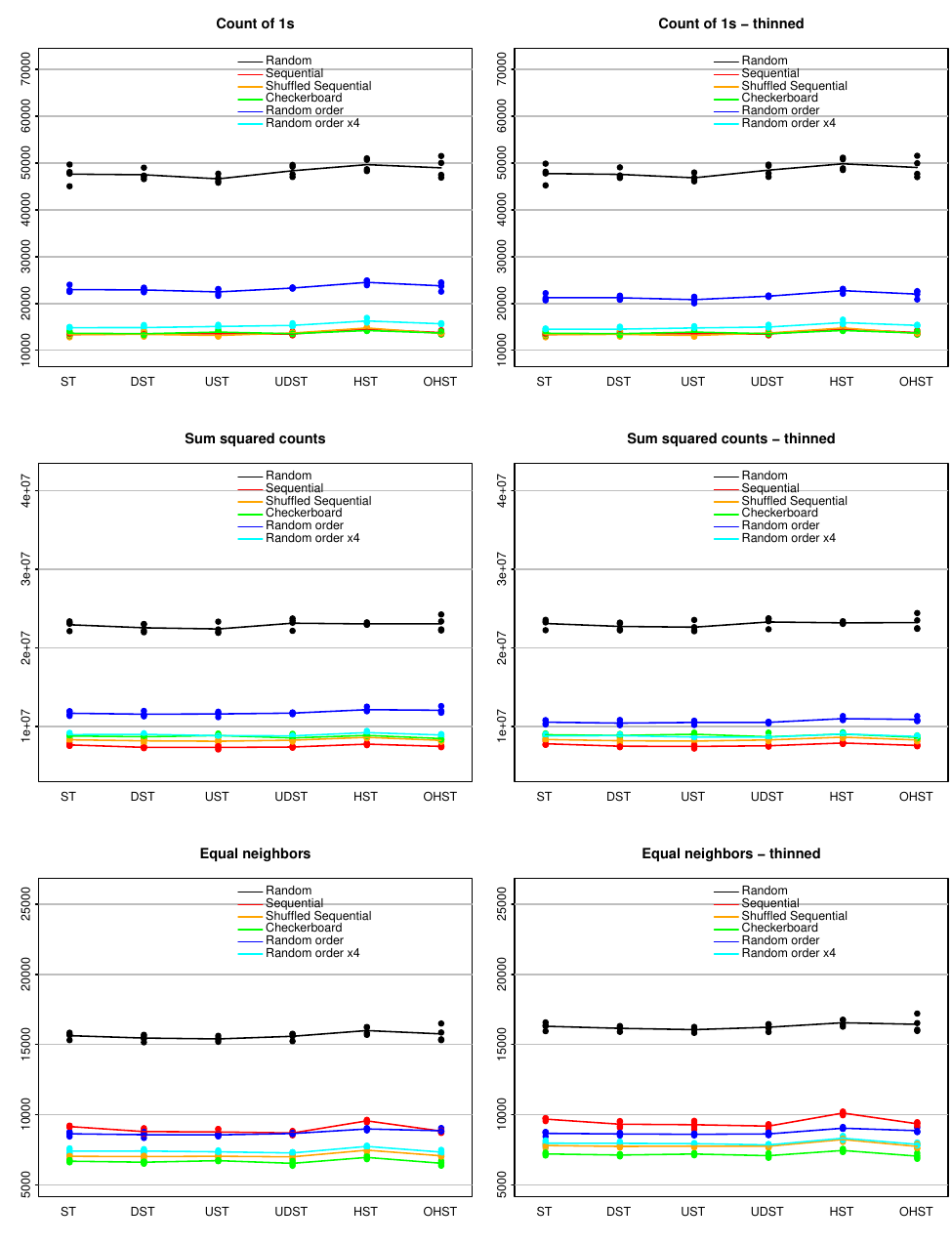}
\end{center}
\caption{Summaries of autocovariance function estimates for the $8\times8$
Potts model, for the second group of methods.}\label{fig-8x8-g2}
\end{figure}

\begin{figure}[p]
\begin{center}
\includegraphics[scale=1]{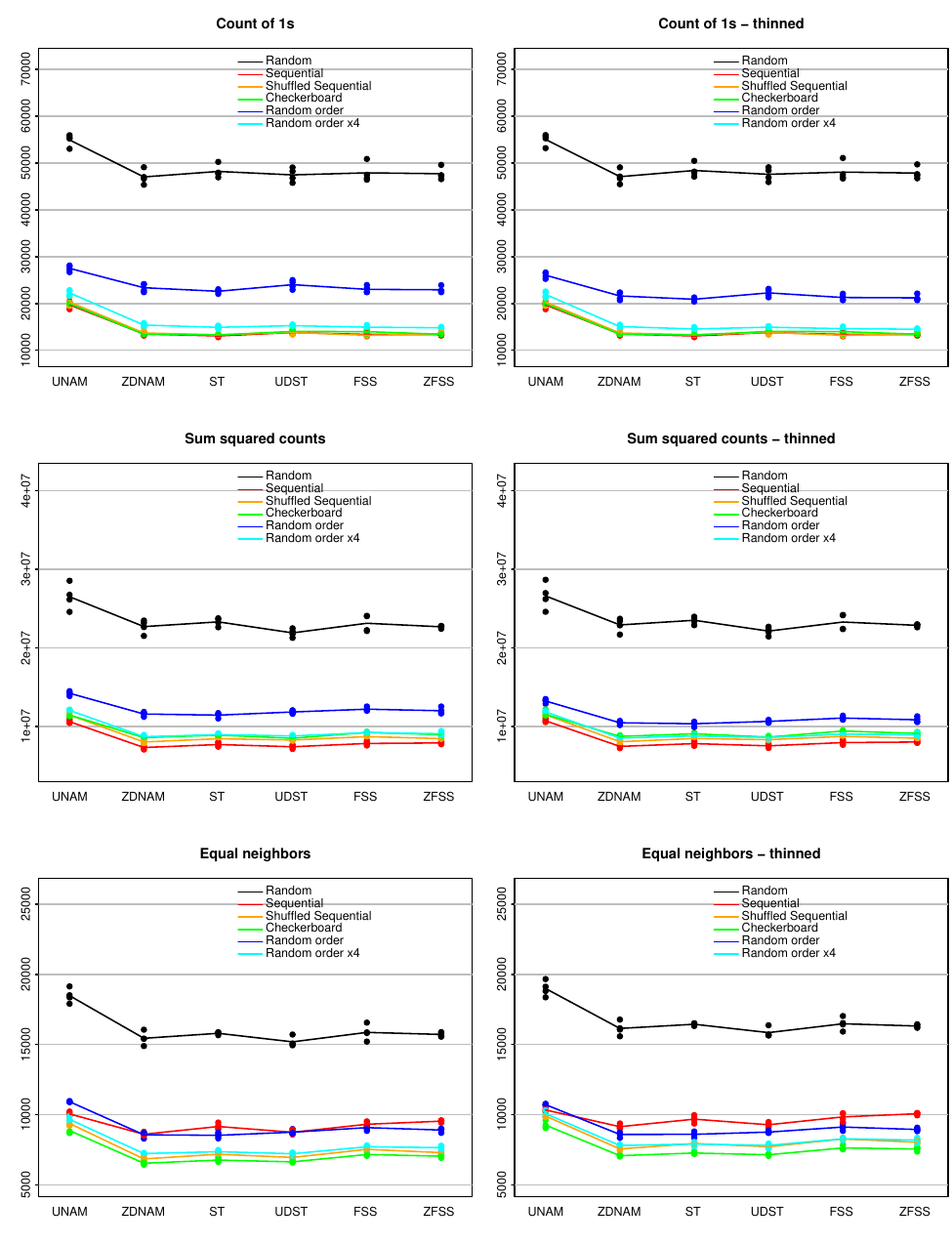}
\end{center}
\caption{Summaries of autocovariance function estimates for the $8\times8$
Potts model, for the third group of methods.}\label{fig-8x8-g3}
\end{figure}

Summaries of asymptotic variance estimates for all three functions
looked at, for all groups of methods, and all scan orders, are shown
in Figures~\ref{fig-8x8-g1} through~\ref{fig-8x8-g3}.  The summary
plots show both the asymptotic variance estimates from the four individual
runs, as dots, and the average of these estimates, as lines connecting
average estimates for the various methods (for each scan order, as
indicated by colour).

It is evident from these figures that random selection of the variable
to update (black dots and lines) is greatly inferior to the other scan
orders.  With a few exceptions, this will prove to also be true for
the problems looked at later.  One disadvantage of random selection is
that by chance some variables will not be updated for many iterations.
This may suffice to explain why it usually performs poorly.  It is,
however, the only scan order for which the theoretical analysis
presented earlier applies (apart from some of the non-dominance results).

The results when the variable to be updated is selected at random
(black dots and lines) are consistent with these theoretical results.
Theory says that MHGS, UNAM, DNAM, UDNAM, and ZDNAM
efficiency-dominate GS, and for the three functions looked at, we do
see in Figure~\ref{fig-8x8-g1} that these methods have substantially
lower asymptotic variance than GS.  Theory also says that UNAM should
efficiency-dominate MHGS, but in this case the differences in
asymptotic variance are quite small, and for the sum of squared
counts, the average estimate for asymptotic variance for UNAM is
actually slightly greater than for MHGS --- though from the spread in
results of the four individual runs, this can be attributed to chance.

With random selection of variable to update, DNAM and ZDNAM perform
somewhat better than UNAM or UDNAM, though there is no theoretical
guarantee of this.  DNAM, ZDNAM, the shifted tower methods (see
Figure~\ref{fig-8x8-g2}) and the slice sampling methods (see
Figure~\ref{fig-8x8-g3}) all perform very similarly.

Theory also says that for reversible methods, with random selection of
variable to update, thinning (looking only at the state after every
$n$ updates) should produce worse estimates (Geyer 1992, Section 3.6).
The results on the $8\times8$ Potts model for the methods in
Figure~\ref{fig-8x8-g1} (all reversible) and for the reversible UDST,
HST, and OHST methods in Figure~\ref{fig-8x8-g2} are consistent with
this, but a higher asymptotic variance with thinning (after
multiplying by $n$ to account for computation time) is only noticeable
for the ``equal neighbors'' function, and even there the difference is
small.  This is expected when, as here, autocovariances are high.

For this problem, thinning also has a very small effect on efficiency
for non-reversible methods and scan orders other than random
selection, with one surprising exception --- when each scan updates
all variables in a random order (different for each scan), thinning
often gives a noticeable \textit{reduction} in asympotic variance.
See the blue dots and lines in Figures~\ref{fig-8x8-g1}
through~\ref{fig-8x8-g3}.  This is true for all methods, and all three
functions, though it is less noticeable for the `equal neighbors''
function than for the other two.

The same phenomenon will be seen later for other problems.  A possible
explanation can be seen by considering an extreme circumstance in
which we are estimating the expectation of a function that depends on
only a single variable, which is independent of the other variables.
When each scan updates variables in a random order, this variable will
sometimes be updated early in the order, and sometimes late in the
order.  If a scan in which it is updated late is followed by a scan in
which it is updated early, the newly sampled value will be present for
only a few iterations (much less than $n$), whereas in the opposite
case, the newly sampled value could be present for almost $2n$
iterations.  When all iterations are used for estimation, this
introduces random variation into how much each sampled value affects
the estimate, which reduces estimation efficiency.  However, a thinned
estimate will look only at the last iteration of each scan, and use
each sampled value equally, giving an estimate with lower variance.
This effect should also be present to some extent in less extreme
circumstances.

Though usually better than random selection, a random scan order is
usually worse than all the other scan orders, for both this problem
and for ones considered later.  Repeating the same random order for
four scans before generating a new order (see the cyan dots and lines)
is almost always an improvement on using a random order for just one
scan.  This is understandable, since using the same random order four
times reduces random variation in intervals between updates of the
same variable, which plausibly is beneficial in most circumstances,
though there is no theoretical guarantee of this.  The ``shuffled
sequential'' order takes this further, generating one random order
that is used for all scans (the same order for all runs).  This is
almost always better that repeating the same scan only four times.

For the Potts model, two other scans are also tried --- a sequential
raster scan across each row, then across the next row, etc., and the
``checkerboard'' scan, of first all ``black'' variables and then all
``white'' variables, as described earlier.  For the $8\times8$ Potts
model, one or the other of these is always the best scan, for the
functions tested, but which is best depends on the function.  The
sequential scan is best for the sum of squared counts, the
checkerboard scan is best for the number of equal neighbors, and these
two are almost the same (and better than the others) for the count of
1s.  Somewhat surprisingly, the sequential raster scan is worse than
the shuffled sequential scan when estimating the expected number of
equal neighbors (though better for the other two functions).

For the most part, the choice of scan order does not affect which of
the modified Gibbs sampling methods is best.  GS, MHGS, UNAM, and
UDNAM are uniformly worse than the other methods.  Very little
difference is seen amongst the shifted tower methods
(Figure~\ref{fig-8x8-g2}), except that HST is perhaps slightly worse
than the others.  DNAM and FSS do not minimize self transition
probabilities, but for this problem their self transition
probabilities are only slightly greater than the minimum, and they
perform only slightly worse than ZDNAM and ZFSS.  The performances of
the ZDNAM, ST, DST, UST UDST, OHST, and ZFSS methods are difficult to
distinguish, but they equal or exceed the performance of the other
methods for all scan orders.

The $5\times5$ Potts models used a negative value for $b$ of
$-0.4$, so neighboring sites will tend to have different values.  Five
arrays of values sampled from this distribution are shown in
Figure~\ref{fig-5x5}.  

\begin{figure}

\begin{verbatim}
 1  2  1  3  2  |  3  2  2  4  4  |  2  1  2  1  3  |  1  1  1  1  2  |  3  2  1  4  3
 3  3  1  3  1  |  1  4  3  3  4  |  4  3  2  2  1  |  3  2  2  2  1  |  2  3  1  2  1
 4  1  2  3  2  |  2  3  2  4  3  |  3  2  3  3  4  |  2  1  4  1  3  |  2  2  2  3  3
 2  3  2  3  1  |  3  1  3  1  2  |  4  4  1  4  2  |  4  2  3  4  1  |  3  2  1  4  2
 2  1  1  4  3  |  2  3  3  3  4  |  2  4  3  1  3  |  3  4  1  4  2  |  2  3  1  4  1
\end{verbatim}

\caption{Five $5\times5$ arrays of values sampled from the Potts distribution 
         with $m=4$ and $b=-0.4$.}\label{fig-5x5}

\end{figure}

For this distribution, the average count of 1 values is exactly 6.25,
from symmetry, with a variance of approximately 3.37. The sum of
squared counts of the four possible values has an average of
approximately $170$ and variance of approximately $116$.  The average
number of neighbor pairs with equal values is approximately 9.09,
less than 12.5, which it would be if values for sites were drawn
uniformly and independently, as expected with a negative value for
$b$.  The variance is approximately 7.7.

Each run for the $5\times5$ Potts model consisted of $K=1000000$ scans,
each with $n=RC=25$ variable updates.  For each of the three groups of
methods, four independent runs of this length were done for each
method in the group.

\begin{figure}[p]
\begin{center}\vspace{-8pt}
\hspace*{-6pt}\includegraphics[scale=1]{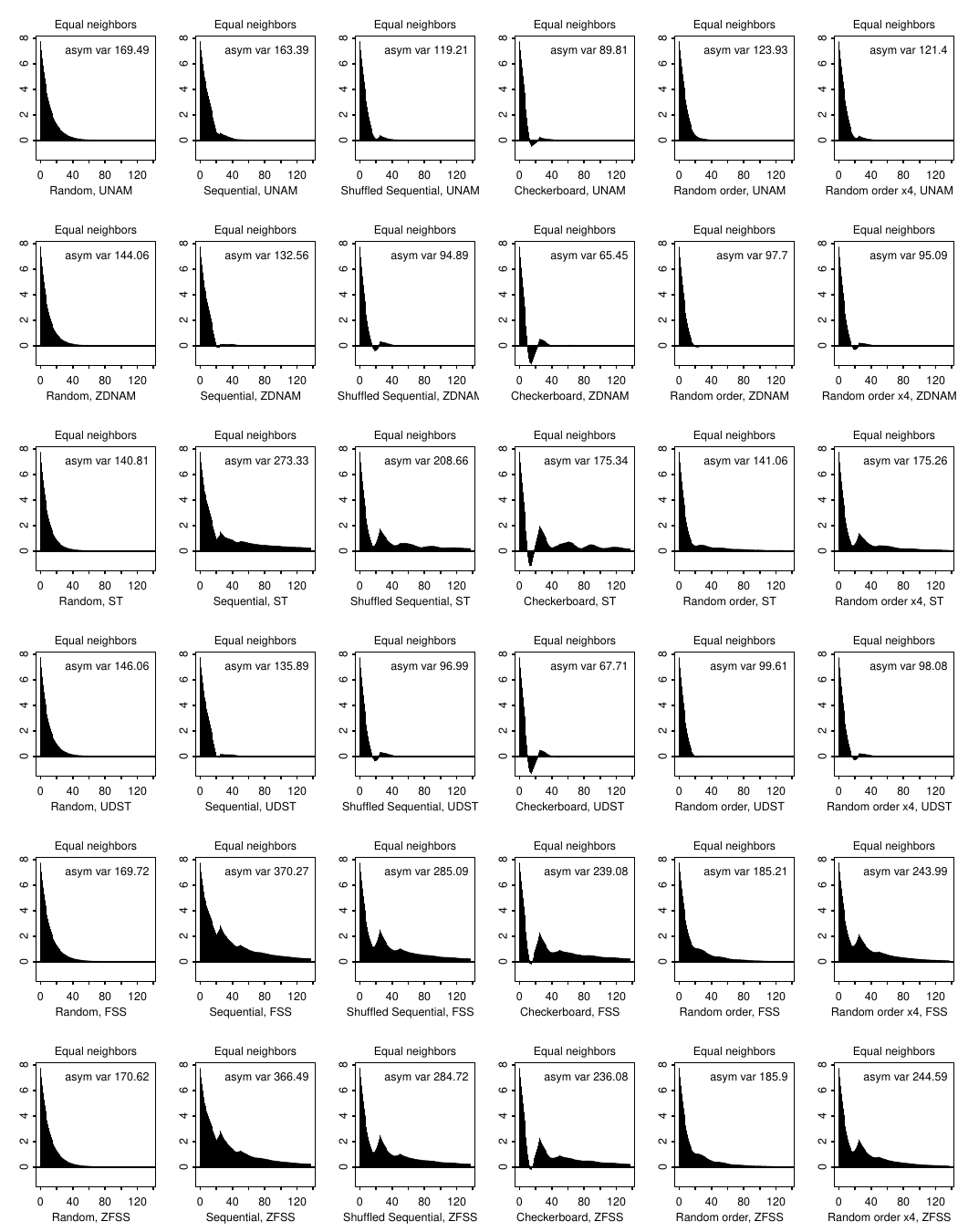}\vspace{-8pt}
\end{center}
\caption{Autocovariance function estimates for the number of equal
neighbors, from one set of runs for the $5\times5$ Potts model, 
for methods in the third group.}\label{fig-5x5-autocov}

\end{figure}

Estimates of autocovariance functions for the number of equal
neighbors (proportional to the energy) based on one of the four sets
of runs done for the third group of methods are shown in
Figure~\ref{fig-5x5-autocov}.  In contrast to the $8\times8$ model
with positive $b$, this $5\times5$ model with negative $b$ has
negative autocovariances for some combinations of update method and
scan order.  Of particular note are the negative autocovariances for
ZDNAM and UDST when the checkerboard scan order is used, which result
in the smallest asymptotic variances for this function.

The antithetic effects of modified Gibbs sampling updates have more
scope to produce negative autocovariances when $b$ is negative, since
\textit{avoiding} the value of a neighboring variable can (with $m=4$)
be done in more than one way, so an antithetic method can switch
between them, whereas \textit{matching} a neighboring value can be
done in only one way.

This effect shows up in the frequencies of self transitions for the various 
methods, which are:\vspace{-6pt}
\begin{quotation}\noindent
  GS: 0.274,\ \ \ MHGS: 0.064,\ \ \ UNAM: 0.031,\ \ \ DNAM: 0.011,\ \ \ 
  UDNAM: 0.021,\ \ \ FSS: 0 \\[4pt]
  ZDNAM, ST, DST, UST, UDST, HST, OHST, ZFSS: 0\vspace{-6pt}
\end{quotation}
The maximum conditional probability for an update was never half or more,
and hence the minimum self transition probability is zero, ensuring
that there is an antithetic aspect to the sampling.

\begin{figure}[p]
\begin{center}
\includegraphics[scale=0.65]{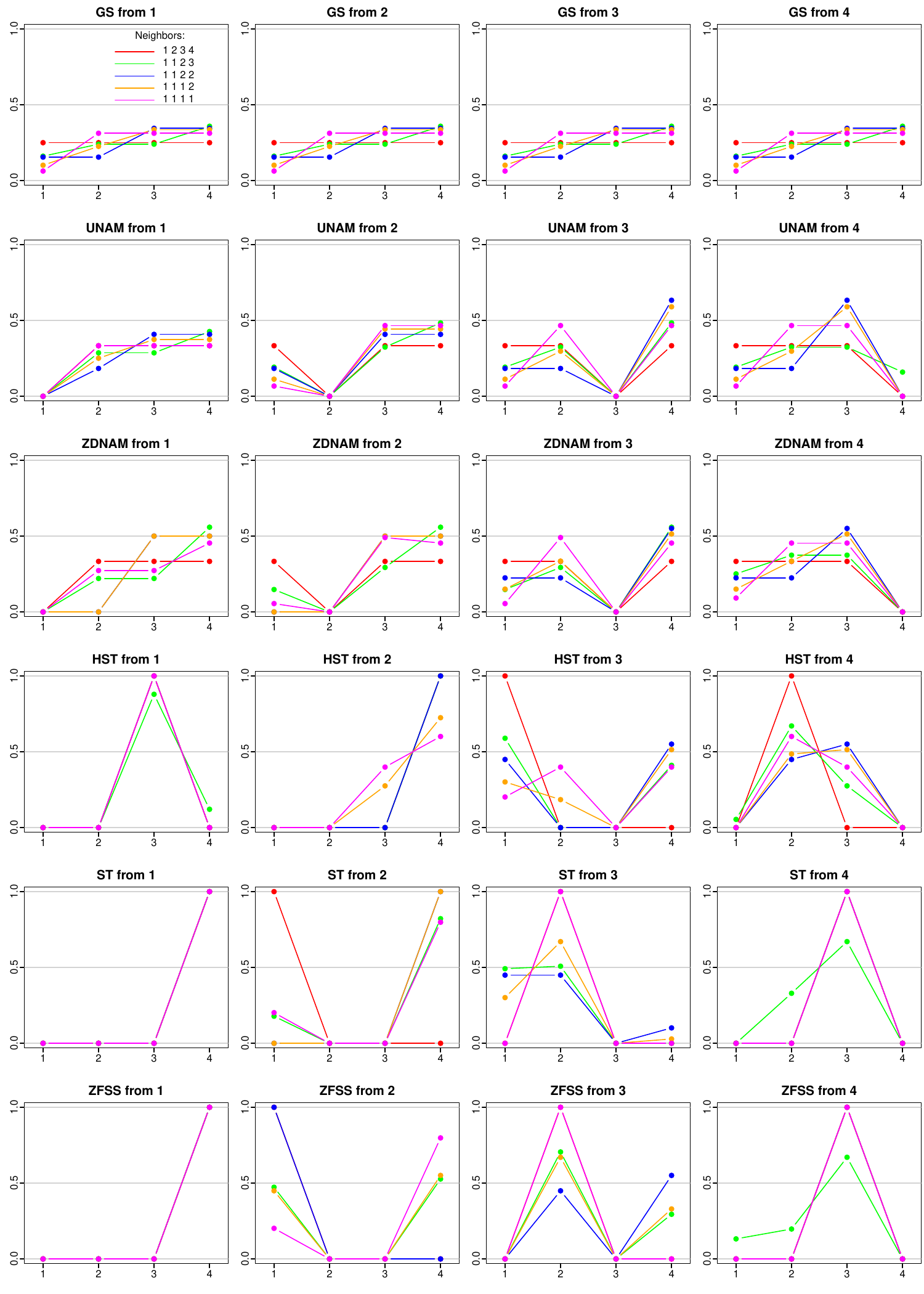}
\vspace*{-12pt}
\end{center}
\caption{Transition probabilities in different contexts for the $5\times5$
Potts model.}\label{fig-5x5-contexts}
\end{figure}

For the $5\times5$ model with negative $b$,
Figure~\ref{fig-5x5-contexts} shows, for each method, how the
transition probabilities vary, depending on the current value of the
variable being updated and on the context of values for its neighbors.
This may be compared to Figure~\ref{fig-8x8-contexts} for the
$8\times8$ model with positive $b$.  One thing to note for the
$5\times5$ model is that ZDNAM, HST, ST, and ZFSS all have zero self
transition probability in all contexts, but ZDNAM differs from the
others in almost always having non-zero transition probabilities to values
other than the current value.  The HST, ST, and ZFSS methods have many
zero transition probabilities, both for the $5\times5$ and $8\times8$
models.  

\begin{figure}[p]
\begin{center}
\includegraphics[scale=1]{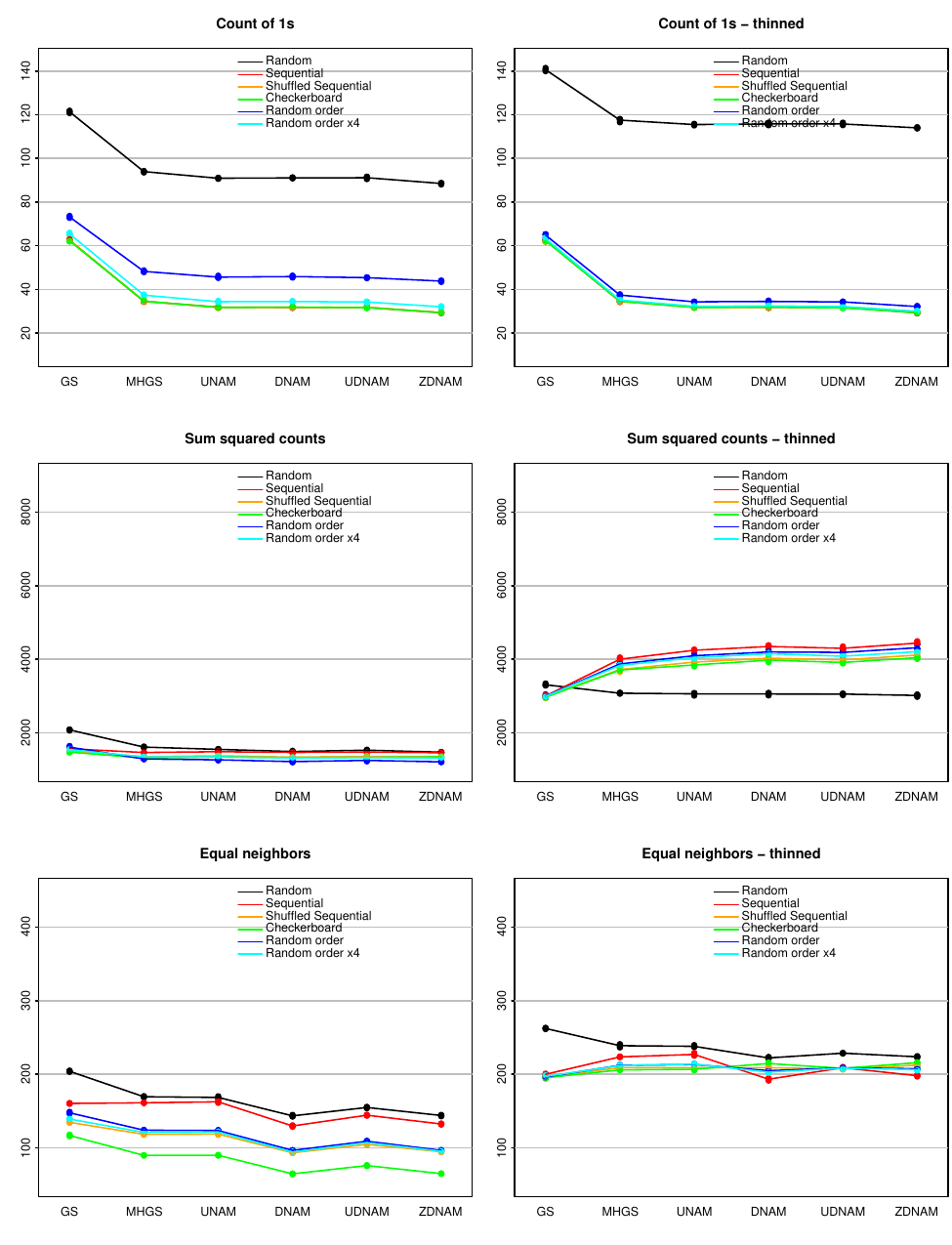}
\end{center}
\caption{Summaries of autocovariance function estimates for the $5\times5$
Potts model, for the first group of methods.}\label{fig-5x5-g1}
\end{figure}

\begin{figure}[p]
\begin{center}
\includegraphics[scale=1]{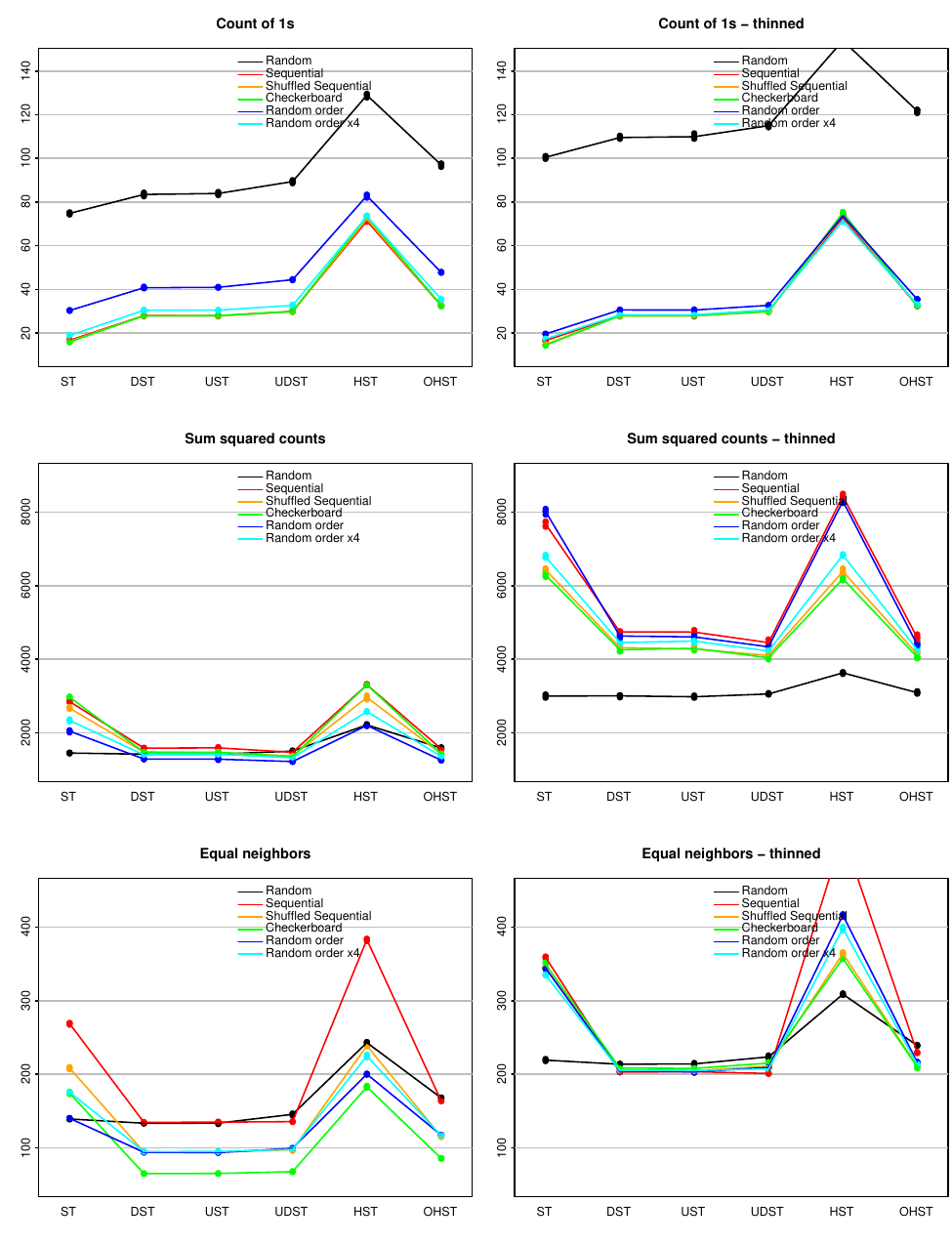}
\end{center}
\caption{Summaries of autocovariance function estimates for the $5\times5$
Potts model, for the second group of methods.}\label{fig-5x5-g2}
\end{figure}

\begin{figure}[p]
\begin{center}
\includegraphics[scale=1]{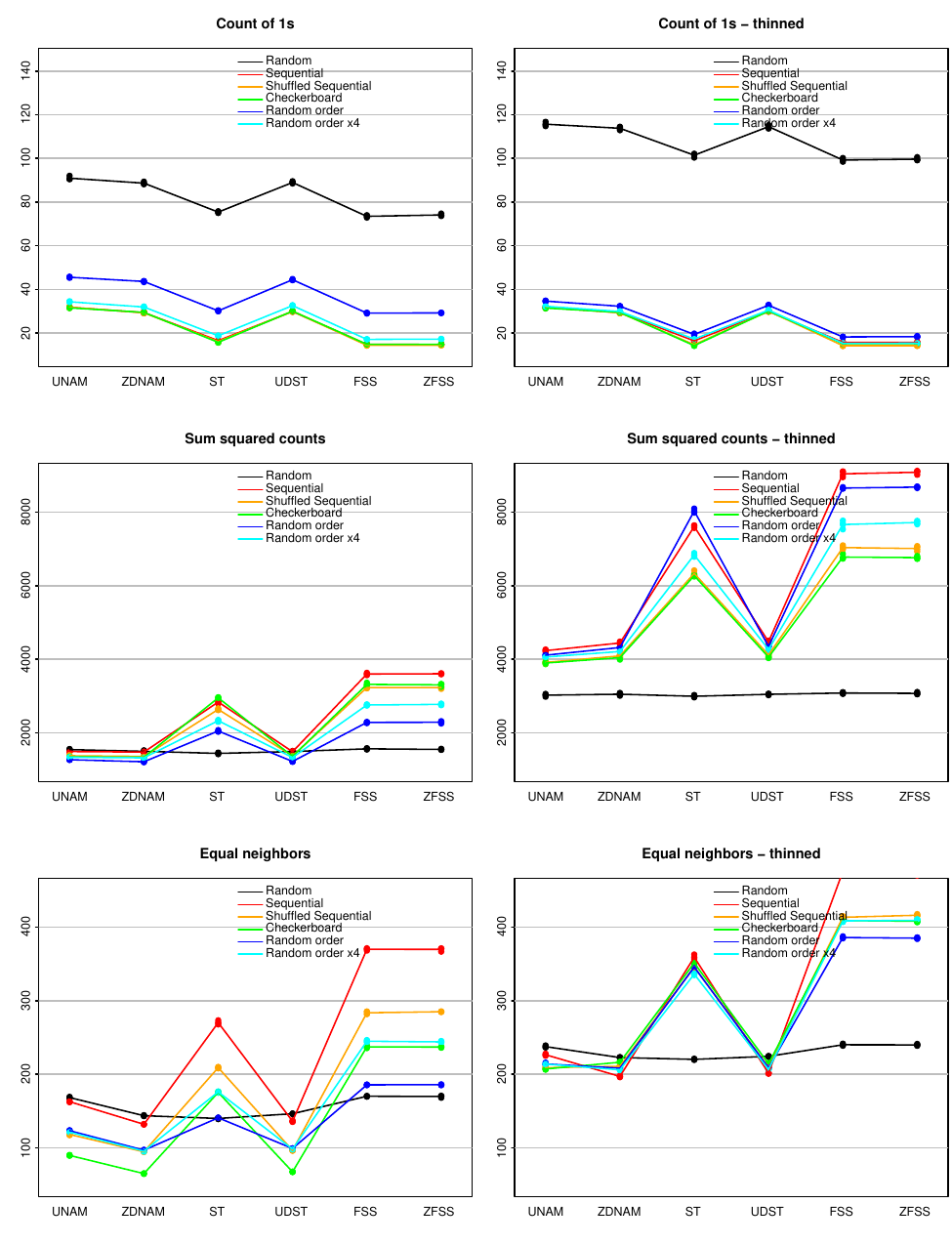}
\end{center}
\caption{Summaries of autocovariance function estimates for the $5\times5$
Potts model, for the third group of methods.}\label{fig-5x5-g3}
\end{figure}

These zero transition probabilities may be responsible for the
somewhat erratic performance of these methods on the $5\times5$ model,
as can be seen in the summaries of asymptotic variance estimates in
Figures~\ref{fig-5x5-g1} through~\ref{fig-5x5-g3}.  (Note that, in
these figures, the dots for the four runs with each method and scan
order are close enough to mostly appear as one dot.)

For the methods deriving from Gibbs sampling, results without
thinning, shown on the left in Figure~\ref{fig-5x5-g1}, are similar to
those for the $8\times8$ Potts model.  GS has the highest asymptotic
variances, followed by MHGS, with asymptotic variances for UNAM
slightly lower than MHGS.  This is as expected by theory for a random
scan.  The DNAM, UDNAM, and ZDNAM methods are somewhat better than
UNAM, with ZDNAM performing best.  One difference from the $8\times8$
model is that the sequential scan is never the best --- the shuffled
sequential scan (which uses a fixed random order rather than a
systematic raster scan) is always better.  For the ``equal neighbors''
function, the checkerboard scan is best of all.

The results with thinning are shown on the right in
Figure~\ref{fig-5x5-g1}.  For the random order scan, thinning reduces
asymptotic variance for the ``count of 1s'' function, a phenomenon
discussed earlier for the $8\times8$ Potts model.  For all other
functions, scans, and methods, thinning increases asymptotic variance.
This is as expected, but for the ``sum squared counts'' function, the
amount of increase varies substantially with scan order, so much so
that the random scan is better than all other scan orders for all
methods except Gibbs sampling, an unusual occurrence for practical
problems.  A similar but less pronounced effect is visible for the
``equal neighbors'' function.

I speculate that combining a scan other than random selection of a
variable with a method other than Gibbs sampling (one having an
antithetic aspect) can induce somewhat periodic movement, which when
sampled only every $n$ iterations can produce inefficient estimates.
One would usually expect this to occur only for fairly easy problems,
such as this one.  For difficult problems, one expects that many scans
will be needed to move to an almost independent state, and
autocovariances for most functions of interest will be strongly
positive.  A modification to Gibbs sampling that introduces antithetic
aspects would then only be expected to somewhat reduce the magnitude
of these autocovariances, not make them negative.  Thinning would then
behave more in the expected way.

The shifted tower methods (Figure~\ref{fig-5x5-g2}) and slice sampling
methods (FSS and ZFSS in Figure~\ref{fig-5x5-g3}) show the same
surprising behaviours.  In addition, the ST, HST, FSS, and ZFSS
methods show large variation in performance.  For the ``count of 1s''
function, the ST, FSS, and ZFSS methods have nearly the same
asymptotic variance for all scan orders, which is lower than that of
all other methods.  However, for the other two functions, these
methods perform very poorly.  The ZDNAM, DST, UST, and UDST methods
show the best overall performance, with OHST behaving similarly, but
with slightly higher asymptotic variance.  

Note that the erratic ST, HST, FSS, and ZFSS methods are the ones that
often have some zero non-self transition probabilities, and that also
use a fixed ordering of values, so these zero transition probabilities
may apply consistently.  In some circumstances, this could be
beneficial, but from these results, it seems it can also have bad
effects.  As discussed in Section~\ref{sec-slice}, ZFFS was
deliberately designed to preserve this order as much as possible, but
in light of these results, it might be interesting to design a slice
sampling method in which the values do not keep the same order.

Pollet, {\em et al.}\ (2004) have also compared Gibbs sampling with
MHGS and UNAM,\footnote{\rule{0pt}{10pt}%
They refer to GS as "heatbath", to MHGS as "MG", and to UNAM as "Opt".
}
for a $4\times4$ Potts model, with random selection of the variable to
be updated, and also report that UNAM performs significantly better
than Gibbs sampling at estimating the expectation of the energy, and
that MHGS is only slightly worse than UNAM.  They did not consider
sequential updates of variables, or functions of state other than the
energy.

Another comparison of methods on the Potts model was done by Suwa
(2022), who compared GS, MHGS, UNAM, ST, HST, and other shifted tower
methods in which the amount shift varies from 0 to $1/2$ (with
corresponding variation in self transition 
probability).\footnote{\rule{0pt}{10pt}%
Suwa refers to GS as ``heatbath'', MHGS as ``Metropolized Gibbs'',
UNAM as ``iterative Metropolized Gibbs'', and ST as the ``Suwa-Todo
algorithm''; other shifted tower methods were characterized by the
shift amount (with $s=1/2$ corresponding to HST).}  Suwa considers
Potts models with $m$ (their $q$) equal to 2, 3, 4, 5, and 6, with the
temperature set to the value corresponding to a phase transition in an
infinite lattice.  For $m=4$, this corresponds to choosing $b=1.098$
in equation~(\ref{eq-potts}).  They used $R=C=32$, so $n=1024$, and
updated variables in a fixed sequential order, which was not
specified, but presumably corresponded to a simple scan across and
down the lattice (corresponding to what is labeled as "Sequential" in
Figures~\ref{fig-8x8-g1} through~\ref{fig-8x8-g3}).  Suwa evaluated
methods by their ``integrated autocorrelation time'', which is
proportional to asympotic variance, of an ``order parameter''. 

Suwa's results show that MHGS is substantially better than GS, and
that UNAM is only slightly better than MHGS, in agreement with
the results of Pollet (2004), and the results reported here for 
the $8\times8$ Potts model.  Suwa also shows a substantial advantage
of ST over UNAM, again in agreement with results here.

A larger claim by Suwa is that the autocorrelation time is an
exponential function of the frequency of non-self transitions ---
equivalently, that the log of the autocorrelation time (or asymptotic
variance) is a linear function of the frequency of non-self
transitions, as pictured in Fig.\ 2 of (Suwa 2022).  This figure shows
results for shifted tower methods in which the amount of shift is
varied from just above 0 to 1/2 (with the latter value corresponding
to HST), with a consequent variation in self transition probability
from just below 1 to the minimum possible. The results obtained are
fit reasonably well by a linear relationship of log autocorrelation
time to non-self transition probability.  Furthermore, the results
with GS, MHGS, UNAM, and ST (with shift not constant, but equal to the
maximum probability) are also close to this line.

This claim seems misleading, however.  First, note that the non-self
transition probability is upper bounded by a value no greater than
one, so an exponential improvement as it increases does not permit
arbitrarily large improvements in autocorrelation time.  Second, the
autocorrelation time must go to infinity as the non-self transition
probability goes to zero, so the exponential relationship cannot hold
in this limit.  One may question whether the experimental results with
the smallest non-self transition probabilities are accurate,
considering the difficulty of estimating autocorrelation times when
they are very large.  The alternative of autocorrelation time being
proportional to some power of the non-self transition probability is
almost indistinguishable from an exponential relationship over the
range of non-self transition probabilities for which the fit of the
latter is good in Suwa's Fig.\ 2, which is from 0.23 to 0.29 for
$q=4$.

Suwa also compares a sequential scan with a random scan, with results
shown in Fig.\ 4 of (Suwa 2022), which appears to be for $q=4$ (though
this is not stated).  For the ST method, the sequential scan is a
factor of about 3.5 more efficient than a random scan, similar to,
though a bit greater, than the advantage seen here in
Figure~\ref{fig-8x8-g2}.  These results are seen by Suwa as following
a relationship in which the autocorrelation time with a random scan is
proportional to a power of the non-self transition probability.  While
this is more plausible than an exponential relationship for small
non-self transition probabilities, I think that further research is
needed to elucidate these relationships.  The results for the
$5\times5$ Potts model here show that methods with the same self
transition probability (including those that minimize it) can have
substantially different efficiencies (for example, ST, UDST, and HST
in Figure~\ref{fig-5x5-g2}).

\section{\hspace*{-8pt}
  Comparisons for a Bayesian mixture model}\label{sec-mix}\vspace{-11pt}

Mixture models are commonly used for data that comes from several
distinct sources, for example, data on symptoms of patients suffering
from different diseases.  In a Bayesian modeling approach (Neal 1992a), the
parameters of the mixture model are integrated over, with respect to a
prior distribution, leaving as the only unknowns which component of
the mixture is associated with each data point (e.g., which disease
each patient has).  Sampling for these component indicators can be
done by Gibbs sampling, which can be modified to avoid self
transitions by the methods discussed in this paper.

A mixture model for independent observations $y_1,\ldots,y_n$ represents their
distribution as a mixture of $m$ component distributions, as 
follows:\vspace{-10pt}
\beq
  P(y_i|\alpha,\theta) & = & \sum_{x_i=1}^m \alpha_{x_i} P(y_i|x_i,\theta_{x_i})
\eeq
Here, $x_i$ indicates which mixture component is associated with observation
$y_i$, $\alpha=[\alpha_1,\ldots,\alpha_m]$ is a vector of mixture weights,
with $\sum_x \alpha_x = 1$, and $\theta_x$ gives the parameters of
mixture component $x$.  For the model used in the experiments here, 
each observation consists
of $H$ binary variables, $y_i=[y_{i1},\ldots,y_{iH}]$ with
$y_{ih}\in\{0,1\}$, and $\theta_x$ contains the probabilities for each of
these binary variables to have the value $1$, so
$\theta_x=[\theta_{x1},\ldots,\theta_{xH}]$ with $\theta_{xh}\in[0,1]$.
Conditional on observation $i$ coming from mixture component $x_i$, the
$H$ binary variables are assumed to be independent, so\vspace{-1pt}
\beq
P(y_i|x_i,\theta) & = & 
 \prod_{h=1}^H \theta_{x_ih}^{y_{ih}}\, (1\!-\!\theta_{x_ih}^{})^{1-y_{ih}}
\eeq\vspace{-8 pt}

The joint probability of all observations, $y_i$, along with all component
indicators, $x_i$, for given values of the model
parameters $\alpha$ and $\theta$, is therefore\vspace{-4pt}
\beq
  P(y_1,\ldots,y_n,x_1,\ldots,x_n|\alpha,\theta) 
  & = & \prod_{i=1}^n\ \alpha_{x_i}
    \prod_{h=1}^H \theta_{x_ih}^{y_{ih}}\, (1\!-\!\theta_{x_ih}^{})^{1-y_{ih}} 
  \\[5pt]
  & = & \left[\prod_{x=1}^m \alpha_x^{C_x}\right]\,
        \left[ \prod_{x=1}^m \prod_{h=1}^H
           \theta_{xh}^{S_{xh}}\, (1-\theta_{xh})^{C_x-S_{xh}} \right]
\eeq
where $C_x$ is the number of $x_i$ for $i=1,\ldots,n$ that are equal to $x$, 
and $S_{xh}$ is the sum of $y_{ih}$ for those $i$ for which $x_i$ equals $x$.

In a Bayesian treatment of this problem, a prior distribution for the
unknown parameters $\alpha$ and $\theta$ is specified.  If in this
prior the $\theta_{xh}$ and $\alpha$ parameters are
independent, with each $\theta_{xh}$ uniform over $(0,1)$ and $\alpha$
uniform over the simplex with $\alpha_x>0$ and $\sum_x \alpha_x=1$, it
is possible to analytically integrate over the prior for these
parameters (Neal 1992a), giving a joint distribution for the
observations and component indicators alone:\vspace{-2pt}
\beq 
  P(y_1,\ldots,y_n,x_1,\ldots,x_n) & = & 
  \left[ {(m\!-\!1)! \over (n\!+\!m\!-\!1)!}\, \prod_{x=1}^m C_x!\, \right]
  \left[ \prod_{x=1}^m \prod_{h=1}^H 
    {S_{xh}!\,(C_x-S_{xh})! \over (C_x+1)!} \right]
\eeq\label{eq-mix-joint}\vspace{-7pt}

When we have observed $y_1,\ldots,y_n$, we may wish to sample from the
conditional distribution of the component indicators $x_1,\ldots,x_n$
given these observations, both because this distribution is of
interest in itself (giving possible ``clusterings'' of the
observations), and because it assists inference for the parameters and
predictions for future observations.  This can be done using Gibbs sampling.
The conditional distribution for $x_i$ given $x_{-i}$ can be obtained from
equation~(\ref{eq-mix-joint}), as\vspace{-4pt}
\beq
 P(x_i=x|y_1,\ldots,y_n,x_1,\ldots,x_{i-1},x_{i+1},\ldots,x_n) & \propto & 
  (C^-_x+1)\,\prod_{h=1}^H 
  {(S^-_{xh}+1)^{y_{ih}} (C^-_x-S^-_{xh}+1)^{1-y_{ih}} \over C^-_x+\,2}
  \ \ \ \ \ \ \ \
\eeq
where $C^-_x=C_x-I(x_i=x)$ is the number of $x_j$ for $j \ne i$ that 
are equal to $x$, and $S^-_{xh}=S_{xh}-y_{ih}I(x_i=x)$ is the sum 
of $y_{jh}$ for those $j \ne i$ for which $x_j$ equals $x$.  

The experiments in this section compare use of Gibbs sampling with the
modified Gibbs sampling methods, on a problem in which there are
$n=30$ observations, each consisting of $H=10$ binary variables.  The
model used has $m=9$ mixture components.  The data, shown in
Figure~\ref{fig-mix-data}, was manually constructed to have five
clusters of observations, which would be expected to correspond to
mixture components, so we anticipate that several of the mixture
components will be associated with few or no observations in typical
samples from the posterior distribution

\begin{figure}\small

\noindent
\hspace{30pt}\parbox[t]{1.75in}{\tt
\mbox{~}1: 1 1 1 1 0 0 0 0 1 0 \\
\mbox{~}2: 1 1 1 1 0 0 0 0 0 0 \\
\mbox{~}3: 1 1 1 1 0 0 0 0 1 0 \\
\mbox{~}4: 1 0 1 1 0 0 0 0 1 0 \\
\mbox{~}5: 1 1 1 1 0 0 0 0 0 1 \\
\mbox{~}6: 1 1 1 1 0 0 1 0 1 1 \\
\mbox{~}7: 0 1 1 1 0 0 0 0 0 0 \\
}\hspace{30pt}%
\parbox[t]{1.75in}{\tt
\mbox{~}8: 0 0 0 0 1 1 1 1 1 0 \\
\mbox{~}9: 0 0 0 0 1 1 1 1 1 0 \\
10: 0 0 0 0 1 1 1 1 1 1 \\
11: 0 0 0 1 1 1 1 1 0 0 \\
12: 0 0 0 0 0 1 1 1 1 1 \\
13: 0 0 1 0 1 1 1 0 1 0 \\
}\hspace{30pt}%
\parbox[t]{1.75in}{\tt
14: 1 0 1 1 0 0 1 1 0 1 \\
15: 0 0 1 1 0 0 1 1 1 1 \\
16: 0 0 1 1 0 0 1 1 1 0 \\
17: 0 0 1 1 0 1 1 1 1 0 \\
18: 0 0 1 1 0 0 1 1 0 0 \\
19: 0 0 1 1 0 0 1 1 0 1 \\
}

\noindent
\hspace{30pt}\parbox[t]{1.75in}{\tt
20: 1 1 0 0 1 1 0 0 0 0 \\
21: 1 1 0 0 1 1 0 0 1 1 \\
22: 1 1 0 0 1 1 0 0 1 0 \\
23: 1 1 0 0 1 1 0 0 0 1 \\
24: 1 1 1 0 1 1 0 0 1 1 \\
25: 1 1 0 0 1 1 0 0 1 0 \\
}\hspace{30pt}%
\parbox[t]{1.75in}{\tt
26: 1 0 0 0 1 0 0 0 0 0 \\
27: 0 0 0 0 0 1 0 0 0 1 \\
28: 0 0 0 1 0 0 0 0 0 0 \\
29: 0 1 0 0 0 0 0 0 1 0 \\
30: 0 0 0 0 0 0 1 0 0 0
}\vspace{-2pt}

\caption{The $n=30$ observations used for the mixture model example. The 
observations are here grouped by the five manually-created clusters.
The order is randomized in the runs done, so this 
``true'' clustering does not affect the results.}\label{fig-mix-data}

\end{figure}

\pagebreak

The expectations of the following functions of state were 
estimated:\vspace{-6pt}
\begin{itemize}
\item[1)] \textbf{Obs 1 in cluster 1.} The indicator function for
          whether $x_1=1$.  Since the mixture components (clusters)
          are treated symmetrically in the model, the true expectation of
          this function must be $1/m = 1/9 = 0.111111$, and its variance must
          be $(1/m)(1\!-\!1/m) = 8/81 = 0.098765$.
\item[2)] \textbf{Obs 10 cluster size.} The number of observations 
          in the cluster associated with observation 10 --- that is,
          $\sum_{i=1}^{30} I(x_i=x_{10})$.  The expectation of this
          function is approximately 5.56 and its variance is approximately
          3.26.
\item[3)] \textbf{Obs 30 cluster size.} The number of observations 
          in the cluster associated with observation 30.  The expectation
          of this function is approximately 4.35 and its variance is
          approximately 6.38.\vspace{-6pt}
\end{itemize}

Each run consisted of $K=200000$ scans, each with $n=30$ updates
component indicators.  For each group of methods, four independent
runs of this length were done, for each method in the group, and each
scan order.  

The frequencies of self transitions for the various methods 
are as follows:\vspace{-7pt}
\begin{quotation}\noindent
  GS: 0.69,\ \ \ MHGS: 0.65,\ \ \ UNAM: 0.64,\ \ \ DNAM: 0.61,\ \ \ 
  UDNAM: 0.62,\ \ \ FSS: 0.61 \\[4pt]
  ZDNAM, ST, DST, UST, UDST, HST, OHST, ZFSS: 0.61\vspace{-7pt}
\end{quotation}
The maximum conditional probability for an update was half or more
86\% of the time.

\begin{figure}[p]
\begin{center}
\includegraphics[scale=1]{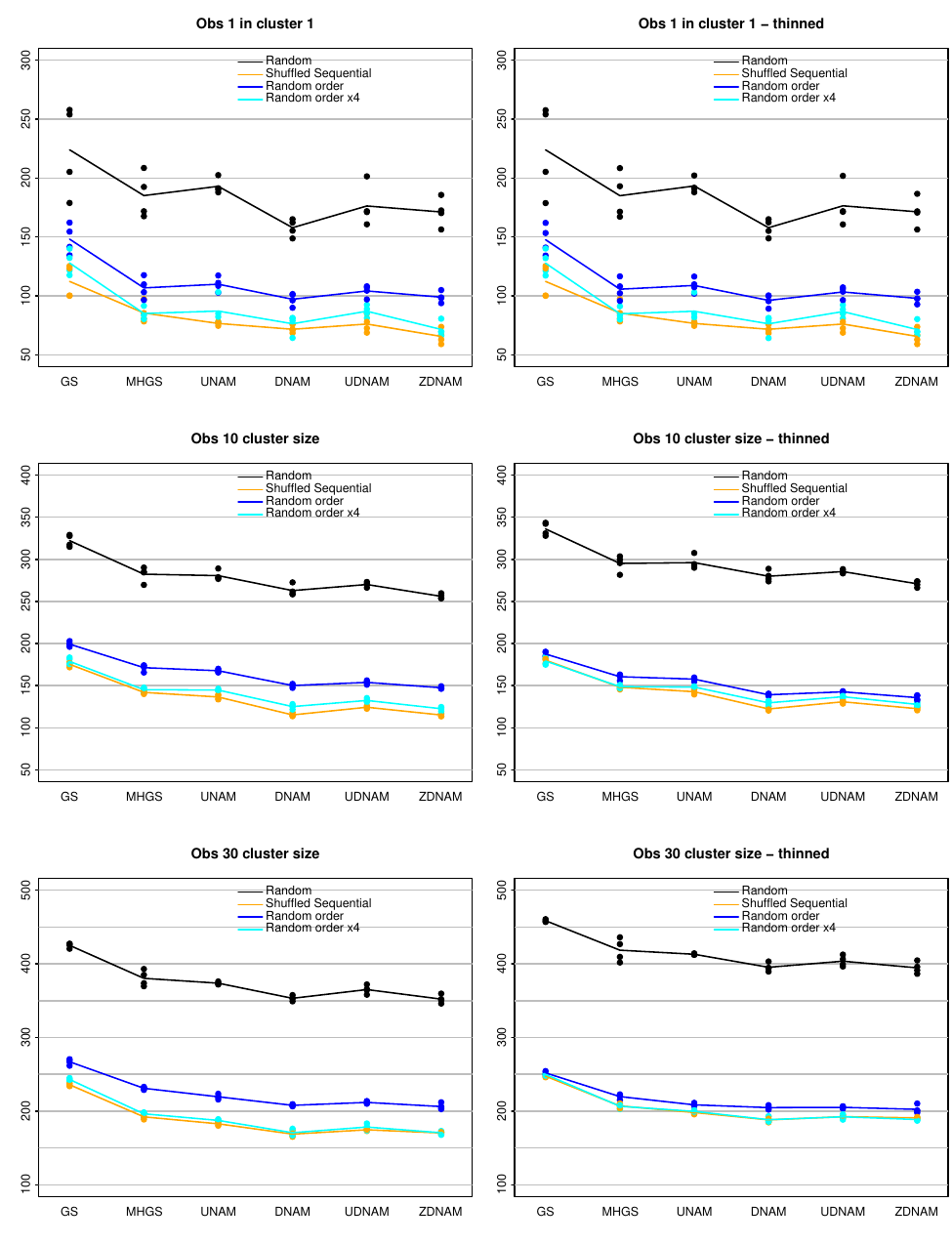}
\end{center}
\caption{Summaries of autocovariance function estimates for the Bayesian
mixture model, for the first group of methods.}\label{fig-mix-g1}
\end{figure}

\begin{figure}[p]
\begin{center}
\includegraphics[scale=1]{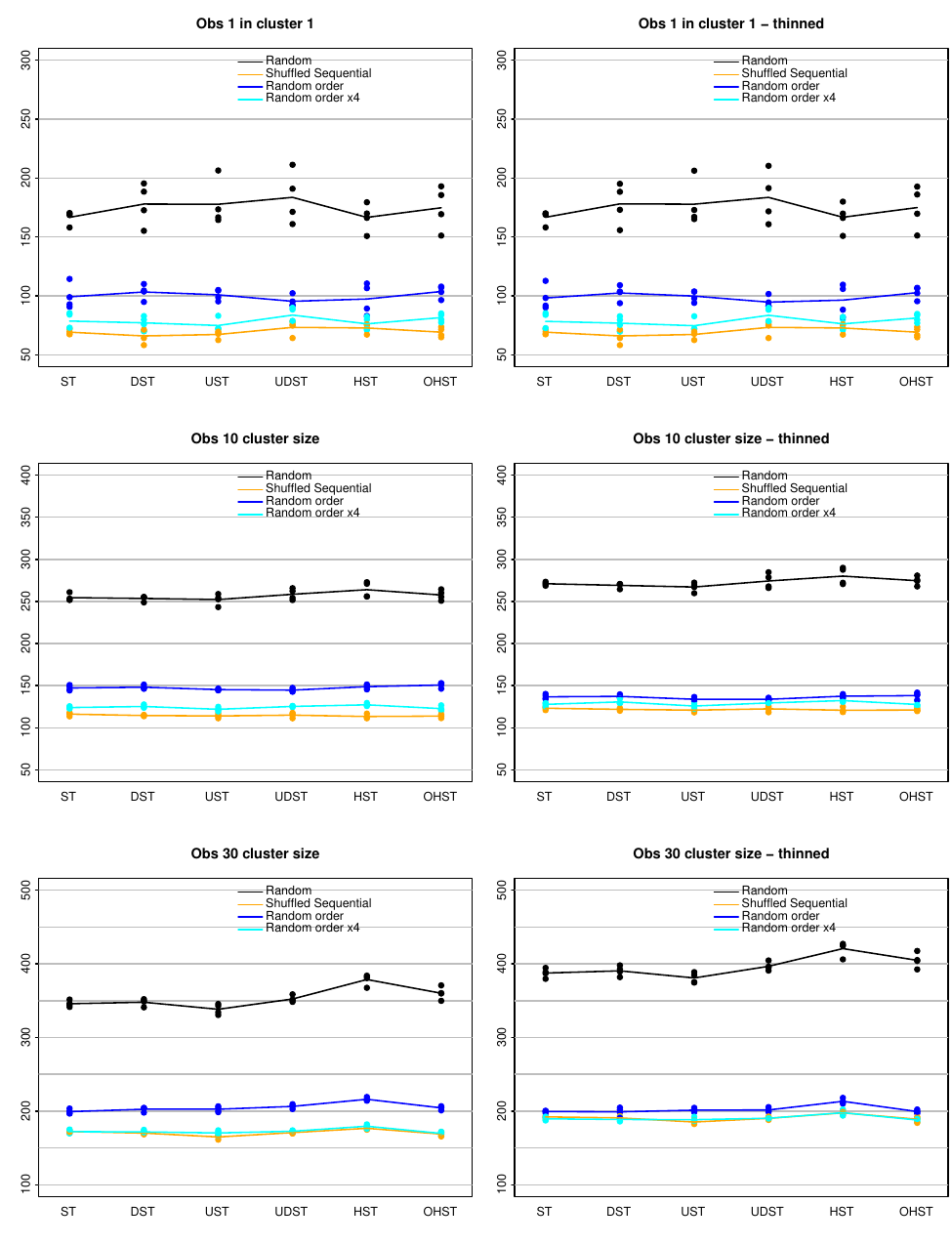}
\end{center}
\caption{Summaries of autocovariance function estimates for the Bayesian
mixture model, for the second group of methods.}\label{fig-mix-g2}
\end{figure}

\begin{figure}[p]
\begin{center}
\includegraphics[scale=1]{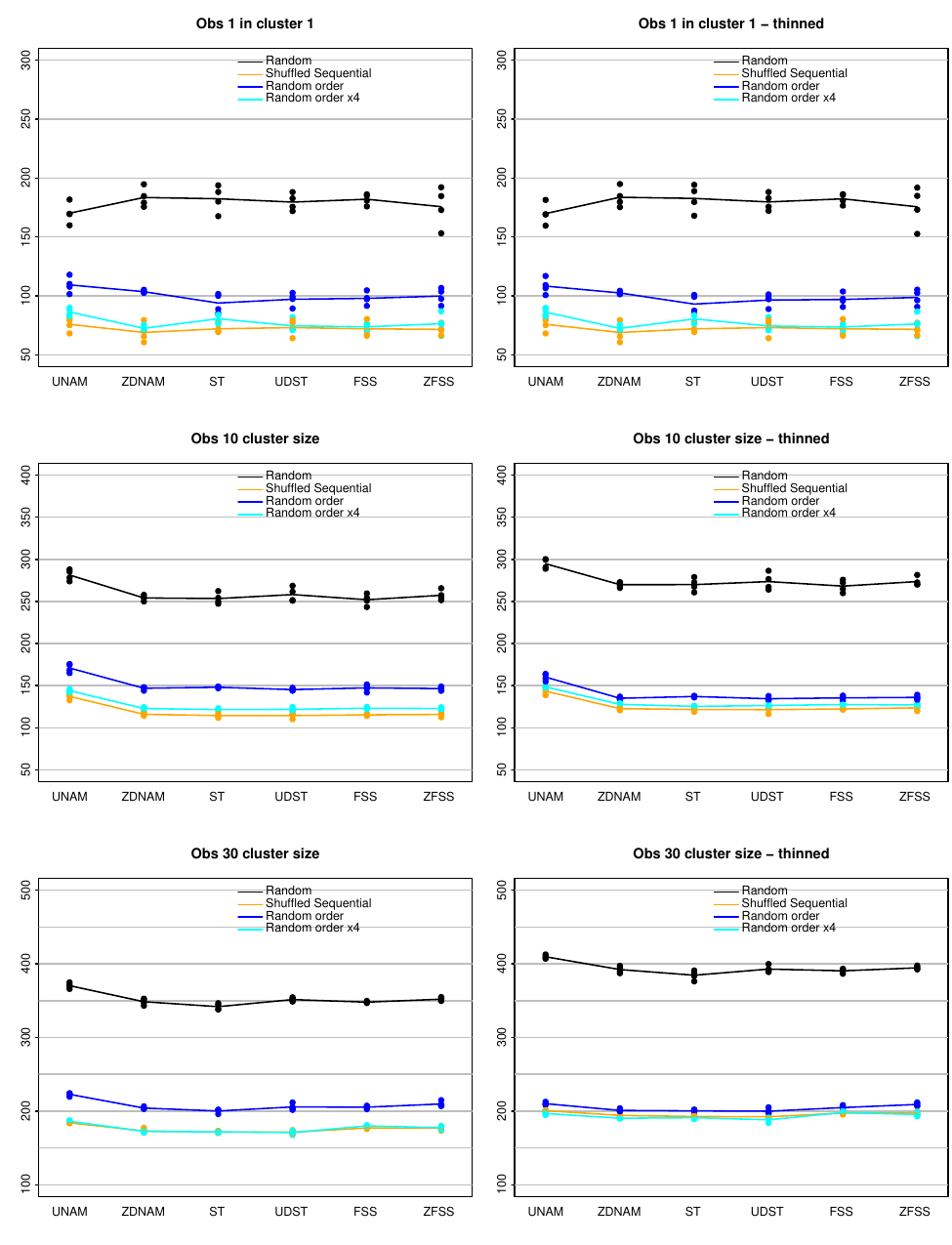}
\end{center}
\caption{Summaries of autocovariance function estimates for the Bayesian
mixture model, for the third group of methods.}\label{fig-mix-g3}
\end{figure}

Summaries of asymptotic variance estimates for the three function
above, for all groups of methods, are shown in
Figures~\ref{fig-mix-g1} through~\ref{fig-mix-g3}.  Note that there is
no meaningful original order for the variables, so there is no
``sequential'' scan order.

The results for the mixture model problem are qualitatively
similar to those for the $8\times8$ Potts model.  The shuffled
sequential scan order gives the best results.  Thinning increases
asymptotic variance, except for the random scan, for which thinning is
beneficial. Amongst the methods deriving from Gibbs sampling
(Figure~\ref{fig-mix-g1}), ZDNAM performs best.  The shifted tower
methods (Figure~\ref{fig-mix-g2}) all perform about equally well,
except that HST may be slightly worse than the others.  FSS and ZFSS
(see Figure~\ref{fig-mix-g3}) also perform well.

\section{\hspace*{-8pt}
  Comparisons for a belief network}\label{sec-beliefnet}\vspace{-11pt}

A joint distribution for random variables $x_1,\ldots,x_n$ can be
written as a product of successive conditional distributions:\vspace{-8pt}
\beq
  \pi(x) & = & 
    \pi(x_1)\,\pi(x_2|x_1)\,\pi(x_3|x_1,x_2)\cdots\pi(x_n|x_1,\ldots,x_{n-1})
\eeq
A belief network (sometimes called a ``Bayesian network'' or
``directed graphical model'') is a directed graph with arrows that go
from some $x_i$ to some $x_j$ with $j>i$, which summarizes how the
representation above can be simplified by omitting some conditioning
variables --- in the factor $\pi(x_j|\ldots)$, the only $x_i$ that
need be conditioned on are those (the ``parents'' of $x_j$) for
which there is an arrow from $x_i$ to $x_j$ in the network.

Consider, for example, the network below:\vspace{-3pt}

\setlength{\unitlength}{2.45in}
\begin{picture}(3,0.75)
 \put(1.5,0.7){\circle{0.115}}
 \put(1.455,0.645){\makebox(0.1,0.1){$x_1$}}
 \put(1.45,0.65){\vector(-1,-1){0.1}}
 \put(1.55,0.65){\vector(1,-1){0.1}}
 \put(1.3,0.5){\circle{0.115}}
 \put(1.255,0.445){\makebox(0.1,0.1){$x_2$}}
 \put(1.35,0.45){\vector(1,-1){0.1}}
 \put(1.7,0.5){\circle{0.115}}
 \put(1.655,0.445){\makebox(0.1,0.1){$x_3$}}
 \put(1.65,0.45){\vector(-1,-1){0.1}}
 \put(1.5,0.3){\circle{0.115}}
 \put(1.455,0.245){\makebox(0.1,0.1){$x_4$}}
\end{picture}\vspace{-0.6in}

\noindent
The absence of arrows from $x_1$ to $x_4$ and from $x_2$ to $x_3$
means that the joint distribution can be written as
\beq
  \pi(x) & = & \pi(x_1)\,\pi(x_2|x_1)\,\pi(x_3|x_1)\,\pi(x_4|x_2,x_3)
  \\[-24pt] \nonumber
\eeq

Many common statistical models (e.g., state space time series models)
can be seen as belief networks.  They have also been used to 
represent knowledge elicited from experts (Pearl 1988;
Lauritzen and Spiegelhalter 1988), and as models in the style of
neural networks that can be learned from data (Neal 1992b).

If some variables in a belief network are known, the conditional
distribution of the other variables given these known variables can be
sampled from using Gibbs sampling.  The unknown variables are updated
in some systematic or random order.  An update to variable $x_i$ is
done by sampling a new value from its conditional distribution given
current values of other variables, including both ones with known
values (which are fixed) and the current values of other unknown
variables.  

The conditional distribution for $x_i$ needed for Gibbs sampling
depends only on the parents of $x_i$, the children of $x_i$, and the
parents of the children of $x_i$, with the conditional probabilities
being proportional to the product of factors of the joint distribution
that involve $x_i$.  For the example network above,
\beq
  \pi(x_1|x_{-1}) & \propto & \pi(x_1)\,\pi(x_2|x_1)\,\pi(x_3|x_1) \\[3pt]
  \pi(x_2|x_{-2}) & \propto & \pi(x_2|x_1)\,\pi(x_4|x_2,x_3) \\[3pt]
  \pi(x_3|x_{-3}) & \propto & \pi(x_3|x_1)\,\pi(x_4|x_2,x_3) \\[3pt]
  \pi(x_4|x_{-4}) & \propto & \pi(x_4|x_2,x_3)
\eeq

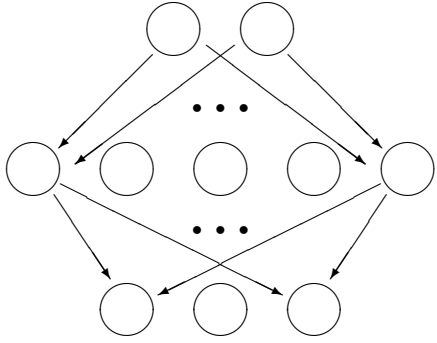
\begin{figure}[t]

\begin{picture}(3,0.75)

\put(1.4,0.7){\circle{0.115}}
\put(1.6,0.7){\circle{0.115}}

\put(1.355,0.645){\vector(-1,-1){0.2}}
\put(1.645,0.645){\vector(1,-1){0.2}}
\put(1.53,0.665){\vector(-4,-3){0.34}}
\put(1.47,0.665){\vector(4,-3){0.34}}
\put(1.45,0.53){\circle*{0.015}}
\put(1.5,0.53){\circle*{0.015}}
\put(1.55,0.53){\circle*{0.015}}

\put(1.1,0.4){\circle{0.115}}
\put(1.3,0.4){\circle{0.115}}
\put(1.5,0.4){\circle{0.115}}
\put(1.7,0.4){\circle{0.115}}
\put(1.9,0.4){\circle{0.115}}

\put(1.145,0.345){\vector(2,-3){0.12}}
\put(1.855,0.345){\vector(-2,-3){0.12}}
\put(1.158,0.368){\vector(2,-1){0.475}}
\put(1.842,0.368){\vector(-2,-1){0.475}}
\put(1.45,0.27){\circle*{0.015}}
\put(1.5,0.27){\circle*{0.015}}
\put(1.55,0.27){\circle*{0.015}}

\put(1.3,0.1){\circle{0.115}}
\put(1.5,0.1){\circle{0.115}}
\put(1.7,0.1){\circle{0.115}}

\end{picture}\vspace{-0.1in}

\caption{The belief network used for the experiments. The variables
represented by the circles at the top have possible values in
$\{1,2,3,4,5\}$, those in the middle have possible values $\{1,2,3,4\}$, and
those at the bottom have possible values $\{1,2,3\}$.}\label{bnet}

\end{figure}

The belief network used for the experiments reported here is shown in
Figure~\ref{bnet}.  The 10 variables in the model, represented as
circles in the network, are arranged in three layers --- a layer at
the top of two variables (each with five possible values), a layer of
five variables in the middle (each with four possible values), and a
layer of three variables at the bottom (each with three possible
values). Arrows go from every variable in the top layer to every
variable in the middle layer, and from every variable in the middle
layer to every variable in the bottom layer.

The marginal distributions for the two variables at the top (which
have no parents) are determined by parameters $\alpha_{i,u}$, where
$i\in\{1,2\}$ identifies the variable and $u \in \{1,2,3,4,5\}$ is a
possible value for the variable.  The probability for variable $i$ in
this top group having value $u$ is $\exp(\alpha_{i,u})\,/\,\sum_{u'}
\exp(\alpha_{i,u'})$. (Note that this and other parameterizations for
this model are redundant, with multiple parameter values producing the
same distribution.)

The conditional distribution for a middle variable given values for its
parent variables is defined using a multinomial logit model (also
known as a ``softmax'' model).  For each possible value, $v$, of
variable $j$ in the middle layer, a summed input, $s_{j,v} = \sum_i
\exp(\beta_{ij,x_iv})$ is computed, where $\beta_{ij,uv}$ are
parameters giving the influence of variable $i$ having value $u$ on
variable $j$ having value $v$.  The probability that variable $j$ has
value $v$ is then $\exp(s_{j,v})\,/\,\sum_{v'} \exp(s_{j,v'})$.  In
similar fashion, parameters $\gamma_{jk,vw}$ define multinomial logit
models for the values of variables in the bottom layer, given values
for variables in the middle layer.

For the experiments, a single set of parameters, $\alpha_{i,u}$,
$\beta_{ij,uv}$, and $\gamma_{jk,vw}$, were randomly sampled,
independently, from the $t$ distribution with four degrees of freedom.
Gibbs sampling and the other methods were then used to sample from the
distribution for all $n=2+5+3=10$ variables.  This is not the typical
usage --- one would usually condition on known values for some of the
variables --- and if one did want to sample from this distribution, it
can be done more easily by sampling top down from the conditional
distribution of each node given its parents.  However, it is a useful
test of sampling methods, since moving around the whole unconstrained
distribution should be more challenging.

Estimates of the expectations of the following functions of state were
found:\vspace{-6pt}
\begin{itemize}
\item[1)] \textbf{Unit 1 of layer 1 is 1.}  The indicator that the
first variable in the middle layer (1) of variables has the value 1.
The expectation of this function is 0.2109 (and consequently
its variance is 0.1664).
\item[2)] \textbf{Unit 1 of layer 2 is 1.}  The indicator that the
first variable in the bottom layer (2) of variables has the value 1.
The expectation of this function is 0.07353 (and its 
variance is 0.06812).
\item[1)] \textbf{Unit 1 layer 0 and unit 1 layer 2 is 1.}  The indicator
that the logical ``and'' of the first variable of the top layer (0) and
the first variable of the bottom layer (2) is 1 (i.e., that both
variables are 1). The expectation of this function is
0.04950 (and its variance is 0.04705).\vspace{-6pt}
\end{itemize}
The expectations above were computed by brute-force marginalization
over all possible combinations of values for the ten variables.

Four runs with $K=1000000$ scans, each with $n=10$ variable updates,
were done for each scan order and each method within each group of
methods. 

The frequencies of self transitions for the various methods are:\vspace{-6pt}
\begin{quotation}\noindent
  GS: 0.68,\ \ \ MHGS: 0.59,\ \ \ UNAM: 0.58,\ \ \ DNAM: 0.56,\ \ \ 
  UDNAM: 0.57,\ \ \ FSS: 0.56 \\[4pt]
  ZDNAM, ST, DST, UST, UDST, HST, OHST, ZFSS: 0.56\vspace{-6pt}
\end{quotation}
The maximum conditional probability for an update was half or more
89\% of the time.

\begin{figure}[p]
\begin{center}
\includegraphics[scale=1]{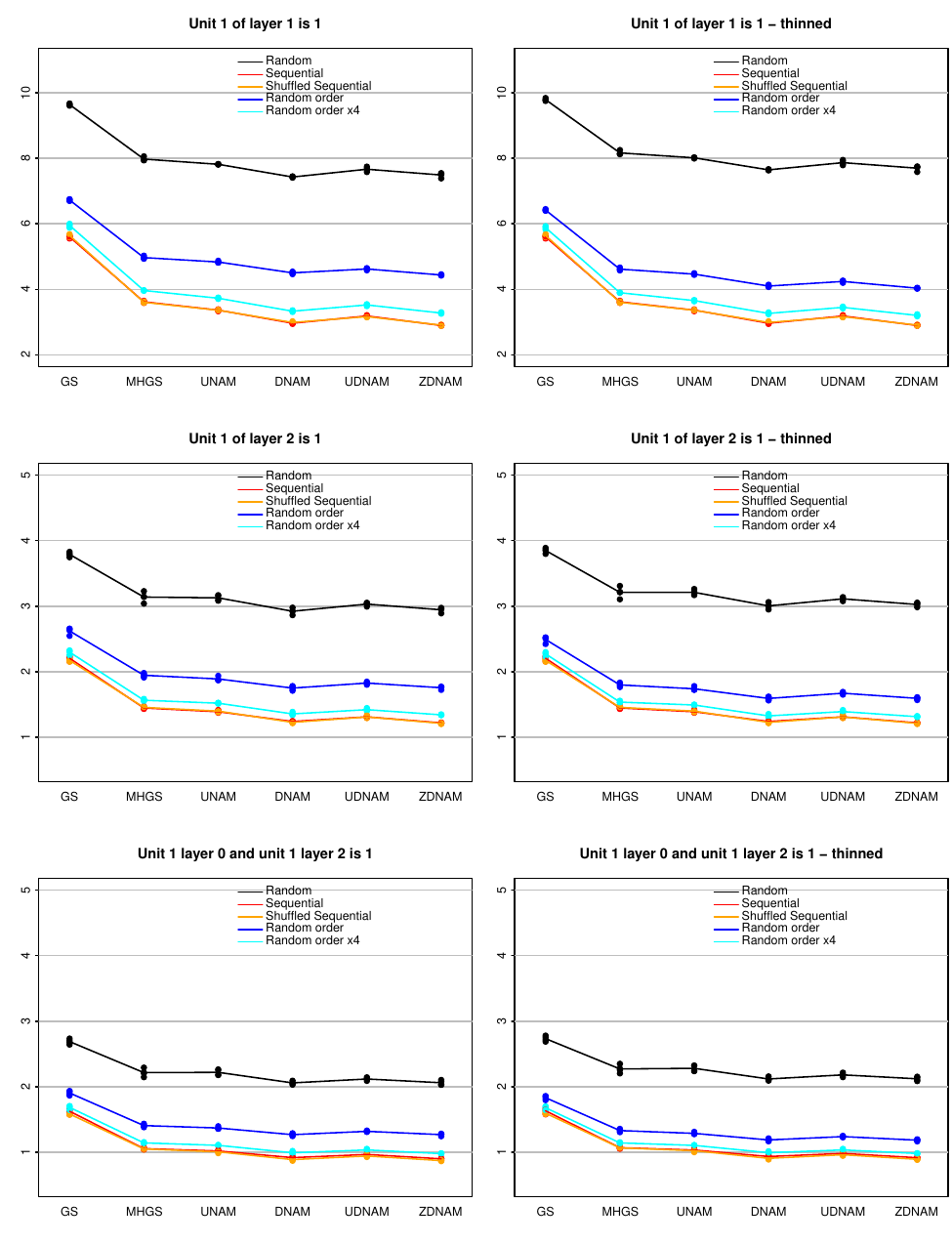}
\end{center}
\caption{Summaries of autocovariance function estimates for the belief
network, for the first group of methods.}\label{fig-bn-g1}
\end{figure}

\begin{figure}[p]
\begin{center}
\includegraphics[scale=1]{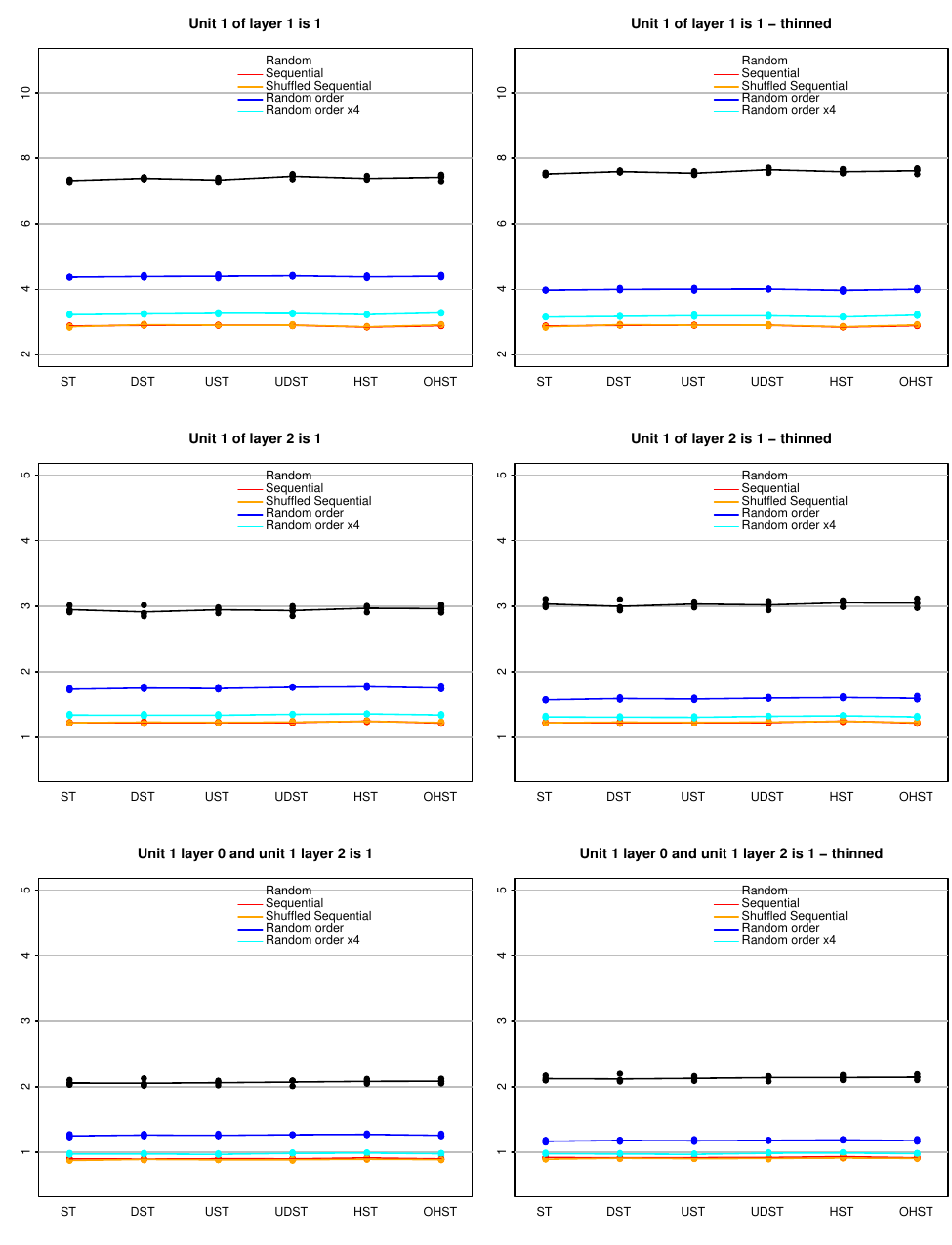}
\end{center}
\caption{Summaries of autocovariance function estimates for the belief
network, for the second group of methods.}\label{fig-bn-g2}
\end{figure}

\begin{figure}[p]
\begin{center}
\includegraphics[scale=1]{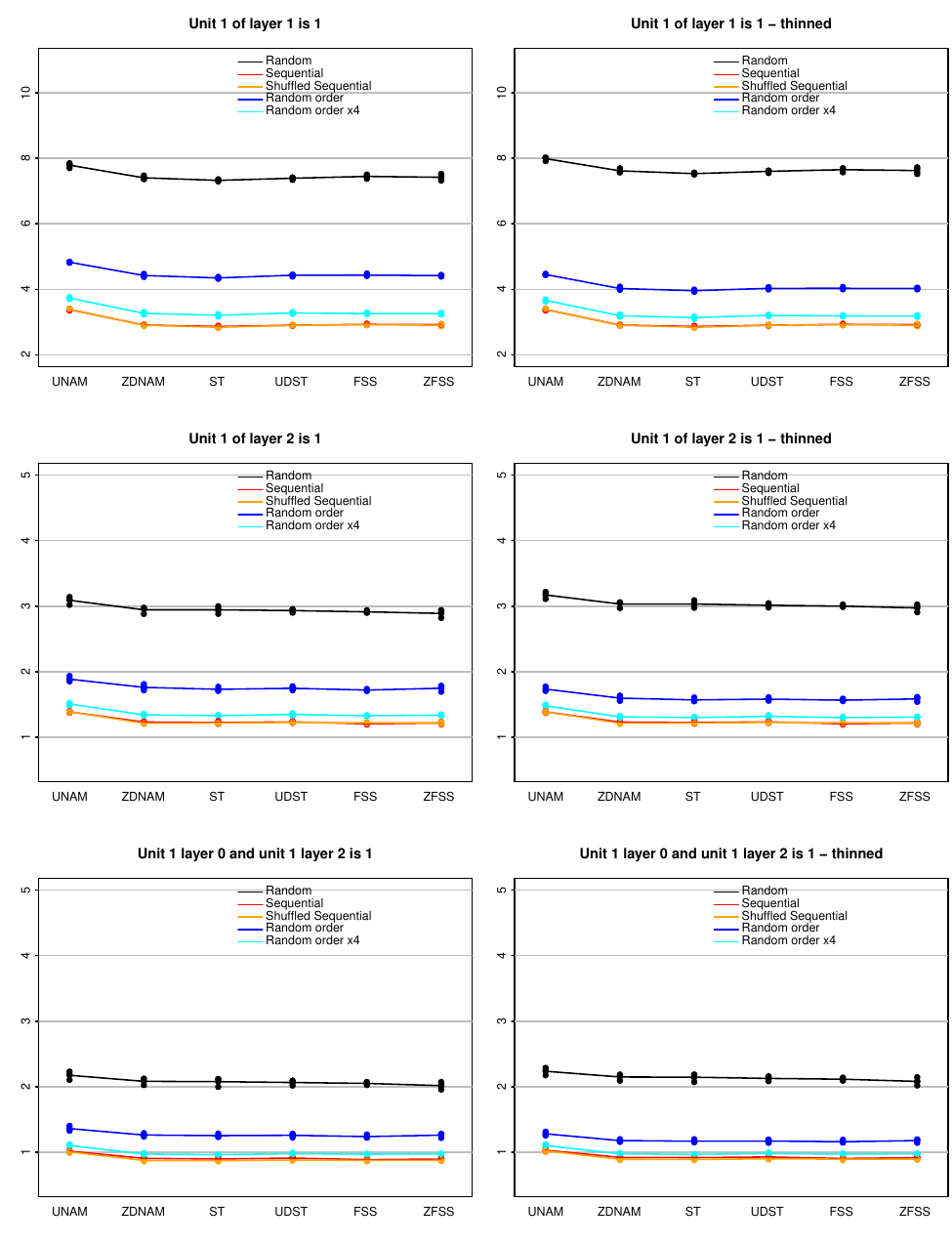}
\end{center}
\caption{Summaries of autocovariance function estimates for the belief
network, for the third group of methods.}\label{fig-bn-g3}
\end{figure}

Summaries of asymptotic variance estimates for the three function
above, for all groups of methods, are shown in Figures~\ref{fig-bn-g1}
through~\ref{fig-bn-g3}.  Note that the results for the sequential
scan (red) and shuffled sequential scan (orange) are almost identical
(with the latter overlaying the former).

The results on the belief network problem are qualitatively quite
similar to those for both the $8\times8$ Potts model and the mixture
model.  The sequential and shuffled sequential scan orders gives the
best results.  Thinning increases asymptotic variance, except for the
random scan, for which thinning is beneficial.  For all scan orders,
with and without thinning, there is almost no difference in asymptotic
variance between DNAM, ZDNAM, FSS, ZFSS, and the shifted tower
methods, all of which are noticeably better than GS, MHGS, UNAM, and
UDNAM.

It is not surprising that all the methods minimizing self transition
probability have nearly the same performance on this problem.  As
noted above, the maximum conditional probability for this problem is
one half or more 89\% of the time, and as shown at the end of
Section~\ref{sec-lim}, in such situations any method that minimizes
self transition probability must have the same transition
probabilities.  There is therefore little scope for differences
amongst these methods, or with DNAM and FSS, both of which almost
minimize self transition probability for this problem.

\section{\hspace*{-8pt}
  Conclusions}\vspace{-11pt}

Liu's (1996) MHGS modification of Gibbs sampling and the
UNAM method due to Frigessi, Hwang, and Younes (1992) and Tjelmeland
(2004) can both be justified as improvements to Gibbs sampling by
applying Peskun's (1973) theorem.  In this paper, I have introduced a
more general class of methods based on nested antithetic modification
(NAM), which can also be shown to efficiency-dominate Gibbs sampling,
using a more general theory, presented in a companion paper (Neal and
Rosenthal 2023).  The DNAM method in this class appears in the
experimental evaluations to usually be superior to UNAM, though this
is not theoretically guaranteed.  The ZDNAM modification to DNAM
reduces self transitions to the minimum possible, and can also be
shown to efficiency-dominate Gibbs sampling, when the variable to
update is chosen randomly.

The minimum possible self transition probability can also be achieved
with the ST method (Suwa and Todo 2010) and the HST method (Suwa
2022).  In this paper, I also consider UST, DST, UDST, and OHST
variations on these methods.  One can show, using theory developed in
(Neal and Rosenthal 2023), that the reversible methods in this class
(UDST, HST, and OHST) cannot be efficiency-dominated by any reversible
method (within the framework of randomly-selected variable updates).
However, unlike ZDNAM, these methods do not always efficiency-dominate
Gibbs sampling.

In this paper, I have also introduced two new non-reversible methods
based on slice sampling, FSS and ZFSS, with the latter minimizing self
transitions.

The experimental evaluations here show that, with Gibbs sampling and
its modifications, random selection of a variable to update is usually
(but not quite always) worse than using other scan orders, such as
sequential updates.  Random updating is necessary for the overall
updates to be reversible, when the modification of Gibbs sampling used
is reversible.  Unfortunately, the theoretical justifications in this
paper apply only to reversible methods, so the practical choice of
method to use when a non-random scan is used must be largely based on
experiment.  However, the experiments do show that the relative
performance of different methods is usually (but not always) similar
for random and systematic scans, so theoretical results for random
scans are still of some interest.

On the four problems looked at, the best overall performance was
achieved using the DNAM, ZDNAM, DST, UST, UDST, and OHST methods (with
DNAM and OHST perhaps being slightly worse than the others).  The ST,
FSS and ZFSS methods also performed well in most circumstances, but
had erratic performance for the $5\times5$ Potts model with
negative~$b$.  The problems with these methods, as well as HST, may be
due to the zero non-self transition probabilities that they can
produce.  DST, UST, UDST, and OHST can also have zero non-self
transition probabilities, but any bad effect of them may be mitigated
by the changing order of values with these methods.  The ZDNAM method
produces zero non-self transition probabilities only in the context of
moving to or from a higher-probability value (as in
equation~(\ref{eq-highpr})), which seems less problematic.  For this
reason, I at present recommend ZDNAM as most suitable for general use.

Amongst the methods minimizing self transitions, these experiments
give no evidence that a non-reversible update method, such as DST or
UST, provides an advantage over reversible methods, such as ZDNAM or
UDST.  In contrast, using a scan order that leads to the overall
method being non-reversible usually has a large advantage.  This
highlights the need for better theoretical tools for analysing
non-reversible methods.

Efficient algorithms for all the methods evaluated are given in this
paper, which I hope will facilitate their use in applications and in
general-purpose MCMC software.  The programs used for the experimental
evaluations, written in R, along with the output files,
are available at \texttt{github.com/radfordneal/gibbsmod}.

\section*{References}\vspace{-11pt}

\leftmargini 0.2in

\begin{description}

\item[\hspace{-5pt}]
  Ackley, D.~H., Hinton, G.~E., and Sejnowski, T.~J.\ (1985)
  ``A learning algorithm for Boltzmann machines'', \textit{Cognitive Science},
  vol.~9, pp.~147--169.

\item[\hspace{-5pt}]
  Devroye, L.\ (1986) \textit{Non-Uniform Random Variate Generation},
  Springer-Verlag.

\item[\hspace{-5pt}]
  Frigessi, A., Hwang, C.-R., and Younes, L.\ (1992) ``Optimal spectral
  structure of reversible stochastic matrices, Monte carlo methods and
  the simulation of Markov random fields'', \textit{The Annals of Applied
  Probability}, vol.~2, pp.~610--628.

\item[\hspace{-5pt}]
  Gelfand, A.~E.~and Smith, A.~F.~M.\ (1990) ``Sampling-based
  approaches to calculating marginal densities'', \textit{Journal
  of the American Statistical Association}, vol.~85, pp.~398--409.

\item[\hspace{-5pt}]
  Geman, S.~and Geman, D.\ (1984) ``Stochastic relaxation, Gibbs  
  distributions and the Bayesian restoration of images'', \textit{IEEE
  Transactions on Pattern Analysis and Machine Intelligence}, vol.~6, 
  pp.~721--741.

\item[\hspace{-5pt}]
  Geyer, C.~J.\ (1992) ``Practical Markov chain Monte Carlo'',
  \textit{Statistical Science}, vol.~7, pp.~473-511.

\item[\hspace{-5pt}]
  Hastings, W.~K.\ (1970) ``Monte Carlo sampling methods using Markov chains 
  and their applications'', \textit{Biometrika}, vol.~57, pp.~97--109.

\item[\hspace{-5pt}]
  He, B., De Sa, C., Mitliagkas, I., and R\'e, C.\ (2016) ``Scan order
  in Gibbs sampling: Models in which it matters and bounds on how much'',
  \texttt{https://arxiv.org/abs/1606.03432}

\item[\hspace{-5pt}]
  Hoffman, M.~D.\ and Gelman, A.\ (2014) ``The No-U-Turn Sampler:
  Adaptively setting path lengths in Hamiltonian Monte Carlo'',
  \textit{Journal of Machine Learning Research}, vol.~15, pp.~1593-1623. 

\item[\hspace{-5pt}]
  Horn, R.~A.\ and Johnson, C.~R.\ (2013) \textit{Matrix Analysis}, 2nd edition,
  Cambridge University Press.

\item[\hspace{-5pt}]
  Landau, D.~P.\ and Binder, K.\ (2009) \textit{A Guide to
  Monte Carlo Simulations in Statistical Physics}, Third Edition,
  Cambridge University Press.

\item[\hspace{-5pt}] Lauritzen, S.~L.\ and Spiegelhalter, D.~J.\ (1988)
  ``Local computations with probabilities on graphical structures and
  their application to expert systems'' (with discussion), {\em Journal
  of the Royal Statistical Society~B}, vol.~50, pp.~157-224.

\item[\hspace{-5pt}]
  Liu, J.~S.\ (1996) ``Peskun's theorem and a modified discrete-state Gibbs
  sampler'', \textit{Biometrika}, vol.~83, pp.~681--682.

\item[\hspace{-5pt}]
  Mira, A.\ and Geyer, C.~J. (1999), ``Ordering Monte Carlo Markov chains''.
  Technical Report No.\ 632, School of Statistics, University of Minnesota.

\item[\hspace{-5pt}]
  Neal, R.~M.\ (1992a) ``Bayesian mixture modelling'', in 
  C.~R.~Smith, G.~J.~Erickson, and P.~O.~Neudorfer (editors) {\em Maximum 
  Entropy and Bayesian Methods: Proceedings of the 11th International Workshop
  on Maximum Entropy and Bayesian Methods of Statistical Analysis,
  Seattle 1991}, pp. 197-211, Dordrecht: Kluwer Academic Publishers.

\item[\hspace{-5pt}]
  Neal, R.~M.\ (1992b) ``Connectionist learning of belief networks'',
  {\em Artificial Intelligence}, vol.~56, pp.~71-113.

\item[\hspace{-5pt}]
  Neal, R.~M.\ (2003) ``Slice sampling'' (with discussion),
  {\em Annals of Statistics}, vol.~31, pp.~705-767.

\item[\hspace{-5pt}]
  Neal, R.~M.\ (2004) ``Improving asymptotic variance of MCMC
  estimators: Non-reversible chains are better'',
  \texttt{https://arxiv.org/abs/math/0407281}


\item[\hspace{-5pt}] Neal, R.~M.\ and Rosenthal, J.~S.\ (2023)
  ``Efficiency of reversible MCMC methods: Elementary derivations and
  applications to composite methods'', 
  \texttt{https://arxiv.org/abs/2305.18268} (revised version of March 2024).

\item[\hspace{-5pt}] Pearl,\ J.\ (1988) {\em 
  Probabilistic Reasoning in Intelligent
  System: Networks of Plausible Inference}, San Mateo,
  California: Morgan Kaufmann.

\item[\hspace{-5pt}]
  Peskun, P.~H.\ (1973) ``Optimum Monte-Carlo sampling using Markov chains'',
  \textit{Biometrika}, vol.~60, pp.~607--612.

\item[\hspace{-5pt}]
  Pollet, L., Rombouts, S.~M.A., Van Houcke, K., and Heyde, K.\ (2004)
  ``Optimal Monte Carlo updating'', 
  \texttt{https://arxiv.org/abs/cond-mat/0405150}

\item[\hspace{-5pt}]
  Suwa, H.\ (2022) ``Reducing rejection exponentially improves Markov
  chain Monte Carlo sampling'', \texttt{https://arxiv.org/abs/2208.03935}

\item[\hspace{-5pt}]
  Suwa, H.\ and Todo S.\ (2010) ``Markov chain Monte Carlo method without
  detailed balance'',\linebreak \texttt{https://arxiv.org/abs/1007.2262}

\item[\hspace{-5pt}]
  Thomas, A., Spiegelhalter, D.~J., and Gilks, W.~R.\ (1992) ``BUGS: A
  program to perform Bayesian inference using Gibbs sampling'', in J.~ M.\
  Bernardo, J.~O.\ Berger, A.~P.\ Dawid, and A.~F.~M.\ Smith (editors),
  \textit{Bayesian Statistics 4}, pp.~837--842, Oxford University Press.

\item[\hspace{-5pt}]
  Tjelmeland, H.\ (2004) ``Using all Metropolis-Hastings proposals to 
  estimate mean values'', Statistics Preprint No.\ 4/2004, Norwegian University
  of Science and Technology.

\end{description}

\end{document}